\documentclass[longauth]{aa}

\usepackage{graphicx}
\usepackage{natbib}
\usepackage{scalerel}
\usepackage{multicol}

\usepackage[table]{xcolor}

\usepackage{txfonts}
\usepackage[pdfencoding=auto,psdextra]{hyperref}
\hypersetup{
    colorlinks=true,
    linkcolor=blue,
    filecolor=magenta,      
    urlcolor=blue,
    citecolor=blue
}
\makeatletter
\renewcommand*\aa@pageof{, page \thepage{} of \pageref*{LastPage}}
\makeatother

\usepackage[utf8]{inputenc}

\usepackage[switch, modulo]{lineno}

\usepackage{euclid}

\begin{document}
%
%

\title{\Euclid\/ preparation} \subtitle{XLVI. The Near-IR Background Dipole Experiment with \Euclid}


\newcommand{\orcid}[1]{} 
\author{Euclid Collaboration: A.~Kashlinsky\thanks{\email{Alexander.Kashlinsky@nasa.gov}}\inst{\ref{aff1},\ref{aff2},\ref{aff3}}
\and R.~G.~Arendt\orcid{0000-0001-8403-8548}\inst{\ref{aff4},\ref{aff1},\ref{aff5}}
\and M.~L.~N.~Ashby\orcid{0000-0002-3993-0745}\inst{\ref{aff6}}
\and F.~Atrio-Barandela\orcid{0000-0002-2130-2513}\inst{\ref{aff7}}
\and R.~Scaramella\orcid{0000-0003-2229-193X}\inst{\ref{aff8},\ref{aff9}}
\and M.~A.~Strauss\orcid{0000-0002-0106-7755}\inst{\ref{aff10}}
\and B.~Altieri\orcid{0000-0003-3936-0284}\inst{\ref{aff11}}
\and A.~Amara\inst{\ref{aff12}}
\and S.~Andreon\orcid{0000-0002-2041-8784}\inst{\ref{aff13}}
\and N.~Auricchio\orcid{0000-0003-4444-8651}\inst{\ref{aff14}}
\and M.~Baldi\orcid{0000-0003-4145-1943}\inst{\ref{aff15},\ref{aff14},\ref{aff16}}
\and S.~Bardelli\orcid{0000-0002-8900-0298}\inst{\ref{aff14}}
\and R.~Bender\orcid{0000-0001-7179-0626}\inst{\ref{aff17},\ref{aff18}}
\and C.~Bodendorf\inst{\ref{aff17}}
\and E.~Branchini\orcid{0000-0002-0808-6908}\inst{\ref{aff19},\ref{aff20},\ref{aff13}}
\and M.~Brescia\orcid{0000-0001-9506-5680}\inst{\ref{aff21},\ref{aff22},\ref{aff23}}
\and J.~Brinchmann\orcid{0000-0003-4359-8797}\inst{\ref{aff24}}
\and S.~Camera\orcid{0000-0003-3399-3574}\inst{\ref{aff25},\ref{aff26},\ref{aff27}}
\and V.~Capobianco\orcid{0000-0002-3309-7692}\inst{\ref{aff27}}
\and C.~Carbone\orcid{0000-0003-0125-3563}\inst{\ref{aff28}}
\and J.~Carretero\orcid{0000-0002-3130-0204}\inst{\ref{aff29},\ref{aff30}}
\and S.~Casas\orcid{0000-0002-4751-5138}\inst{\ref{aff31}}
\and M.~Castellano\orcid{0000-0001-9875-8263}\inst{\ref{aff8}}
\and S.~Cavuoti\orcid{0000-0002-3787-4196}\inst{\ref{aff22},\ref{aff23}}
\and A.~Cimatti\inst{\ref{aff32}}
\and G.~Congedo\orcid{0000-0003-2508-0046}\inst{\ref{aff33}}
\and C.~J.~Conselice\inst{\ref{aff34}}
\and L.~Conversi\orcid{0000-0002-6710-8476}\inst{\ref{aff35},\ref{aff11}}
\and Y.~Copin\orcid{0000-0002-5317-7518}\inst{\ref{aff36}}
\and L.~Corcione\orcid{0000-0002-6497-5881}\inst{\ref{aff27}}
\and F.~Courbin\orcid{0000-0003-0758-6510}\inst{\ref{aff37}}
\and H.~M.~Courtois\orcid{0000-0003-0509-1776}\inst{\ref{aff38}}
\and A.~Da~Silva\orcid{0000-0002-6385-1609}\inst{\ref{aff39},\ref{aff40}}
\and H.~Degaudenzi\orcid{0000-0002-5887-6799}\inst{\ref{aff41}}
\and A.~M.~Di~Giorgio\inst{\ref{aff42}}
\and J.~Dinis\inst{\ref{aff40},\ref{aff39}}
\and F.~Dubath\orcid{0000-0002-6533-2810}\inst{\ref{aff41}}
\and X.~Dupac\inst{\ref{aff11}}
\and S.~Dusini\orcid{0000-0002-1128-0664}\inst{\ref{aff43}}
\and A.~Ealet\inst{\ref{aff44}}
\and M.~Farina\orcid{0000-0002-3089-7846}\inst{\ref{aff42}}
\and S.~Farrens\orcid{0000-0002-9594-9387}\inst{\ref{aff45}}
\and S.~Ferriol\inst{\ref{aff36}}
\and M.~Frailis\orcid{0000-0002-7400-2135}\inst{\ref{aff46}}
\and E.~Franceschi\orcid{0000-0002-0585-6591}\inst{\ref{aff14}}
\and S.~Galeotta\orcid{0000-0002-3748-5115}\inst{\ref{aff46}}
\and B.~Gillis\orcid{0000-0002-4478-1270}\inst{\ref{aff33}}
\and C.~Giocoli\orcid{0000-0002-9590-7961}\inst{\ref{aff14},\ref{aff47}}
\and A.~Grazian\orcid{0000-0002-5688-0663}\inst{\ref{aff48}}
\and F.~Grupp\inst{\ref{aff17},\ref{aff49}}
\and S.~V.~H.~Haugan\orcid{0000-0001-9648-7260}\inst{\ref{aff50}}
\and I.~Hook\inst{\ref{aff51}}
\and F.~Hormuth\inst{\ref{aff52}}
\and A.~Hornstrup\orcid{0000-0002-3363-0936}\inst{\ref{aff53},\ref{aff54}}
\and K.~Jahnke\orcid{0000-0003-3804-2137}\inst{\ref{aff55}}
\and E.~Keih\"anen\orcid{0000-0003-1804-7715}\inst{\ref{aff56}}
\and S.~Kermiche\orcid{0000-0002-0302-5735}\inst{\ref{aff57}}
\and A.~Kiessling\orcid{0000-0002-2590-1273}\inst{\ref{aff58}}
\and M.~Kilbinger\orcid{0000-0001-9513-7138}\inst{\ref{aff59}}
\and B.~Kubik\inst{\ref{aff36}}
\and M.~Kunz\orcid{0000-0002-3052-7394}\inst{\ref{aff60}}
\and H.~Kurki-Suonio\orcid{0000-0002-4618-3063}\inst{\ref{aff61},\ref{aff62}}
\and S.~Ligori\orcid{0000-0003-4172-4606}\inst{\ref{aff27}}
\and P.~B.~Lilje\orcid{0000-0003-4324-7794}\inst{\ref{aff50}}
\and V.~Lindholm\orcid{0000-0003-2317-5471}\inst{\ref{aff61},\ref{aff62}}
\and I.~Lloro\inst{\ref{aff63}}
\and D.~Maino\inst{\ref{aff64},\ref{aff28},\ref{aff65}}
\and E.~Maiorano\orcid{0000-0003-2593-4355}\inst{\ref{aff14}}
\and O.~Mansutti\orcid{0000-0001-5758-4658}\inst{\ref{aff46}}
\and O.~Marggraf\orcid{0000-0001-7242-3852}\inst{\ref{aff66}}
\and K.~Markovic\orcid{0000-0001-6764-073X}\inst{\ref{aff58}}
\and N.~Martinet\orcid{0000-0003-2786-7790}\inst{\ref{aff67}}
\and F.~Marulli\orcid{0000-0002-8850-0303}\inst{\ref{aff68},\ref{aff14},\ref{aff16}}
\and R.~Massey\orcid{0000-0002-6085-3780}\inst{\ref{aff69}}
\and S.~Maurogordato\inst{\ref{aff70}}
\and H.~J.~McCracken\orcid{0000-0002-9489-7765}\inst{\ref{aff71}}
\and E.~Medinaceli\orcid{0000-0002-4040-7783}\inst{\ref{aff14}}
\and S.~Mei\orcid{0000-0002-2849-559X}\inst{\ref{aff72}}
\and Y.~Mellier\inst{\ref{aff73},\ref{aff71}}
\and M.~Meneghetti\orcid{0000-0003-1225-7084}\inst{\ref{aff14},\ref{aff16}}
\and G.~Meylan\inst{\ref{aff37}}
\and M.~Moresco\orcid{0000-0002-7616-7136}\inst{\ref{aff68},\ref{aff14}}
\and L.~Moscardini\orcid{0000-0002-3473-6716}\inst{\ref{aff68},\ref{aff14},\ref{aff16}}
\and E.~Munari\orcid{0000-0002-1751-5946}\inst{\ref{aff46}}
\and S.-M.~Niemi\inst{\ref{aff74}}
\and C.~Padilla\orcid{0000-0001-7951-0166}\inst{\ref{aff29}}
\and S.~Paltani\inst{\ref{aff41}}
\and F.~Pasian\inst{\ref{aff46}}
\and K.~Pedersen\inst{\ref{aff75}}
\and W.~J.~Percival\orcid{0000-0002-0644-5727}\inst{\ref{aff76},\ref{aff77},\ref{aff78}}
\and S.~Pires\orcid{0000-0002-0249-2104}\inst{\ref{aff45}}
\and G.~Polenta\orcid{0000-0003-4067-9196}\inst{\ref{aff79}}
\and M.~Poncet\inst{\ref{aff80}}
\and L.~A.~Popa\inst{\ref{aff81}}
\and F.~Raison\orcid{0000-0002-7819-6918}\inst{\ref{aff17}}
\and A.~Renzi\orcid{0000-0001-9856-1970}\inst{\ref{aff82},\ref{aff43}}
\and J.~Rhodes\inst{\ref{aff58}}
\and G.~Riccio\inst{\ref{aff22}}
\and E.~Romelli\orcid{0000-0003-3069-9222}\inst{\ref{aff46}}
\and M.~Roncarelli\orcid{0000-0001-9587-7822}\inst{\ref{aff14}}
\and E.~Rossetti\inst{\ref{aff15}}
\and R.~Saglia\orcid{0000-0003-0378-7032}\inst{\ref{aff18},\ref{aff17}}
\and D.~Sapone\orcid{0000-0001-7089-4503}\inst{\ref{aff83}}
\and B.~Sartoris\inst{\ref{aff18},\ref{aff46}}
\and M.~Schirmer\orcid{0000-0003-2568-9994}\inst{\ref{aff55}}
\and P.~Schneider\orcid{0000-0001-8561-2679}\inst{\ref{aff66}}
\and T.~Schrabback\orcid{0000-0002-6987-7834}\inst{\ref{aff84}}
\and A.~Secroun\orcid{0000-0003-0505-3710}\inst{\ref{aff57}}
\and G.~Seidel\orcid{0000-0003-2907-353X}\inst{\ref{aff55}}
\and M.~Seiffert\orcid{0000-0002-7536-9393}\inst{\ref{aff58}}
\and S.~Serrano\orcid{0000-0002-0211-2861}\inst{\ref{aff85},\ref{aff86},\ref{aff87}}
\and C.~Sirignano\orcid{0000-0002-0995-7146}\inst{\ref{aff82},\ref{aff43}}
\and G.~Sirri\orcid{0000-0003-2626-2853}\inst{\ref{aff16}}
\and L.~Stanco\orcid{0000-0002-9706-5104}\inst{\ref{aff43}}
\and C.~Surace\inst{\ref{aff67}}
\and P.~Tallada-Cresp\'{i}\orcid{0000-0002-1336-8328}\inst{\ref{aff88},\ref{aff30}}
\and A.~N.~Taylor\inst{\ref{aff33}}
\and H.~I.~Teplitz\orcid{0000-0002-7064-5424}\inst{\ref{aff89}}
\and I.~Tereno\inst{\ref{aff39},\ref{aff90}}
\and R.~Toledo-Moreo\orcid{0000-0002-2997-4859}\inst{\ref{aff91}}
\and F.~Torradeflot\orcid{0000-0003-1160-1517}\inst{\ref{aff30},\ref{aff88}}
\and I.~Tutusaus\orcid{0000-0002-3199-0399}\inst{\ref{aff92}}
\and L.~Valenziano\orcid{0000-0002-1170-0104}\inst{\ref{aff14},\ref{aff93}}
\and T.~Vassallo\orcid{0000-0001-6512-6358}\inst{\ref{aff18},\ref{aff46}}
\and A.~Veropalumbo\orcid{0000-0003-2387-1194}\inst{\ref{aff13},\ref{aff20}}
\and Y.~Wang\orcid{0000-0002-4749-2984}\inst{\ref{aff89}}
\and G.~Zamorani\orcid{0000-0002-2318-301X}\inst{\ref{aff14}}
\and J.~Zoubian\inst{\ref{aff57}}
\and E.~Zucca\orcid{0000-0002-5845-8132}\inst{\ref{aff14}}
\and A.~Biviano\orcid{0000-0002-0857-0732}\inst{\ref{aff46},\ref{aff94}}
\and E.~Bozzo\orcid{0000-0002-8201-1525}\inst{\ref{aff41}}
\and C.~Burigana\orcid{0000-0002-3005-5796}\inst{\ref{aff95},\ref{aff93}}
\and C.~Colodro-Conde\inst{\ref{aff96}}
\and D.~Di~Ferdinando\inst{\ref{aff16}}
\and G.~Fabbian\orcid{0000-0002-3255-4695}\inst{\ref{aff97},\ref{aff98}}
\and R.~Farinelli\inst{\ref{aff14}}
\and J.~Graci\'{a}-Carpio\inst{\ref{aff17}}
\and G.~Mainetti\inst{\ref{aff99}}
\and M.~Martinelli\orcid{0000-0002-6943-7732}\inst{\ref{aff8},\ref{aff9}}
\and N.~Mauri\orcid{0000-0001-8196-1548}\inst{\ref{aff32},\ref{aff16}}
\and C.~Neissner\inst{\ref{aff29},\ref{aff30}}
\and Z.~Sakr\orcid{0000-0002-4823-3757}\inst{\ref{aff100},\ref{aff92},\ref{aff101}}
\and V.~Scottez\inst{\ref{aff73},\ref{aff102}}
\and M.~Tenti\orcid{0000-0002-4254-5901}\inst{\ref{aff16}}
\and M.~Viel\orcid{0000-0002-2642-5707}\inst{\ref{aff94},\ref{aff46},\ref{aff103},\ref{aff104},\ref{aff105}}
\and M.~Wiesmann\orcid{0009-0000-8199-5860}\inst{\ref{aff50}}
\and Y.~Akrami\orcid{0000-0002-2407-7956}\inst{\ref{aff106},\ref{aff107}}
\and V.~Allevato\orcid{0000-0001-7232-5152}\inst{\ref{aff22}}
\and S.~Anselmi\orcid{0000-0002-3579-9583}\inst{\ref{aff43},\ref{aff82},\ref{aff108}}
\and C.~Baccigalupi\orcid{0000-0002-8211-1630}\inst{\ref{aff103},\ref{aff46},\ref{aff104},\ref{aff94}}
\and M.~Ballardini\orcid{0000-0003-4481-3559}\inst{\ref{aff109},\ref{aff110},\ref{aff14}}
\and A.~Blanchard\orcid{0000-0001-8555-9003}\inst{\ref{aff92}}
\and S.~Borgani\orcid{0000-0001-6151-6439}\inst{\ref{aff111},\ref{aff94},\ref{aff46},\ref{aff104}}
\and A.~S.~Borlaff\orcid{0000-0003-3249-4431}\inst{\ref{aff112},\ref{aff113},\ref{aff114}}
\and S.~Bruton\orcid{0000-0002-6503-5218}\inst{\ref{aff115}}
\and R.~Cabanac\orcid{0000-0001-6679-2600}\inst{\ref{aff92}}
\and A.~Cappi\inst{\ref{aff14},\ref{aff70}}
\and C.~S.~Carvalho\inst{\ref{aff90}}
\and G.~Castignani\orcid{0000-0001-6831-0687}\inst{\ref{aff68},\ref{aff14}}
\and T.~Castro\orcid{0000-0002-6292-3228}\inst{\ref{aff46},\ref{aff104},\ref{aff94},\ref{aff105}}
\and G.~Ca\~{n}as-Herrera\orcid{0000-0003-2796-2149}\inst{\ref{aff74},\ref{aff116}}
\and K.~C.~Chambers\orcid{0000-0001-6965-7789}\inst{\ref{aff117}}
\and S.~Contarini\inst{\ref{aff68},\ref{aff16},\ref{aff14}}
\and J.~Coupon\inst{\ref{aff41}}
\and G.~De~Lucia\orcid{0000-0002-6220-9104}\inst{\ref{aff46}}
\and G.~Desprez\inst{\ref{aff118}}
\and S.~Di~Domizio\orcid{0000-0003-2863-5895}\inst{\ref{aff19},\ref{aff20}}
\and H.~Dole\orcid{0000-0002-9767-3839}\inst{\ref{aff119}}
\and A.~D\'{i}az-S\'{a}nchez\orcid{0000-0003-0748-4768}\inst{\ref{aff120}}
\and J.~A.~Escartin~Vigo\inst{\ref{aff17}}
\and I.~Ferrero\orcid{0000-0002-1295-1132}\inst{\ref{aff50}}
\and F.~Finelli\orcid{0000-0002-6694-3269}\inst{\ref{aff14},\ref{aff93}}
\and L.~Gabarra\inst{\ref{aff121}}
\and J.~Garc\'ia-Bellido\orcid{0000-0002-9370-8360}\inst{\ref{aff106}}
\and V.~Gautard\inst{\ref{aff122}}
\and E.~Gaztanaga\orcid{0000-0001-9632-0815}\inst{\ref{aff86},\ref{aff85},\ref{aff123}}
\and K.~George\orcid{0000-0002-1734-8455}\inst{\ref{aff18}}
\and F.~Giacomini\orcid{0000-0002-3129-2814}\inst{\ref{aff16}}
\and G.~Gozaliasl\orcid{0000-0002-0236-919X}\inst{\ref{aff124},\ref{aff61}}
\and A.~Gregorio\orcid{0000-0003-4028-8785}\inst{\ref{aff111},\ref{aff46},\ref{aff104}}
\and A.~Hall\orcid{0000-0002-3139-8651}\inst{\ref{aff33}}
\and H.~Hildebrandt\orcid{0000-0002-9814-3338}\inst{\ref{aff125}}
\and J.~J.~E.~Kajava\orcid{0000-0002-3010-8333}\inst{\ref{aff126},\ref{aff127}}
\and V.~Kansal\inst{\ref{aff128},\ref{aff129},\ref{aff130}}
\and C.~C.~Kirkpatrick\inst{\ref{aff56}}
\and L.~Legrand\orcid{0000-0003-0610-5252}\inst{\ref{aff60},\ref{aff131}}
\and A.~Loureiro\orcid{0000-0002-4371-0876}\inst{\ref{aff132},\ref{aff133}}
\and M.~Magliocchetti\orcid{0000-0001-9158-4838}\inst{\ref{aff42}}
\and F.~Mannucci\orcid{0000-0002-4803-2381}\inst{\ref{aff134}}
\and R.~Maoli\orcid{0000-0002-6065-3025}\inst{\ref{aff135},\ref{aff8}}
\and C.~J.~A.~P.~Martins\orcid{0000-0002-4886-9261}\inst{\ref{aff136},\ref{aff24}}
\and S.~Matthew\inst{\ref{aff33}}
\and L.~Maurin\orcid{0000-0002-8406-0857}\inst{\ref{aff119}}
\and R.~B.~Metcalf\orcid{0000-0003-3167-2574}\inst{\ref{aff68},\ref{aff14}}
\and M.~Migliaccio\inst{\ref{aff137},\ref{aff138}}
\and P.~Monaco\orcid{0000-0003-2083-7564}\inst{\ref{aff111},\ref{aff46},\ref{aff104},\ref{aff94}}
\and G.~Morgante\inst{\ref{aff14}}
\and S.~Nadathur\orcid{0000-0001-9070-3102}\inst{\ref{aff123}}
\and Nicholas~A.~Walton\orcid{0000-0003-3983-8778}\inst{\ref{aff139}}
\and L.~Patrizii\inst{\ref{aff16}}
\and V.~Popa\inst{\ref{aff81}}
\and D.~Potter\orcid{0000-0002-0757-5195}\inst{\ref{aff140}}
\and M.~P\"{o}ntinen\orcid{0000-0001-5442-2530}\inst{\ref{aff61}}
\and P.-F.~Rocci\inst{\ref{aff119}}
\and M.~Sahl\'en\inst{\ref{aff141}}
\and A.~Schneider\orcid{0000-0001-7055-8104}\inst{\ref{aff140}}
\and E.~Sefusatti\orcid{0000-0003-0473-1567}\inst{\ref{aff46},\ref{aff94},\ref{aff104}}
\and M.~Sereno\orcid{0000-0003-0302-0325}\inst{\ref{aff14},\ref{aff16}}
\and J.~Steinwagner\inst{\ref{aff17}}
\and G.~Testera\inst{\ref{aff20}}
\and R.~Teyssier\orcid{0000-0001-7689-0933}\inst{\ref{aff10}}
\and S.~Toft\inst{\ref{aff54},\ref{aff142},\ref{aff143}}
\and S.~Tosi\orcid{0000-0002-7275-9193}\inst{\ref{aff19},\ref{aff20},\ref{aff13}}
\and A.~Troja\orcid{0000-0003-0239-4595}\inst{\ref{aff82},\ref{aff43}}
\and M.~Tucci\inst{\ref{aff41}}
\and J.~Valiviita\orcid{0000-0001-6225-3693}\inst{\ref{aff61},\ref{aff62}}
\and D.~Vergani\orcid{0000-0003-0898-2216}\inst{\ref{aff14}}
\and G.~Verza\orcid{0000-0002-1886-8348}\inst{\ref{aff144},\ref{aff97}}
\and G.~Hasinger\orcid{0000-0002-0797-0646}\inst{\ref{aff145},\ref{aff146}}}
										   
\institute{Code 665, NASA/GSFC, 8800 Greenbelt Road, Greenbelt, MD 20771, USA\label{aff1}
\and
SSAI, Lanham, MD 20706, USA\label{aff2}
\and
Department of Astronomy, University of Maryland, College Park, MD 20742, USA\label{aff3}
\and
Center for Space Sciences and Technology, University of Maryland, Baltimore County, Baltimore, MD 21250, USA\label{aff4}
\and
Center for Research and Exploration in Space Science and Technology, NASA/GSFC, Greenbelt, MD 20771, USA\label{aff5}
\and
Center for Astrophysics | Harvard \& Smithsonian, 60 Garden St., Cambridge, MA 02138, USA\label{aff6}
\and
Departamento de F{\'\i}sica Fundamental. Universidad de Salamanca.Plaza de la Merced s/n. 37008 Salamanca, Spain\label{aff7}
\and
INAF-Osservatorio Astronomico di Roma, Via Frascati 33, 00078 Monteporzio Catone, Italy\label{aff8}
\and
INFN-Sezione di Roma, Piazzale Aldo Moro, 2 - c/o Dipartimento di Fisica, Edificio G. Marconi, 00185 Roma, Italy\label{aff9}
\and
Department of Astrophysical Sciences, Peyton Hall, Princeton University, Princeton, NJ 08544, USA\label{aff10}
\and
ESAC/ESA, Camino Bajo del Castillo, s/n., Urb. Villafranca del Castillo, 28692 Villanueva de la Ca\~nada, Madrid, Spain\label{aff11}
\and
School of Mathematics and Physics, University of Surrey, Guildford, Surrey, GU2 7XH, UK\label{aff12}
\and
INAF-Osservatorio Astronomico di Brera, Via Brera 28, 20122 Milano, Italy\label{aff13}
\and
INAF-Osservatorio di Astrofisica e Scienza dello Spazio di Bologna, Via Piero Gobetti 93/3, 40129 Bologna, Italy\label{aff14}
\and
Dipartimento di Fisica e Astronomia, Universit\`a di Bologna, Via Gobetti 93/2, 40129 Bologna, Italy\label{aff15}
\and
INFN-Sezione di Bologna, Viale Berti Pichat 6/2, 40127 Bologna, Italy\label{aff16}
\and
Max Planck Institute for Extraterrestrial Physics, Giessenbachstr. 1, 85748 Garching, Germany\label{aff17}
\and
Universit\"ats-Sternwarte M\"unchen, Fakult\"at f\"ur Physik, Ludwig-Maximilians-Universit\"at M\"unchen, Scheinerstrasse 1, 81679 M\"unchen, Germany\label{aff18}
\and
Dipartimento di Fisica, Universit\`a di Genova, Via Dodecaneso 33, 16146, Genova, Italy\label{aff19}
\and
INFN-Sezione di Genova, Via Dodecaneso 33, 16146, Genova, Italy\label{aff20}
\and
Department of Physics "E. Pancini", University Federico II, Via Cinthia 6, 80126, Napoli, Italy\label{aff21}
\and
INAF-Osservatorio Astronomico di Capodimonte, Via Moiariello 16, 80131 Napoli, Italy\label{aff22}
\and
INFN section of Naples, Via Cinthia 6, 80126, Napoli, Italy\label{aff23}
\and
Instituto de Astrof\'isica e Ci\^encias do Espa\c{c}o, Universidade do Porto, CAUP, Rua das Estrelas, PT4150-762 Porto, Portugal\label{aff24}
\and
Dipartimento di Fisica, Universit\`a degli Studi di Torino, Via P. Giuria 1, 10125 Torino, Italy\label{aff25}
\and
INFN-Sezione di Torino, Via P. Giuria 1, 10125 Torino, Italy\label{aff26}
\and
INAF-Osservatorio Astrofisico di Torino, Via Osservatorio 20, 10025 Pino Torinese (TO), Italy\label{aff27}
\and
INAF-IASF Milano, Via Alfonso Corti 12, 20133 Milano, Italy\label{aff28}
\and
Institut de F\'{i}sica d'Altes Energies (IFAE), The Barcelona Institute of Science and Technology, Campus UAB, 08193 Bellaterra (Barcelona), Spain\label{aff29}
\and
Port d'Informaci\'{o} Cient\'{i}fica, Campus UAB, C. Albareda s/n, 08193 Bellaterra (Barcelona), Spain\label{aff30}
\and
Institute for Theoretical Particle Physics and Cosmology (TTK), RWTH Aachen University, 52056 Aachen, Germany\label{aff31}
\and
Dipartimento di Fisica e Astronomia "Augusto Righi" - Alma Mater Studiorum Universit\`a di Bologna, Viale Berti Pichat 6/2, 40127 Bologna, Italy\label{aff32}
\and
Institute for Astronomy, University of Edinburgh, Royal Observatory, Blackford Hill, Edinburgh EH9 3HJ, UK\label{aff33}
\and
Jodrell Bank Centre for Astrophysics, Department of Physics and Astronomy, University of Manchester, Oxford Road, Manchester M13 9PL, UK\label{aff34}
\and
European Space Agency/ESRIN, Largo Galileo Galilei 1, 00044 Frascati, Roma, Italy\label{aff35}
\and
University of Lyon, Univ Claude Bernard Lyon 1, CNRS/IN2P3, IP2I Lyon, UMR 5822, 69622 Villeurbanne, France\label{aff36}
\and
Institute of Physics, Laboratory of Astrophysics, Ecole Polytechnique F\'ed\'erale de Lausanne (EPFL), Observatoire de Sauverny, 1290 Versoix, Switzerland\label{aff37}
\and
UCB Lyon 1, CNRS/IN2P3, IUF, IP2I Lyon, 4 rue Enrico Fermi, 69622 Villeurbanne, France\label{aff38}
\and
Departamento de F\'isica, Faculdade de Ci\^encias, Universidade de Lisboa, Edif\'icio C8, Campo Grande, PT1749-016 Lisboa, Portugal\label{aff39}
\and
Instituto de Astrof\'isica e Ci\^encias do Espa\c{c}o, Faculdade de Ci\^encias, Universidade de Lisboa, Campo Grande, 1749-016 Lisboa, Portugal\label{aff40}
\and
Department of Astronomy, University of Geneva, ch. d'Ecogia 16, 1290 Versoix, Switzerland\label{aff41}
\and
INAF-Istituto di Astrofisica e Planetologia Spaziali, via del Fosso del Cavaliere, 100, 00100 Roma, Italy\label{aff42}
\and
INFN-Padova, Via Marzolo 8, 35131 Padova, Italy\label{aff43}
\and
Univ Claude Bernard Lyon 1, CNRS, IP2I Lyon, UMR 5822, 69622 Villeurbanne, France\label{aff44}
\and
Universit\'e Paris-Saclay, Universit\'e Paris Cit\'e, CEA, CNRS, AIM, 91191, Gif-sur-Yvette, France\label{aff45}
\and
INAF-Osservatorio Astronomico di Trieste, Via G. B. Tiepolo 11, 34143 Trieste, Italy\label{aff46}
\and
Istituto Nazionale di Fisica Nucleare, Sezione di Bologna, Via Irnerio 46, 40126 Bologna, Italy\label{aff47}
\and
INAF-Osservatorio Astronomico di Padova, Via dell'Osservatorio 5, 35122 Padova, Italy\label{aff48}
\and
University Observatory, Faculty of Physics, Ludwig-Maximilians-Universit{\"a}t, Scheinerstr. 1, 81679 Munich, Germany\label{aff49}
\and
Institute of Theoretical Astrophysics, University of Oslo, P.O. Box 1029 Blindern, 0315 Oslo, Norway\label{aff50}
\and
Department of Physics, Lancaster University, Lancaster, LA1 4YB, UK\label{aff51}
\and
von Hoerner \& Sulger GmbH, Schlo{\ss}Platz 8, 68723 Schwetzingen, Germany\label{aff52}
\and
Technical University of Denmark, Elektrovej 327, 2800 Kgs. Lyngby, Denmark\label{aff53}
\and
Cosmic Dawn Center (DAWN), Denmark\label{aff54}
\and
Max-Planck-Institut f\"ur Astronomie, K\"onigstuhl 17, 69117 Heidelberg, Germany\label{aff55}
\and
Department of Physics and Helsinki Institute of Physics, Gustaf H\"allstr\"omin katu 2, 00014 University of Helsinki, Finland\label{aff56}
\and
Aix-Marseille Universit\'e, CNRS/IN2P3, CPPM, Marseille, France\label{aff57}
\and
Jet Propulsion Laboratory, California Institute of Technology, 4800 Oak Grove Drive, Pasadena, CA, 91109, USA\label{aff58}
\and
AIM, CEA, CNRS, Universit\'{e} Paris-Saclay, Universit\'{e} de Paris, 91191 Gif-sur-Yvette, France\label{aff59}
\and
Universit\'e de Gen\`eve, D\'epartement de Physique Th\'eorique and Centre for Astroparticle Physics, 24 quai Ernest-Ansermet, CH-1211 Gen\`eve 4, Switzerland\label{aff60}
\and
Department of Physics, P.O. Box 64, 00014 University of Helsinki, Finland\label{aff61}
\and
Helsinki Institute of Physics, Gustaf H{\"a}llstr{\"o}min katu 2, University of Helsinki, Helsinki, Finland\label{aff62}
\and
NOVA optical infrared instrumentation group at ASTRON, Oude Hoogeveensedijk 4, 7991PD, Dwingeloo, The Netherlands\label{aff63}
\and
Dipartimento di Fisica "Aldo Pontremoli", Universit\`a degli Studi di Milano, Via Celoria 16, 20133 Milano, Italy\label{aff64}
\and
INFN-Sezione di Milano, Via Celoria 16, 20133 Milano, Italy\label{aff65}
\and
Universit\"at Bonn, Argelander-Institut f\"ur Astronomie, Auf dem H\"ugel 71, 53121 Bonn, Germany\label{aff66}
\and
Aix-Marseille Universit\'e, CNRS, CNES, LAM, Marseille, France\label{aff67}
\and
Dipartimento di Fisica e Astronomia "Augusto Righi" - Alma Mater Studiorum Universit\`a di Bologna, via Piero Gobetti 93/2, 40129 Bologna, Italy\label{aff68}
\and
Department of Physics, Institute for Computational Cosmology, Durham University, South Road, DH1 3LE, UK\label{aff69}
\and
Universit\'e C\^{o}te d'Azur, Observatoire de la C\^{o}te d'Azur, CNRS, Laboratoire Lagrange, Bd de l'Observatoire, CS 34229, 06304 Nice cedex 4, France\label{aff70}
\and
Institut d'Astrophysique de Paris, UMR 7095, CNRS, and Sorbonne Universit\'e, 98 bis boulevard Arago, 75014 Paris, France\label{aff71}
\and
Universit\'e Paris Cit\'e, CNRS, Astroparticule et Cosmologie, 75013 Paris, France\label{aff72}
\and
Institut d'Astrophysique de Paris, 98bis Boulevard Arago, 75014, Paris, France\label{aff73}
\and
European Space Agency/ESTEC, Keplerlaan 1, 2201 AZ Noordwijk, The Netherlands\label{aff74}
\and
Department of Physics and Astronomy, University of Aarhus, Ny Munkegade 120, DK-8000 Aarhus C, Denmark\label{aff75}
\and
Centre for Astrophysics, University of Waterloo, Waterloo, Ontario N2L 3G1, Canada\label{aff76}
\and
Department of Physics and Astronomy, University of Waterloo, Waterloo, Ontario N2L 3G1, Canada\label{aff77}
\and
Perimeter Institute for Theoretical Physics, Waterloo, Ontario N2L 2Y5, Canada\label{aff78}
\and
Space Science Data Center, Italian Space Agency, via del Politecnico snc, 00133 Roma, Italy\label{aff79}
\and
Centre National d'Etudes Spatiales -- Centre spatial de Toulouse, 18 avenue Edouard Belin, 31401 Toulouse Cedex 9, France\label{aff80}
\and
Institute of Space Science, Str. Atomistilor, nr. 409 M\u{a}gurele, Ilfov, 077125, Romania\label{aff81}
\and
Dipartimento di Fisica e Astronomia "G. Galilei", Universit\`a di Padova, Via Marzolo 8, 35131 Padova, Italy\label{aff82}
\and
Departamento de F\'isica, FCFM, Universidad de Chile, Blanco Encalada 2008, Santiago, Chile\label{aff83}
\and
Universit\"at Innsbruck, Institut f\"ur Astro- und Teilchenphysik, Technikerstr. 25/8, 6020 Innsbruck, Austria\label{aff84}
\and
Institut d'Estudis Espacials de Catalunya (IEEC), Carrer Gran Capit\'a 2-4, 08034 Barcelona, Spain\label{aff85}
\and
Institute of Space Sciences (ICE, CSIC), Campus UAB, Carrer de Can Magrans, s/n, 08193 Barcelona, Spain\label{aff86}
\and
Satlantis, University Science Park, Sede Bld 48940, Leioa-Bilbao, Spain\label{aff87}
\and
Centro de Investigaciones Energ\'eticas, Medioambientales y Tecnol\'ogicas (CIEMAT), Avenida Complutense 40, 28040 Madrid, Spain\label{aff88}
\and
Infrared Processing and Analysis Center, California Institute of Technology, Pasadena, CA 91125, USA\label{aff89}
\and
Instituto de Astrof\'isica e Ci\^encias do Espa\c{c}o, Faculdade de Ci\^encias, Universidade de Lisboa, Tapada da Ajuda, 1349-018 Lisboa, Portugal\label{aff90}
\and
Universidad Polit\'ecnica de Cartagena, Departamento de Electr\'onica y Tecnolog\'ia de Computadoras,  Plaza del Hospital 1, 30202 Cartagena, Spain\label{aff91}
\and
Institut de Recherche en Astrophysique et Plan\'etologie (IRAP), Universit\'e de Toulouse, CNRS, UPS, CNES, 14 Av. Edouard Belin, 31400 Toulouse, France\label{aff92}
\and
INFN-Bologna, Via Irnerio 46, 40126 Bologna, Italy\label{aff93}
\and
IFPU, Institute for Fundamental Physics of the Universe, via Beirut 2, 34151 Trieste, Italy\label{aff94}
\and
INAF, Istituto di Radioastronomia, Via Piero Gobetti 101, 40129 Bologna, Italy\label{aff95}
\and
Instituto de Astrof\'isica de Canarias, Calle V\'ia L\'actea s/n, 38204, San Crist\'obal de La Laguna, Tenerife, Spain\label{aff96}
\and
Center for Computational Astrophysics, Flatiron Institute, 162 5th Avenue, 10010, New York, NY, USA\label{aff97}
\and
School of Physics and Astronomy, Cardiff University, The Parade, Cardiff, CF24 3AA, UK\label{aff98}
\and
Centre de Calcul de l'IN2P3/CNRS, 21 avenue Pierre de Coubertin 69627 Villeurbanne Cedex, France\label{aff99}
\and
Institut f\"ur Theoretische Physik, University of Heidelberg, Philosophenweg 16, 69120 Heidelberg, Germany\label{aff100}
\and
Universit\'e St Joseph; Faculty of Sciences, Beirut, Lebanon\label{aff101}
\and
Junia, EPA department, 41 Bd Vauban, 59800 Lille, France\label{aff102}
\and
SISSA, International School for Advanced Studies, Via Bonomea 265, 34136 Trieste TS, Italy\label{aff103}
\and
INFN, Sezione di Trieste, Via Valerio 2, 34127 Trieste TS, Italy\label{aff104}
\and
ICSC - Centro Nazionale di Ricerca in High Performance Computing, Big Data e Quantum Computing, Via Magnanelli 2, Bologna, Italy\label{aff105}
\and
Instituto de F\'isica Te\'orica UAM-CSIC, Campus de Cantoblanco, 28049 Madrid, Spain\label{aff106}
\and
CERCA/ISO, Department of Physics, Case Western Reserve University, 10900 Euclid Avenue, Cleveland, OH 44106, USA\label{aff107}
\and
Laboratoire Univers et Th\'eorie, Observatoire de Paris, Universit\'e PSL, Universit\'e Paris Cit\'e, CNRS, 92190 Meudon, France\label{aff108}
\and
Dipartimento di Fisica e Scienze della Terra, Universit\`a degli Studi di Ferrara, Via Giuseppe Saragat 1, 44122 Ferrara, Italy\label{aff109}
\and
Istituto Nazionale di Fisica Nucleare, Sezione di Ferrara, Via Giuseppe Saragat 1, 44122 Ferrara, Italy\label{aff110}
\and
Dipartimento di Fisica - Sezione di Astronomia, Universit\`a di Trieste, Via Tiepolo 11, 34131 Trieste, Italy\label{aff111}
\and
NASA Ames Research Center, Moffett Field, CA 94035, USA\label{aff112}
\and
Kavli Institute for Particle Astrophysics \& Cosmology (KIPAC), Stanford University, Stanford, CA 94305, USA\label{aff113}
\and
Bay Area Environmental Research Institute, Moffett Field, California 94035, USA\label{aff114}
\and
Minnesota Institute for Astrophysics, University of Minnesota, 116 Church St SE, Minneapolis, MN 55455, USA\label{aff115}
\and
Institute Lorentz, Leiden University, PO Box 9506, Leiden 2300 RA, The Netherlands\label{aff116}
\and
Institute for Astronomy, University of Hawaii, 2680 Woodlawn Drive, Honolulu, HI 96822, USA\label{aff117}
\and
Department of Astronomy \& Physics and Institute for Computational Astrophysics, Saint Mary's University, 923 Robie Street, Halifax, Nova Scotia, B3H 3C3, Canada\label{aff118}
\and
Universit\'e Paris-Saclay, CNRS, Institut d'astrophysique spatiale, 91405, Orsay, France\label{aff119}
\and
Departamento F\'isica Aplicada, Universidad Polit\'ecnica de Cartagena, Campus Muralla del Mar, 30202 Cartagena, Murcia, Spain\label{aff120}
\and
Department of Physics, Oxford University, Keble Road, Oxford OX1 3RH, UK\label{aff121}
\and
CEA Saclay, DFR/IRFU, Service d'Astrophysique, Bat. 709, 91191 Gif-sur-Yvette, France\label{aff122}
\and
Institute of Cosmology and Gravitation, University of Portsmouth, Portsmouth PO1 3FX, UK\label{aff123}
\and
Department of Computer Science, Aalto University, PO Box 15400, Espoo, FI-00 076, Finland\label{aff124}
\and
Ruhr University Bochum, Faculty of Physics and Astronomy, Astronomical Institute (AIRUB), German Centre for Cosmological Lensing (GCCL), 44780 Bochum, Germany\label{aff125}
\and
Department of Physics and Astronomy, Vesilinnantie 5, 20014 University of Turku, Finland\label{aff126}
\and
Serco for European Space Agency (ESA), Camino bajo del Castillo, s/n, Urbanizacion Villafranca del Castillo, Villanueva de la Ca\~nada, 28692 Madrid, Spain\label{aff127}
\and
ARC Centre of Excellence for Dark Matter Particle Physics, Melbourne, Australia\label{aff128}
\and
Centre for Astrophysics \& Supercomputing, Swinburne University of Technology, Victoria 3122, Australia\label{aff129}
\and
W.M. Keck Observatory, 65-1120 Mamalahoa Hwy, Kamuela, HI, USA\label{aff130}
\and
ICTP South American Institute for Fundamental Research, Instituto de F\'{\i}sica Te\'orica, Universidade Estadual Paulista, S\~ao Paulo, Brazil\label{aff131}
\and
Oskar Klein Centre for Cosmoparticle Physics, Department of Physics, Stockholm University, Stockholm, SE-106 91, Sweden\label{aff132}
\and
Astrophysics Group, Blackett Laboratory, Imperial College London, London SW7 2AZ, UK\label{aff133}
\and
INAF-Osservatorio Astrofisico di Arcetri, Largo E. Fermi 5, 50125, Firenze, Italy\label{aff134}
\and
Dipartimento di Fisica, Sapienza Universit\`a di Roma, Piazzale Aldo Moro 2, 00185 Roma, Italy\label{aff135}
\and
Centro de Astrof\'{\i}sica da Universidade do Porto, Rua das Estrelas, 4150-762 Porto, Portugal\label{aff136}
\and
Dipartimento di Fisica, Universit\`a di Roma Tor Vergata, Via della Ricerca Scientifica 1, Roma, Italy\label{aff137}
\and
INFN, Sezione di Roma 2, Via della Ricerca Scientifica 1, Roma, Italy\label{aff138}
\and
Institute of Astronomy, University of Cambridge, Madingley Road, Cambridge CB3 0HA, UK\label{aff139}
\and
Institute for Computational Science, University of Zurich, Winterthurerstrasse 190, 8057 Zurich, Switzerland\label{aff140}
\and
Theoretical astrophysics, Department of Physics and Astronomy, Uppsala University, Box 515, 751 20 Uppsala, Sweden\label{aff141}
\and
Niels Bohr Institute, University of Copenhagen, Jagtvej 128, 2200 Copenhagen, Denmark\label{aff142}
\and
Cosmic Dawn Center (DAWN)\label{aff143}
\and
Center for Cosmology and Particle Physics, Department of Physics, New York University, New York, NY 10003, USA\label{aff144}
\and
Technische Universitat Dresden, Institut f\"ur Kern- und Teilchenphysik, Zellescher Weg 19, 01069 Dresden, Germany\label{aff145}
\and
Deutsches Elektronen-Synchrotron DESY, Platanenallee 6, 15738 Zeuthen, Germany\label{aff146}}    




%
%
\abstract{
Verifying the fully kinematic nature of the long-known cosmic microwave background (CMB) dipole is of fundamental importance in cosmology. In the standard cosmological model with the Friedman--Lemaitre--Robertson--Walker (FLRW) metric from the inflationary expansion the CMB dipole should be entirely kinematic. Any non-kinematic CMB dipole component would thus reflect the preinflationary structure of spacetime probing the extent of the FLRW applicability. Cosmic backgrounds from galaxies after the matter-radiation decoupling, should have kinematic dipole component identical in velocity with the CMB kinematic dipole. Comparing the two can lead to isolating the CMB non-kinematic dipole. It was recently proposed that such measurement can be done using the near-IR cosmic infrared background (CIB) measured with the currently operating \Euclid telescope, and later with {\it Roman}. The proposed method reconstructs the resolved CIB, the Integrated Galaxy Light (IGL), from \Euclid's Wide Survey and probes its dipole, with a kinematic component  amplified over that of the CMB by the Compton--Getting effect. The amplification coupled with the extensive galaxy samples forming the IGL would determine the CIB dipole with an overwhelming signal-to-noise ratio, isolating its direction to sub-degree accuracy. We develop details of the method for the \Euclid's Wide Survey in four bands spanning 0.6 to 2 \micron. We isolate the systematic and other uncertainties and present methodologies to minimize them, after confining the sample to the magnitude range with negligible IGL/CIB dipole from galaxy clustering. These include the required star--galaxy separation, accounting for the extinction correction dipole using the method newly developed here achieving total separation, accounting for the Earth's orbital motion and other systematic effects. Finally, we apply the developed  methodology to the simulated \Euclid galaxy catalogs testing successfully the upcoming applications. With the presented techniques one would indeed  measure the IGL/CIB dipole from \Euclid's Wide Survey with high precision probing the non-kinematic CMB dipole.
    }
%
%
%
%
   \titlerunning{\Euclid preparation. The Near-IR Background Dipole}
   \authorrunning{Euclid Collaboration: A.~Kashlinsky, et al.}
   
   \maketitle
   \tableofcontents
%
%
%
%
   
\section{Motivation}
\label{sec:motivation}

The cosmic microwave background (CMB) dipole is the oldest known CMB anisotropy of $\delta T_{\rm CMB}=3.35$ mK, or $\delta T_{\rm CMB}/T_{\rm CMB}=1.23\times 10^{-3}$, measured with the unprecedented precision of a signal-to-noise ratio of ${\rm S/N}\gtrsim 200$ \citep{Kogut:1993,Fixsen:1994}. See Table 1 in \cite{Lineweaver:1997} for the history of the CMB dipole measurements and discovery throughout the 20th century. It is conventionally interpreted as being entirely of kinematic origin due to the Solar System moving at velocity $V_{\rm CMB}=370$ km s$^{-1}$ in the Galactic direction of $(l,b)_{\rm CMB}=(263\fdg85\pm0\fdg1, 48\fdg25\pm0\fdg04)$.

The fully kinematic origin of the CMB dipole is further motivated theoretically by the fact that any curvature perturbations on superhorizon scales leave zero dipole because the density gradient associated with them is exactly cancelled by that from their gravitational potential \citep{Turner:1991}. However, already prior to the development of inflationary cosmology there were suggestions that the CMB dipole may be, even if in part, primordial \citep{King:1973,Matzner:1980}. Within the inflationary cosmology, which posits the non-Friedmann--Lemaitre--Robertson--Walker (FLRW) metric on sufficiently large scales due to the primeval (preinflationary) structure of space-time \citep{Turner:1991,Grishchuk:1992,Kashlinsky:1994,Das:
2021}, such possibility can arise from isocurvature perturbations induced by the latter \citep{Turner:1991} and/or from entanglement of our Universe with other superhorizon domains of the Multiverse \citep{Mersini-Houghton:2009}. Hence, establishing the nature of the CMB dipole is a problem of fundamental importance in cosmology.

Despite the overwhelming preference of the kinematic CMB dipole interpretation, there have been longstanding observational claims to the contrary \citep{Gunn:1988}. Comparing the gravity dipole with peculiar velocity measurements \citep{Villumsen:1987} indicates an offset \citep{Gunn:1988,Erdogdu:2006,Kocevski:2006,Lavaux:2010,Wiltshire:2013} broadly buttressed by other peculiar velocity data \citep{Mathewson:1992,Lauer:1994,Ma:2011,Colin:2019}. There appears a ``dark flow'' of galaxy clusters in the analysis of the cumulative kinematic Sunyaev--Zeldovich effect extending to at least $\sim$1 Gpc in both WMAP and {\it Planck} data \citep{Kashlinsky:2008,Kashlinsky:2009,Atrio-Barandela:2010,Kashlinsky:2010,Kashlinsky:2012a,Atrio-Barandela:2013,Atrio-Barandela:2015}, which is generally consistent with the radio \citep{Nodland:1997,Jain:1999,Singal:2011} and WISE \citep{Secrest:2021} source count dipoles and the anisotropy in X-ray scaling relations \citep{Migkas:2020}. Dark flow, with its dipole signal extending to at least $\sim 1$ Gpc, in particular hints at the superhorizon non-FLRW structure in the overall space-time metric. See reviews by \cite{Kashlinsky:2012,Aluri:2023}. All of these assertions have achieved only a limited significance of ${\rm S/N}\sim 3$--$5$, with the subsequently significant directional uncertainty, and are debated.

It is important to establish observationally the fully kinematic nature of the CMB dipole and whether the homogeneity in the
Universe as reflected in the FLRW metric models is adequate to describe what we
observe. Since any curvature perturbations must have zero dipole at last scattering such probe would be fundamental to cosmology with the non-kinematic CMB dipole component potentially  providing a probe of the primordial preinflationary structure of spacetime. To this end a technique has been proposed recently by \cite{Kashlinsky:2022} to be applied to the Euclid Wide Survey  to probe the dipole of the resolved part of the CIB, the IGL, at an overwhelming ${\rm S/N}$ thereby settling the issue of the origin of the CMB dipole. 

Here we develop the detailed methodology for this experiment we call NIRBADE (Near IR BAckground Dipole Experiment) dedicated to measuring, at high ${\rm S/N}$, the (amplified) CIB dipole from the Euclid Wide Survey. In Sect. \ref{section2} we discuss the different physics governing CMB and CIB dipoles, pointing out how at the \Euclid-covered wavelengths the expected kinematic CIB dipole will be significantly amplified over that of the CMB. Section \ref{section3} sums up the details of the Euclid Wide Survey and their application to NIRBADE following \cite{Kashlinsky:2022}. Section \ref{section4} is devoted to the required development to achieve the NIRBADE goal covering the overall pipeline. These topics include isolating the needed magnitude range here (AB magnitudes are used throughout this paper), developing the methodology to successfully isolate the dipole from Galactic extinction, and accounting for the Earth's orbital motion. Here, we also discuss a slew of less critical, but still important items, such as photometry, before moving on to quantifying the overall uncertainties expected in the pipeline. Section \ref{section5} then applies the development here to the simulated \Euclid catalog to demonstrate how comparing the measured CIB dipole with the well known CMB dipole will isolate any non-kinematic CMB dipole component down to interestingly low levels. We sum up the prospects for NIRBADE with \Euclid in Sect. \ref{section6}. 

More specifically the outline of the developmental part of the study is as follows:
\begin{itemize}
    \item The procedure of the measurement with the required steps to be implemented here has been designed in Sect. \ref{sec:procedure}. The procedure requires successfully finessing the various items that are subsequently outlined, discussed, and resolved. 
    \item In the following Sect. \ref{sec:widesurvey} we present the pre-launch plan of the Euclid Wide Survey coverage that we use in the computations here. Now that the mission is at L2, the details of the survey may be altered, so this is given as an example used for development in finalizing the details of the methodology. The methodology developed here will be applied to the actual observed coverage. 
    \item We identify the aspects required for selecting galaxies from the Euclid Wide Survey for this measurement in Sect. \ref{sec:m-range} -- Eq. (\ref{eq:mag_range_vis}) for VIS and Eq. (\ref{eq:mag_range_nisp}) for NISP. Throughout we used, in the absence of the forthcoming \Euclid data, the observed galaxy counts presented in Sect. \ref{sec:counts} for JWST measurements \citep{Windhorst:2023} and, when needed, the HRK reconstruction \citep{Helgason:2012}. The range of galaxy magnitudes required to sufficiently reduce the clustering dipole component is isolated in Sect. \ref{sec:m0m1-final}. The prospects of the star--galaxy separation desired in the experiment are given in Sect. \ref{sec:star-galaxy}. 
    \item Section \ref{sec:extinction} discusses how the extinction, using the SFD template \citep{Schlegel:1998}, can affect the measurement and design a method to isolate the contribution due to extinction corrections from that of the IGL/CIB. The method is applicable at small extinction corrections $A\ll1$.
    \item The needed corrections, for the high-precision measurement, from the effects of the Earth's orbital motion are then discussed in Sect. \ref{sec:orbital}. It is shown how the corrections will be incorporated into the designed pipeline.
    \item The potential systematic effects, and how to correct for them are considered in Sect. \ref{sec:systematic} followed by the requirements on the photometric accuracy and zero points, etc. in Sect. \ref{sec:zeropoints}.
    \item Section \ref{section5} then shows the application of the developed methodology to the forthcoming Euclid Wide Survey data. In Sect. \ref{subsection5.1} we evaluate the statistical uncertainties after each year of the \Euclid observation. In Sect. \ref{subsection5.2} we apply the method developed here to correct for extinction using a simulated catalog for \Euclid with available spectral colors, to isolate the contribution from extinction if the need arises in the actual data to finalize the high-precision determination of the IGL/CIB dipole from the Euclid Wide Survey. Section \ref{subsection5.3} discusses and quantifies the identified systematic corrections when converting the measured IGL/CIB dipole into the equivalent velocity, which affect all the velocity components equally thereby being of relevance to its amplitude, and not direction.
\end{itemize}

Such an experiment can, and must, also be done with {\it Roman}
\citep[formerly WFIRST;][]{Akeson:2019}, which would require a separate and significantly different preparation.

\section{On importance of cosmic background dipoles}
\label{section2}
Here we discuss the different physics governing the CMB and CIB dipoles and why and how the CIB kinematic dipole is amplified over that of the CMB.
\subsection{On the intrinsic CMB dipole}
\label{subsec:cmb}
The CMB as observed today originates at the last scattering, which occurred at the cosmic epochs corresponding to redshift $z\simeq 10^3$. Its structure from the quadrupole term ($\ell=2$) to higher-order in $\ell$ multipoles is in very good general agreement with predictions of inflation. The latter posits that the observed Universe originated from a small smooth patch, with the underlying FLRW metric, of the size of or smaller than the horizon scale at the start of inflation, which then quickly inflated to encompass scales well beyond the current cosmological horizon \citep{Kazanas:1980,Guth:1981}. At the same time, on sufficiently large scales the preinflationary spacetime could have preserved its original structure, assumed generally to be inhomogeneous \citep{Turner:1991,Grishchuk:1992,Kashlinsky:1994}. Such preinflationary structures, currently on superhorizon scales, could leave CMB signatures via the Grishchuk--Zeldovich effect \citep{Grishchuk:1978}. The smallness of the measured CMB quadrupole (the relative value of $Q\sim 2\times 10^{-6}$) indicates that preinflationary structures in spacetime were pushed during inflation to scales currently $\gtrsim Q^{-1/2} cH_0^{-1}\sim10^3 cH_0^{-1}$ \citep{Turner:1991,Kashlinsky:1994}.

However, as was shown by \cite{Turner:1991}, the Grishchuk--Zeldovich effect does not produce an observable dipole anisotropy in any superhorizon modes from curvature  perturbations because at the last scattering the linear gradient associated with them is cancelled {\it exactly} by the corresponding dipole anisotropy from their gravitational potential term. This points to the unique importance of probing the fully kinematic nature of the CMB dipole where any non-kinematic dipole would arise from, within the inflationary paradigm, the preinflationary structure of spacetime and potentially provide new information on the details of inflation and the applicability limits of the FLRW metric.

This differentiates the CMB dipole from the dipole components of the cosmic backgrounds, discussed next, which are produced by sources that formed well after decoupling.
\subsection{The Compton--Getting effect and dipole for cosmic backgrounds from galaxies}
\label{subsec_cib}
Cosmic backgrounds produced by luminous sources that formed at $z\ll 10^3$ are subject to a different physics and their spatial distribution is characterized by the matter power spectrum imprinted during the inflationary period which is later modified by the standard gravitational evolution during the radiation-dominated era. In addition if the Solar System moves with respect to the frame defined by distant sources producing the background with mean intensity $\bar{I}_\nu$ at frequency $\nu$, it would have a dipole in the Sun's rest frame
\begin{equation}
\boldsymbol{d}_\nu= (3-\alpha_{\nu,\infty}) \frac{\boldsymbol{V}}{c} \bar{I}_{\nu,\infty}\,,
\label{eq:dipole_alpha}
\end{equation}
where 
\begin{equation}
\alpha_\nu=\frac{\partial{\ln I_\nu}}{\partial{\ln \nu}}
\label{eq:alpha}
\end{equation}
and the subscript $\infty$ implies that the background intensity $I_\nu$ comes from integrating over the entire range of fluxes/magnitudes of the contributing sources. This is known as the Compton--Getting \citep{Compton:1935} effect for cosmic rays \citep[e.g.][]{Gleeson:1968}. Equation (\ref{eq:dipole_alpha}) follows since photons emitted at frequency $\nu_0$ from a source moving at velocity $V\ll c$ forming angle $\Theta$ toward the apex of motion will be received by the observer at rest at frequency $\nu=\nu_0[1+({V}/{c})\cos\Theta]$ and the Lorentz transformation requires that $I_\nu/\nu^3$ remains invariant \citep{Peebles:1968}. Hence the observer at rest will see the direction-dependent specific intensity $I_\nu=(\nu/\nu_0)^3 I_{\nu/[1+({V}/{c})\cos\Theta]}$; see Appendix. The spectral index of the Rayleigh-Jeans spectrum $\alpha_\nu^{\rm RJ}=2$ describes the CMB at mm wavelengths. 

If the CMB is the rest frame of the Universe then $\boldsymbol{V}=\boldsymbol{V}_{\rm CMB}$ for any cosmic background that originates from galaxies. Otherwise, the non-zero non-kinematic part of the CMB dipole would be likely to indicate the existence of superhorizon deviations from the FLRW metric, possibly due to the primordial (preinflationary) structure of spacetime.

At wavelengths where cosmic backgrounds from galaxies have $\alpha_\nu\ll2$, the amplitude of their kinematic dipole in $I_\nu$ is amplified. This is the case for CIB \citep{Kashlinsky:2005} and is also the case at high energies [X-ray \citep{Fabian:1979} and $\gamma$-ray \citep{Maoz:1994,Kashlinsky:2024} backgrounds, and cosmic rays \citep{Kachelriess:2006}]. However, at infrared wavelengths significant pollution to the CIB dipole would come from dust emission and reflection by the Galaxy (cirrus) and the Solar System (zodiacal light) as discussed 
in \cite{Kashlinsky:2022}.

\section{Probing the near-IR background dipole in the Euclid Wide Survey}
\label{section3}

\begin{figure*}[ht] 
	\begin{centering} 
		{\includegraphics[width=5in]{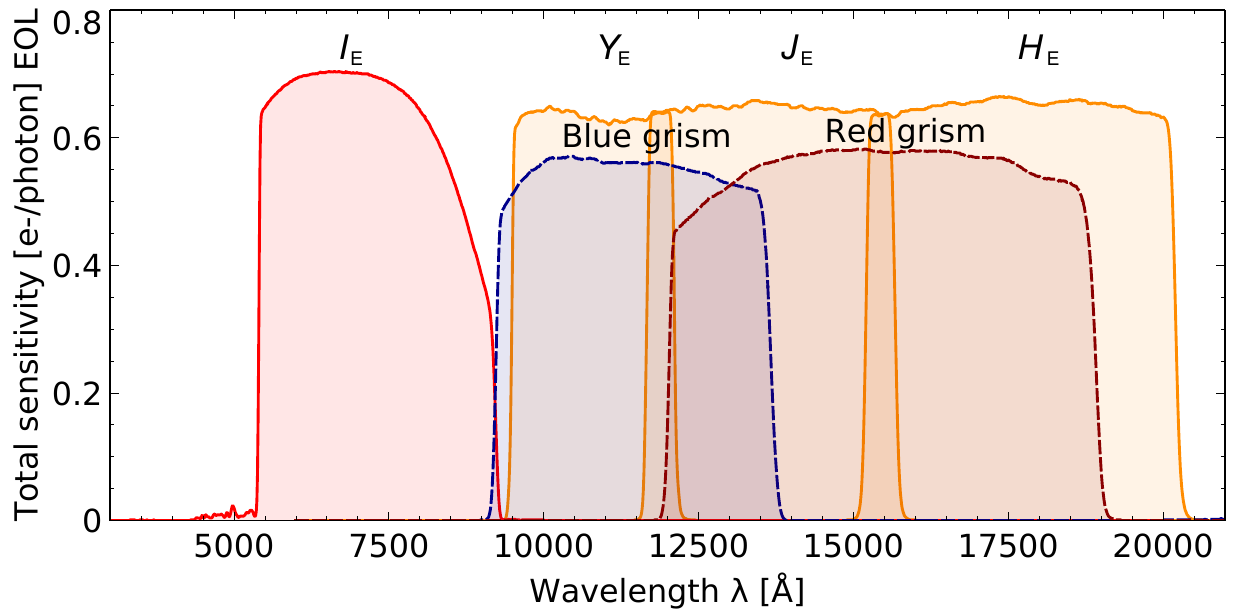}}
		\caption{Total sensitivity of \Euclid's photometric  and spectroscopic bands. (credit:ECSURV/J.-C. Cuillandre)} 
		\label{fig:filter_shapes}
	\end{centering}
\end{figure*}

The \Euclid satellite was successfully launched on July 1, 2023 to the L2 orbit. The photometric bands covered by \Euclid are shown in Fig. \ref{fig:filter_shapes}. The unresolved CIB dipole at the \Euclid bands will be subject to significant contributions from Galactic and Solar System foregrounds, but the foreground dipole contributions can be excluded efficiently by considering the CIB from resolved galaxies. 

\begin{figure*}[ht] 
   \centering
   \includegraphics[width=6.5in]{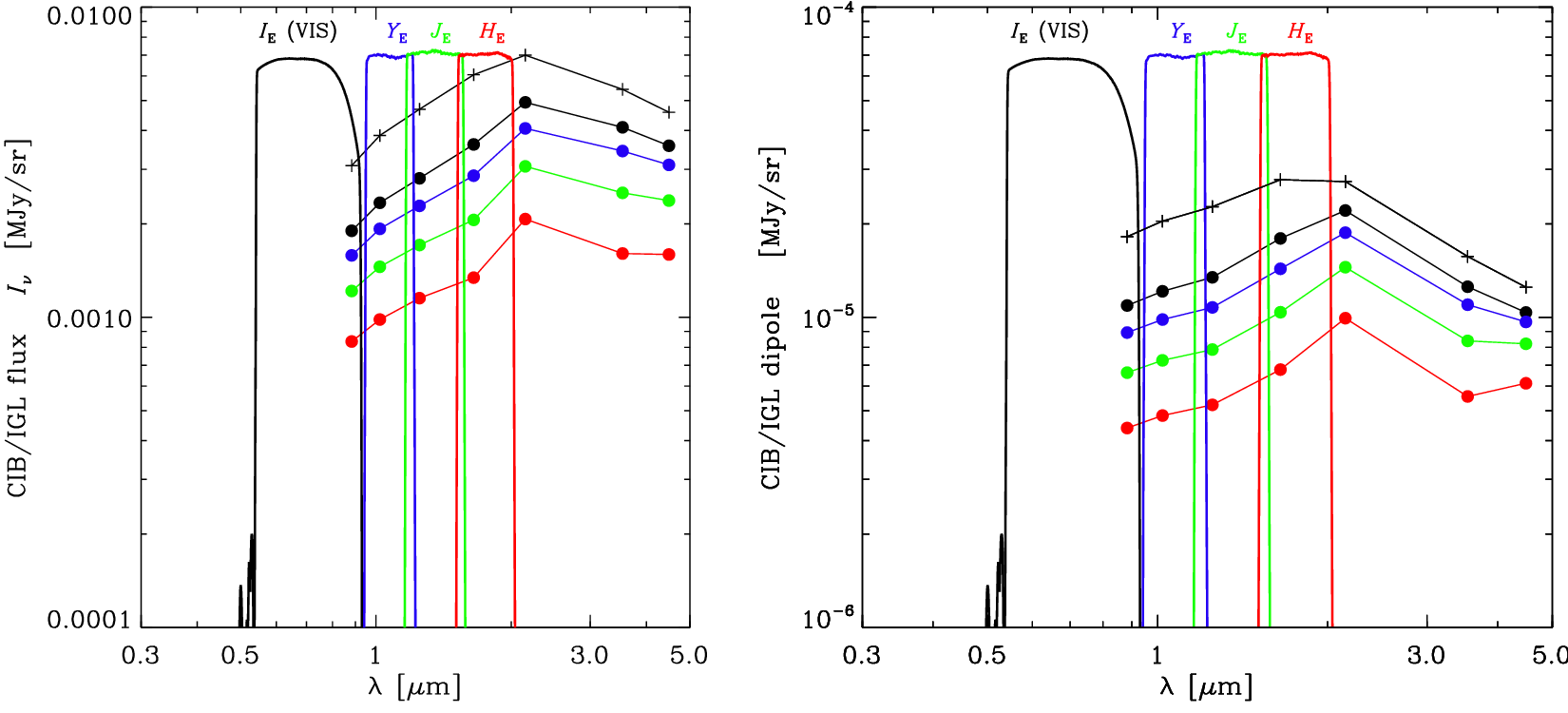} 
   \caption{Expected IGL amplitudes. Left: The mean IGL flux. Right: The IGL dipole per Eqs. (\ref{eq:alpha}) and (\ref{eq:dndm}) and assuming $V=370$ km s$^{-1}$. Black plus signs correspond to the entire range of magnitudes. Black, blue, green, and red circles correspond to IGL from galaxies between $m_1 = 25$ and $m_0 = 18, 19, 20, 21$. The IGL is integrated over the JWST latest counts \citep{Windhorst:2023} at the marked central wavelengths. The four \Euclid filters are shown per \cite{Schirmer-EP18}.}
   \label{fig:dipole}
\end{figure*}
To overcome the obstacles due to the otherwise dominant at near-IR foreground dipoles \cite{Kashlinsky:2022} proposed to use the all-sky part of the background, known as IGL (Integrated Galaxy Light), reconstructed from resolved galaxies in the Euclid Wide Survey \citep{Laureijs:2011}, 
\begin{equation}
I_\nu(\theta,\phi) = 10^{-0.4A_\nu(\theta,\phi)} S_0\int_{m_0}^{m_1} 10^{-0.4m} \left[\frac{\diff N_\nu(\theta,\phi)}{\diff m}\right] \diff m\,,
\label{eq:dndm}
\end{equation}
where $S_0=3631$ Jy and $A_\nu$ is the magnitude extinction in the direction $(\theta,\phi)$. The above expression is a short-hand for the actual procedure outlined in Sec. \ref{sec:procedure}, Eq. (\ref{eq:igl}) which requires no source counts determination. The Euclid Wide Survey galaxy samples will be corrected for extinction, so strictly speaking $A_\nu$ should be interpreted as the magnitude correction remaining after the extinction correction; more on this will be presented later. The IGL is evaluated over a suitably selected $m_0\sim 18$--$21$ required to remove the galaxy clustering dipole and $m_1$ imposed by the sensitivity limits of the Wide 
Survey, which is also below the expected magnitudes of the new populations expected to be present in the CIB source-subtracted anisotropies \citep{Kashlinsky:2018}. 

As discussed in \cite{Kashlinsky:2022} at the \Euclid VIS and NISP bands $\alpha_\nu\sim-1$, so from an all-sky catalog of $N_{\rm gal}$ galaxies and for a fixed direction, one would reach the statistical signal-to-noise ratio in the measured IGL dipole amplitude, $d_\nu/\langle I_\nu\rangle$, of
\begin{equation}
{\rm S/N} \sim 160 \left(\frac{3-\alpha_\nu}{4} \right)\left(\frac{V}{V_{\rm CMB}}\right) \left(\frac{N_{\rm gal}}{10^9}\right)^{1/2}\,.
\label{eq:s2n}
\end{equation}
 An all-sky CMB dipole measured with a signal-to-noise ratio of ${\rm S/N}$ will have its direction probed with directional accuracy of \citep{Fixsen:2011}
\begin{equation}
\Delta\Theta_{\rm dipole}\sim \sqrt{2}({\rm S/N})^{-1}\; {\rm radian}\,.
\label{eq:direction}
\end{equation}
This demonstrates that the directional uncertainty, say $\Delta\Theta_{\rm dipole}\lesssim 1^\circ$, needed to decisively probe the alignment requires ${\rm S/N}\gtrsim 80$. The statistical significance will depend on the actual dipole amplitude, direction and region of the sky observed by \Euclid. For a partial sky coverage the above order of magnitude estimates will be reduced since the three dipole components $(X,Y,Z)$ will have different errors \citep{Atrio-Barandela:2010,Kashlinsky:2022}. A discussion of the error budget is deferred to Sect.~\ref{subsection5.1}.

Equations (\ref{eq:s2n}) and (\ref{eq:direction}) demonstrate why the to-date probes of the kinematic nature of the CMB dipole discussed in sec. \ref{sec:motivation}, which reach ${\rm S/N}\simeq 4$--$5$ by utilizing the cumulative kinematic Sunyaev-Zeldovich \citep{Sunyaev:1980} effect \citep{Kashlinsky:2000} or the relativistic aberration \citep{Ellis:1984}, have poor directional accuracy of $\Delta\Theta_{\rm dipole}\sim 15^\circ$--$20^\circ$ and hence are insufficient to test,  in addition to the dipole amplitude and its convergence with distance, the consistency of the dipole directions. Both will be achieved with NIRBADE as outlined below.
 
Figure \ref{fig:dipole} (left) shows the IGL reconstructed from integrating over magnitudes exceeding some fiducial $m_0$ (see caption) using observed galaxy counts from Figs. 9 and 10 of the JWST counts data by \cite{Windhorst:2023} at the wavelengths similar to the \Euclid bands. The right panel of the figure shows the expected dipole amplitudes evaluated with Eq. (\ref{eq:alpha}) for $V=V_{\rm CMB}=370$ km s$^{-1}$. 
Later we will discuss the selections of $(m_0,m_1)$ required specifically for this measurement.

\section{Required development}
\label{section4}

If the CMB dipole is entirely kinematic, the expected CIB dipole components in the Galactic coordinate system $(X,Y,Z)$ would be
\begin{equation}
\boldsymbol{d}_\nu = 5.2\times 10^{-3} \left(\frac{3-\alpha_\nu}{4} \right) (-0.07, -0.66, 0.75) \langle I_\nu\rangle\,,
\label{eq:dipole_vec}
\end{equation}
with the $X$-component being by far the smallest, contributing just a few percent to the net dipole, and the $Z$-component being the largest, but close in amplitude to the $Y$-component. Hence, if the CMB dipole is purely kinematic, the IGL/CIB dipole, after correcting for the Earth motion, should lie almost entirely in the $(Y,Z)$ plane with nearly equal amplitude $Y$ and $Z$ components.

We are aiming to measure the IGL dipole of dimensionless amplitude of $\simeq 0.5\%$ in each or any of the \Euclid's four bands: 
\IE, \YE, \JE, and \HE.
Two points make this promising: 1) the IGL dipole is amplified by the Compton--Getting effect, and 2) the noise is substantially decreased by the large number of galaxies the Euclid Wide Survey will have.

Below are the items to discuss in order to get this measurement done, and at high precision. In what follows the sought dipole signal, Eq. (\ref{eq:dipole_alpha}), is denoted with a lower case $\boldsymbol{d}$ and the nuisance dipoles with a capital case $\boldsymbol{D}$.

In the absence of the \Euclid data we will use here, for the bulk of estimates, the formulation per Eq. (\ref{eq:dndm}) inputting the latest JWST counts data from \cite{Windhorst:2023}, which are consistent with the reconstruction from \cite{Helgason:2012} used originally by \cite{Kashlinsky:2022}. 

Depending on the context throughout this section we will work with both the absolute CIB dipole amplitude ($d_\nu$ in MJy\,sr$^{-1}$ equivalent to $\delta T$ in mK for the CMB) and its relative amplitude ($d_\nu/I_\nu$ equivalent to $\delta T/T$ for the CMB). The former would be useful when e.g discussing the measurability and overcoming the Galactic components while the latter is useful when estimating the extragalactic non-kinematic terms and converting to velocity.

\subsection{Procedure}
\label{sec:procedure}



The procedure required to apply the method of \cite{Kashlinsky:2022} to 
probe the kinematic component of the IGL/CIB dipole would go through the following steps: 
\begin{enumerate}
\item We will subdivide the \Euclid sky coverage into areas, ${\cal A}$, centered on Galactic coordinates $(l,b)$. ${\cal A}$ could be the size of each FOV (0.5 deg$^2$) or larger.
\item We will collect the photometry on all galaxies in each ${\cal A}$, with care to exclude Galactic stellar sources from the sample.
\item 
We will apply extinction corrections in each band and/or test for contributions from the extinction effects on the dipole by band, latitude etc. Then a method developed below to eliminate the extinction induced dipole will be applied.
\item We will identify, in each of the four \Euclid photometric bands, the uniform upper magnitude limit, $m_1$, that can be applied to all selected regions ${\cal A}$. This would be one of the important criteria for selecting the sky for this measurement.
\item We will select a lower magnitude limit, $m_0$, to ensure the IGL dipole from galaxy clustering is sufficiently negligible. We may choose the same $m_0$ for each bands or leave it band-dependent, provided the clustering dipole contribution is negligible in all four \Euclid bands.
\item We will compute the net IGL flux from the selected galaxies as
\begin{equation}
I_\nu(l,b)=\frac{1}{{\cal A}}S_0\sum_{m_0\leq m\leq m_1} 10^{-0.4m_\nu}
\label{eq:igl}
\end{equation}
over ${\cal A}$ and do this for the entire sky, or a selected part of it. Here $S_0=\num{3631}$ Jy. The magnitudes in the above expression are assumed to be extinction-corrected as will be provided in the course of the Euclid Wide Survey. The remaining extinction effects on the resultant dipole will be removed as discussed below. The residual extinction correction down to the relative accuracy $\epsilon_A$ would introduce a multiplicative factor in the RHS of Eq. \ref{eq:igl} of $10^{-0.4 \epsilon_A A}$ which is incorporated later in the discussion of the elimination of the extinction contribution to the IGL/CIB dipole.
\item We will evaluate the IGL dipole, $d_\nu$, over the selected \Euclid sky in each of the four bands of frequency $\nu$ after dividing the galaxy sample by color to eliminate extinction.
\item We will eliminate the dipole contribution from extinction from a subsample of galaxies with selected IGL spectral index, $\alpha_\nu$, and isolate the kinematic IGL dipole part.
\item We will compute the dipole error.
\item We will evaluate the other systematics discussed below.
\item We will translate into the effective velocity via the refined estimation, for each galaxy subsample, of the IGL spectral index $\alpha_\nu$ and the Compton--Getting amplification using
\begin{equation}
V=(3-\alpha_\nu)^{-1} \left(\frac{d_\nu}{\langle I_\nu\rangle}\right)c\,.
\label{eq:velocity}
\end{equation}
\end{enumerate}

We use HEALPix {\tt remove\_dipole} routine \citep{Gorski:2005} in the computations throughout the paper. 
\subsection{Euclid Wide Survey galaxy samples}
\label{sec:widesurvey}


The Euclid Wide Survey aims to cover most of the best parts of the extragalactic sky in terms of extinction and star density. 
An area larger than $\num{14000}$ deg$^2$ is expected to be covered with a single visit (four exposures via three dithers). In each visit imaging data are acquired over 0.53 deg$^2$ for a wide visible band, \IE\, (sampling $0\farcs1$), and three near infrared bands (\YE, \JE, and \HE), where the sampling is $0\farcs3$. 

\begin{figure*}[ht]
 \centering{
\includegraphics[width=6.5in]{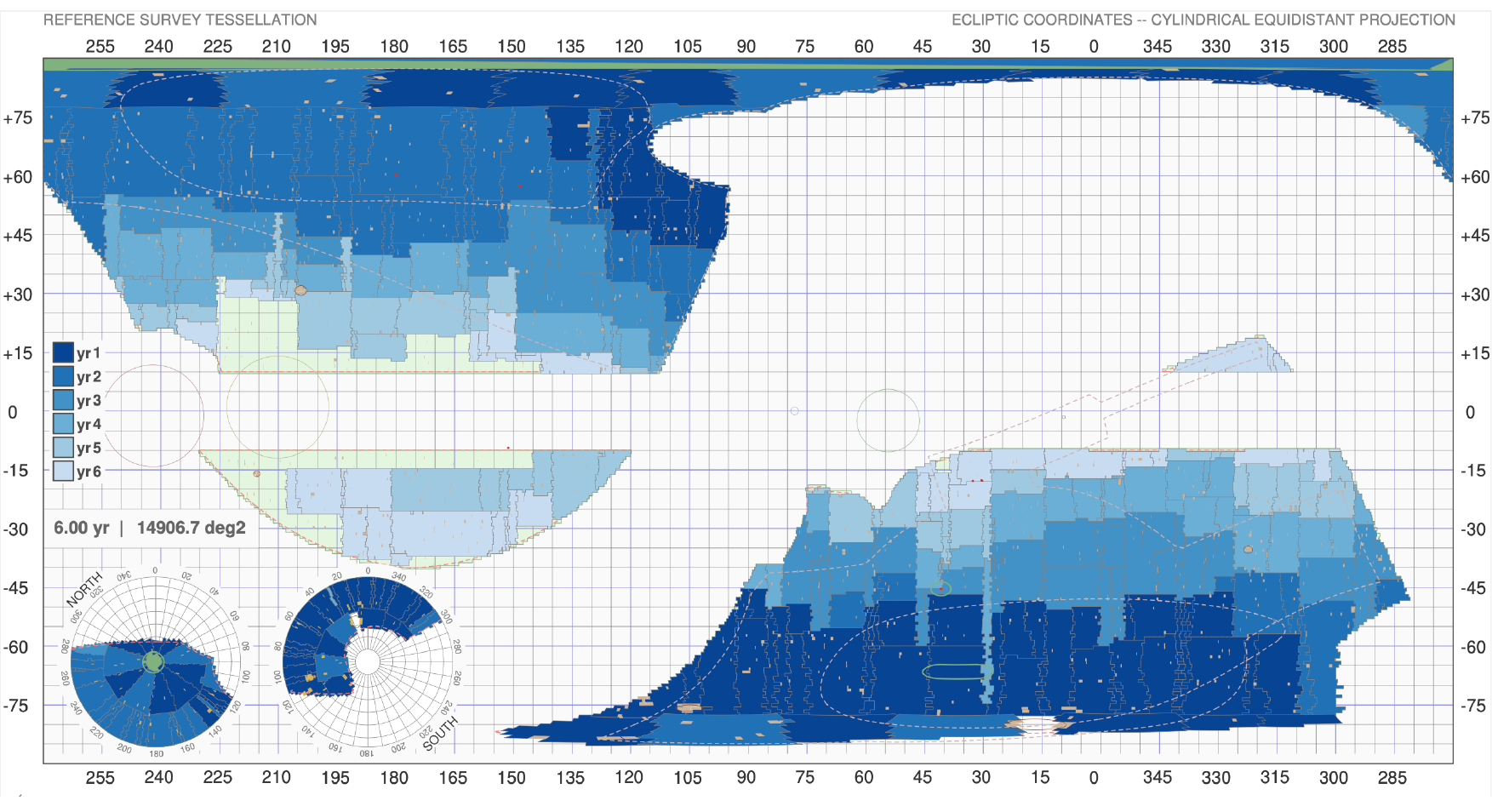} 
\caption{Expected coverage of the sky year by year of the wide survey in ecliptic coordinates.  The ecliptic poles are also shown in a different projection. Circles along the ecliptic denote planet avoidance regions
circa 2029 Sep 27, near the end of the survey. (credit: ECSURV/J.Dinis)}
\label{fig:ecliptic_area_coverage} }
\end{figure*}

In Fig.~\ref{fig:ecliptic_area_coverage} the latest planned sky coverage is shown. There are three main contiguous areas that are covered [the fourth, that was presented in Fig. 45 of \cite{Scaramella-EP1}, is now greatly reduced because of the lack of timely ground based photometry]. Grey regions denote unobserved areas due to the presence of extremely bright stars. Illustration and tabulation of the 
fractional and absolute sky coverage over time is found in Table 9 and Fig. 49 of \cite{Scaramella-EP1}.

We focus in this subsection on the galaxy number density in $H$ band, which is the one least affected by extinction. Deep galaxy counts in $i$ and $K$  bands have been given by studies of the COSMOS field \citep{Laigle:2016,Weaver:2022}. 
Here we are mostly concerned with the intermediate range of magnitudes $20 \leq m_{H} \leq 24$, which is appropriate to get a uniform sample from the Euclid Wide Survey.

%
%
%
These literature estimates, however, are affected by  cosmic variance \citep{Abbott:1984}: the COSMOS area covers only  2 deg$^2$ and therefore is different from the average value taken on much larger areas. Moreover, we will need to work with several sub areas of the wide survey because of cuts to get subsamples and different epochs of increasing coverage.
\begin{figure}[ht!]
	\centering{
		\includegraphics[width=3.5in]{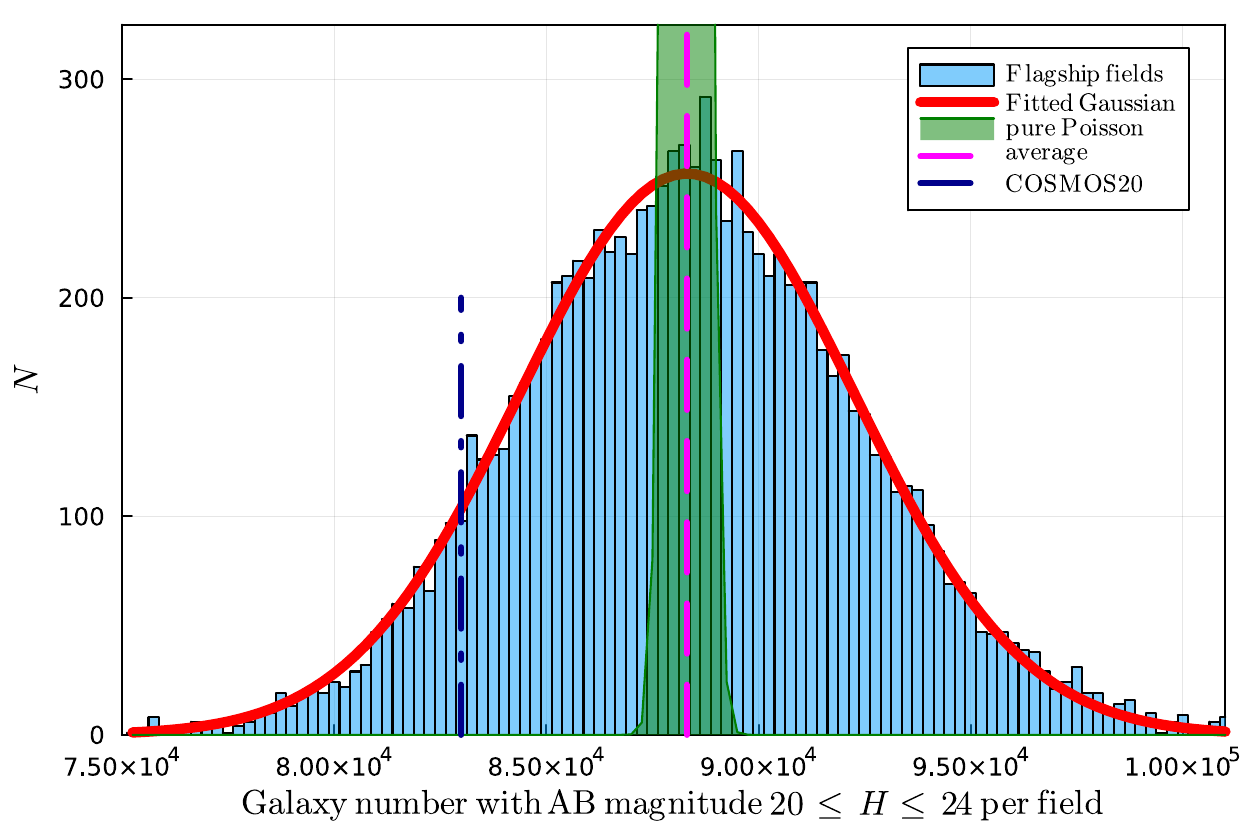} 
		\caption{Distribution of simulated counts per \Euclid field (half square degree).}
	\label{fig:mag24_histo_and_fit.pdf} }
\end{figure}

Therefore we derive, for the time being, the impact of cosmic variance on the \HE\, counts from the Euclid Flagship simulation. 
From the large N-body simulation, the Flagship catalogue of many observables was derived. Of particular importance is the color-color relation and photo-$z$ distribution obtained by imposing spectral energy distributions to the halos identified as galaxies in the simulation.
Therefore the parent spatial halo/galaxy distribution is clustered and so the catalogue 2D sample has an intrinsic angular correlation, which causes the distribution in cells to deviate from the simple Poisson distribution. How large the deviation is would be a function of both the limiting magnitude and the area considered.

In Fig.~\ref{fig:mag24_histo_and_fit.pdf} we show how cosmic variance \citep{Abbott:1984} affects the counts in a single \Euclid field: the standard deviation, $\sigma$, is ${\sim}\,14$ times larger than the simple Poisson one  due to clustering of the sources \citep{Abbott:1984}. We also show the counts from the COSMOS2020 catalog \citep{Weaver:2022}. At this scale the $\sigma$  is still ${\sim}\,5\%$ of the average, but in simply increasing the basic area considered, this ratio will greatly decrease. 

The total expected number density of galaxies is ${\sim}\,1.8\times 10^5$ deg$^{-2}$, which would yield a total of over $2.5$ billion objects from the whole survey.
 



%
%
%
%
\subsection{Selecting the optimal magnitude range \texorpdfstring{$[m_0,m_1]$}{TEXT}}
\label{sec:m-range}

\subsubsection{Galaxy counts}
\label{sec:counts}

\begin{figure*}[ht!] 
   \centering
   \includegraphics[width=6.5in]{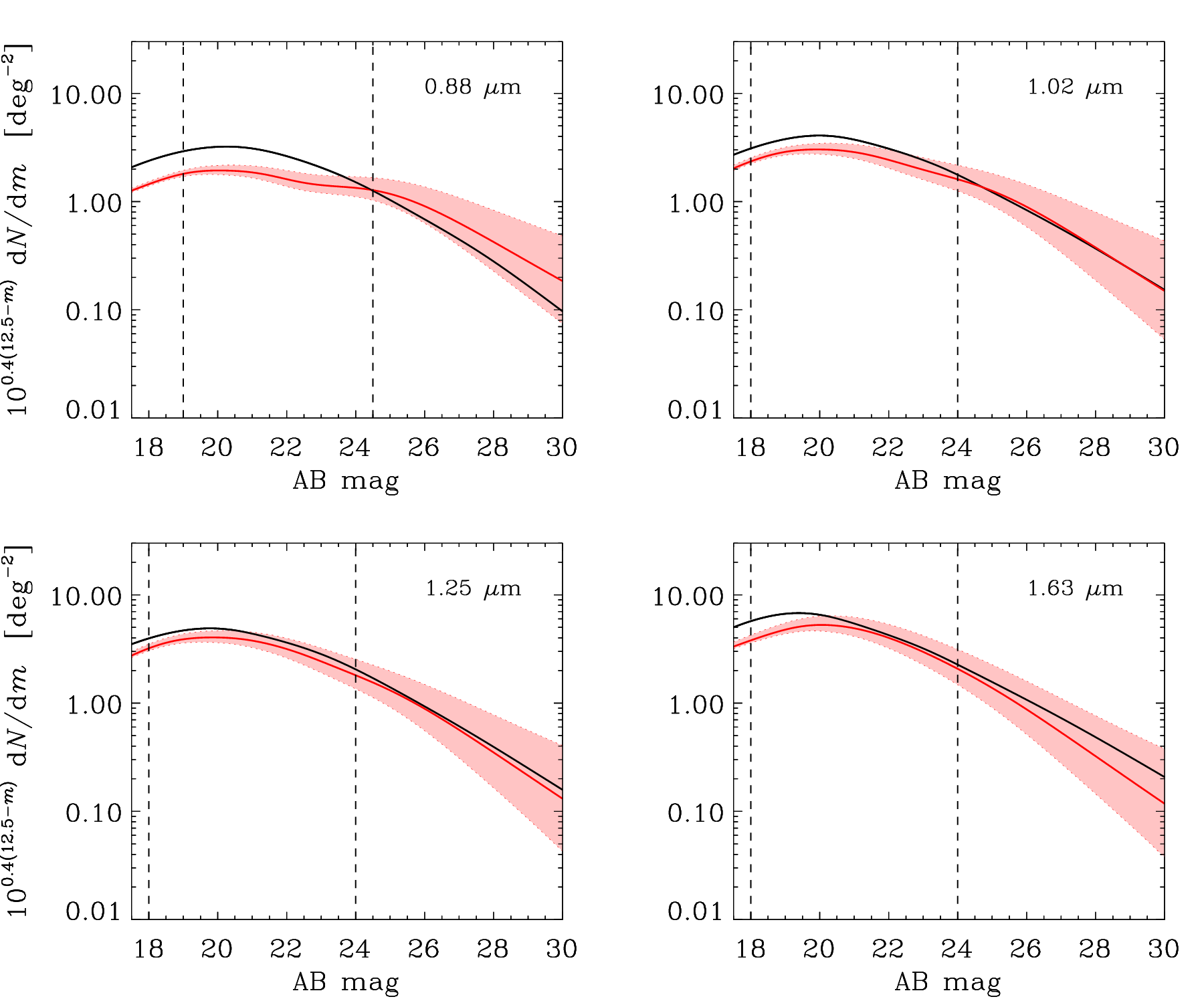} 
   \caption{Galaxy counts per $\diff m$, plotted $\propto \diff I_\nu/\diff m$, at the four sets of \Euclid-related wavelengths. Black lines show the JWST counts at 0.88, 1.02, 1.25, and 1.63 \micron. HRK reconstructions are displayed with red colors, going to the 0.8 \micron\ at the VIS-related end. The spreads in the reconstructions between the LFE (low-faint-end) and HFE (high-faint-end) limits of the extrapolation allowed for the Schechter-type luminosity function \citep{Helgason:2012} are shown with pink shades. Dashed vertical lines mark the range of magnitudes identified for this study in Eq. (\ref{eq:mag_range_nisp}). The JWST counts are filter-transformed to either ground-based VISTA filters or {\it Spitzer} filters at the long wavelengths (S. Tompkins, R. Windhorst, private communication). HRK reconstructions are shown at the wavelengths of 0.8, 1.05, 1.25, and 1.63 \micron.}
   \label{fig:counts}
\end{figure*}
\begin{figure*}[ht] 
   \centering
   \includegraphics[width=7in]{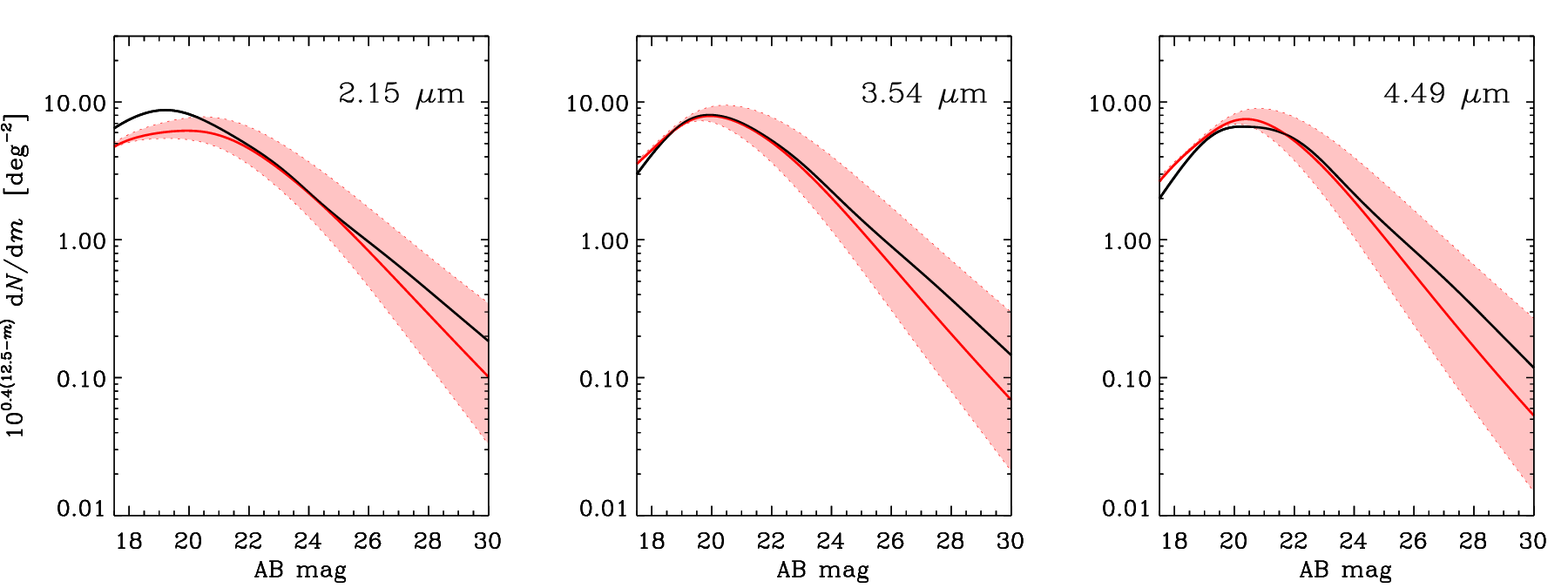} 
   \caption{Same as Fig. \ref{fig:counts}, except at the wavelengths longward of, but adjacent to the NISP bands, which are used for refinements in $\alpha_\nu$ (e.g. Fig. \ref{fig:dipole} and Sect. \ref{section5}). The JWST counts are filter-transformed to either ground-based VISTA filters or {\it Spitzer} filters at the long near-IR wavelengths (S. Tompkins, R. Windhorst, private communication). HRK reconstruction is shown at the wavelengths of 2.2, 3.6, and 4.5 \micron.}
   \label{fig:counts_klm}
\end{figure*}

Throughout this discussion we will need the numbers for total galaxies expected to be available from the Euclid Wide Survey in the given magnitude range at the appropriate wavelengths. Such information is available from the recent JWST counts \citep{Windhorst:2023} and we will be using also the HRK reconstruction \citep{Helgason:2012} used in the pre-JWST era by \cite{Kashlinsky:2022}; we will use both intermittently in our numerical estimates. Figure \ref{fig:counts} shows the comparison at \Euclid-related wavelengths of the HRK reconstruction (red) and the JWST counts for the \Euclid bands. Figure \ref{fig:counts_klm} shows the same for the longer bands adjacent to NISP, which will be used later in Sect. \ref{section5}. The VIS-related numbers are shown at 0.88 \micron\ for JWST data and 0.8 \micron\ for the HRK reconstruction mimicking the fits to the broad \IE\, band. The overall comparison shows good consistency, within the uncertainties, between the HRK reconstruction and JWST data, indicating that the former can be used for the power estimates below. The occasional deviations, seen at bright magnitudes, may stem from the incompleteness of the star-galaxy separation when counts were evaluated and/or from the difference in wavelengths in HRK reconstructions and the \Euclid and JWST bands. Ultimately, for the actual IGL/CIB dipole measurement the real galaxy samples from the Euclid Wide Survey will be used, with the reconstruction, used here for estimates, not required. 

The quantities, $10^{0.4(12.5-m)}\diff N/\diff m$, plotted on the vertical axis in the figures directly reflect the IGL/CIB produced by galaxies in the $\diff m$ range of $m$. The value of 1 deg$^{-2}$ on the vertical axis of Figs. \ref{fig:counts} and \ref{fig:counts_klm} corresponds to $\diff I_\nu/\diff m = 1.2\times 10^{-4}$ MJy\,sr$^{-1}$.

\subsubsection{Star--galaxy separation}
\label{sec:star-galaxy}

Galactic stars need to be excluded from the \Euclid source counts when constructing the IGL.
At wavelengths from 1 to 5\,$\mu$m prior studies indicate that stars outnumber galaxies
at $m \lesssim 18$ \citep[e.g. ][]{Ashby:2013,Windhorst:2022a,Windhorst:2023}. 
Thus, star--galaxy separation is essential if $m_0 < 18$ and still important for sources 
with $m > 18$. 

\begin{figure*}[htbp] 
   \centering
   \includegraphics[width=2.in,angle=90]{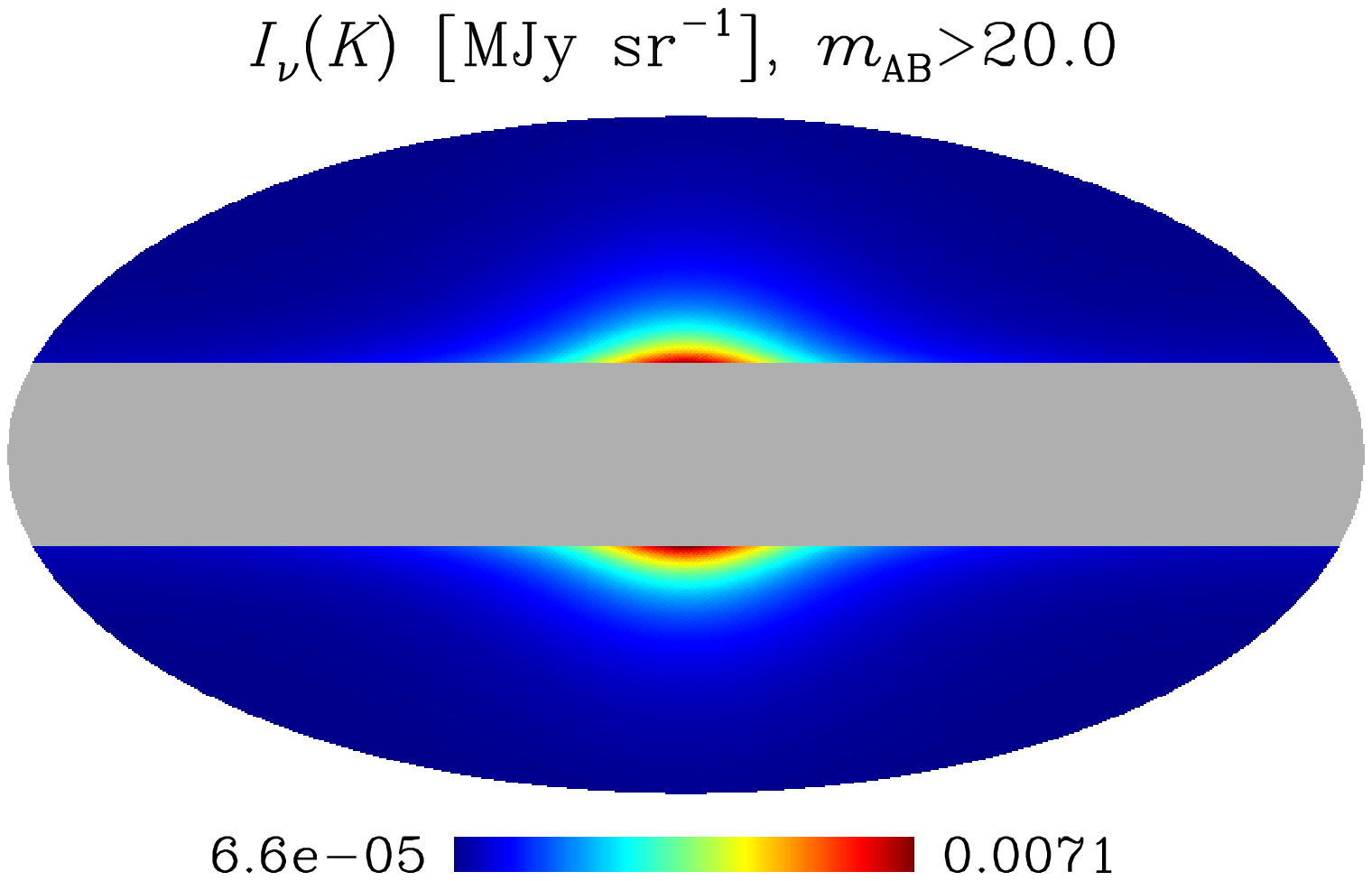}
   \includegraphics[width=2.in,angle=90]{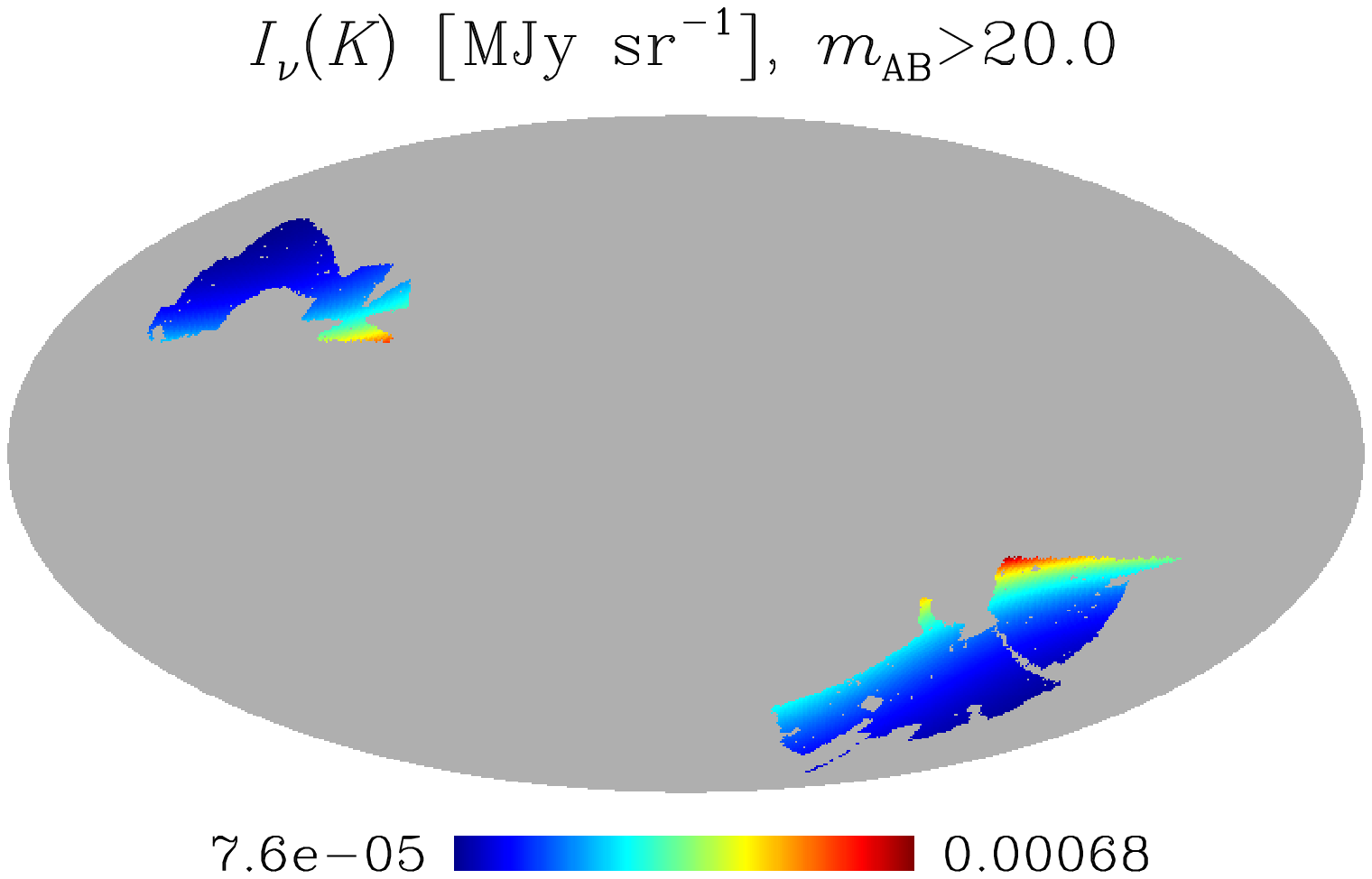}\\
   \includegraphics[width=3.5in]{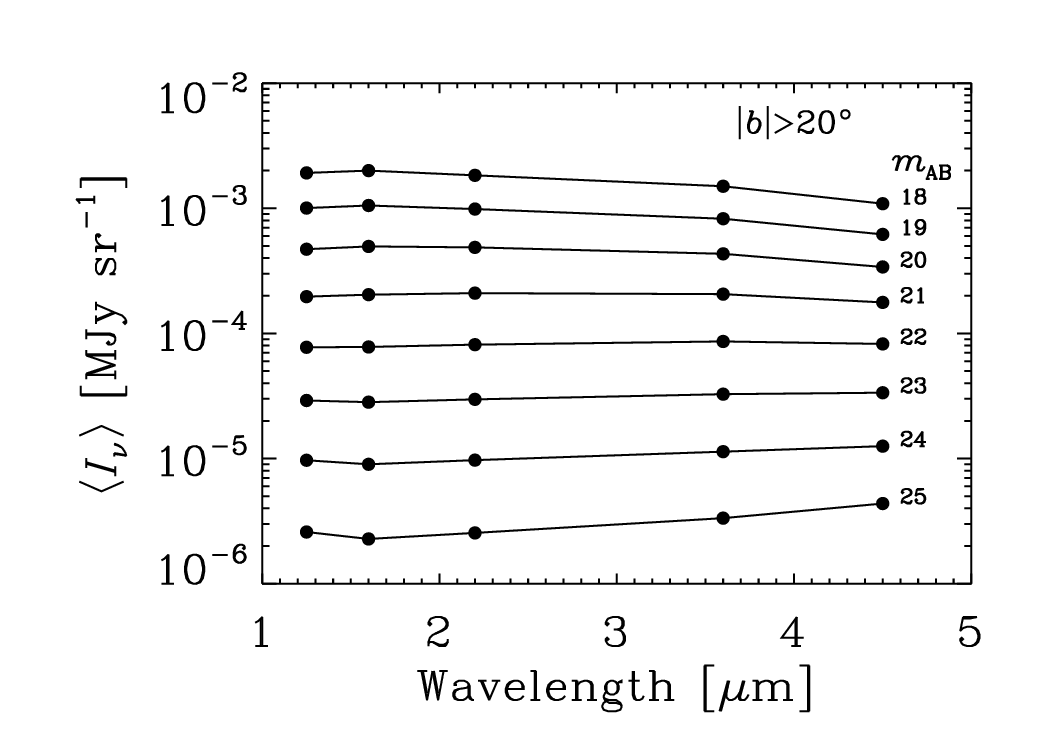}
   \includegraphics[width=3.5in]{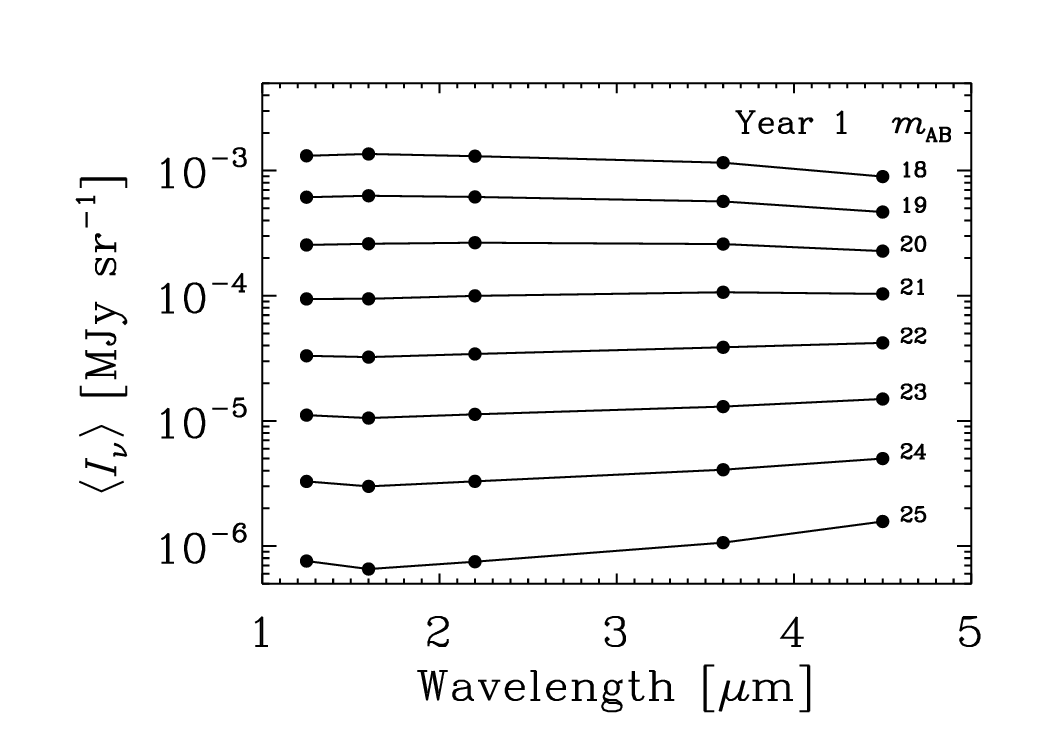}\\
   \includegraphics[width=3.5in]{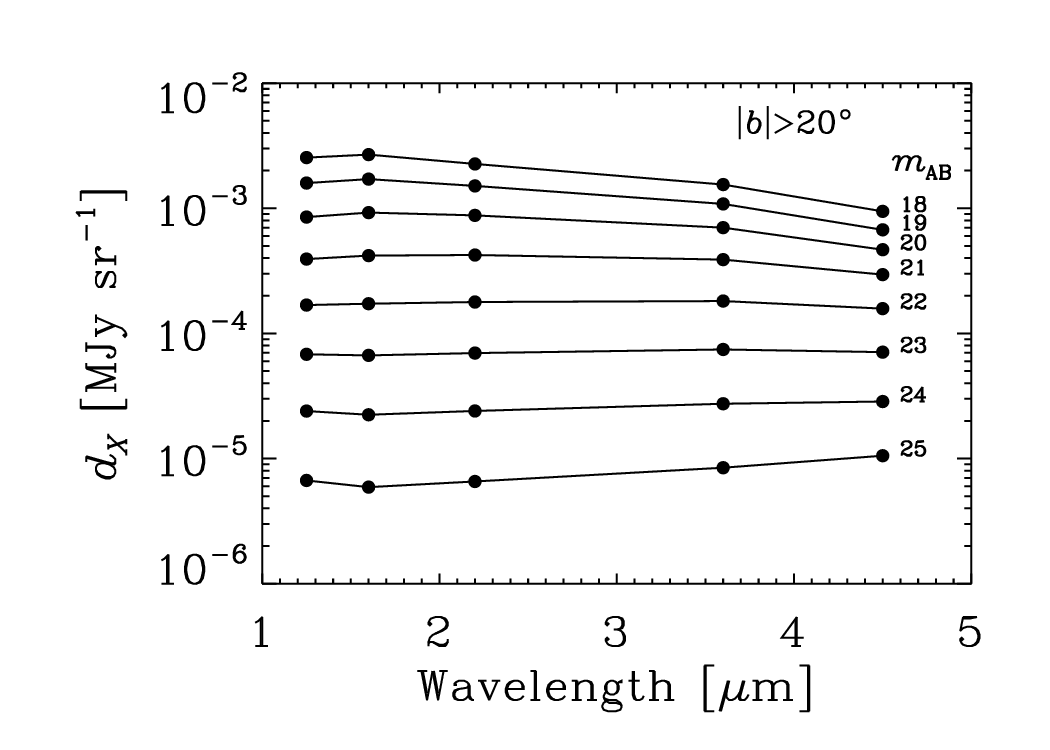}
   \includegraphics[width=3.5in]{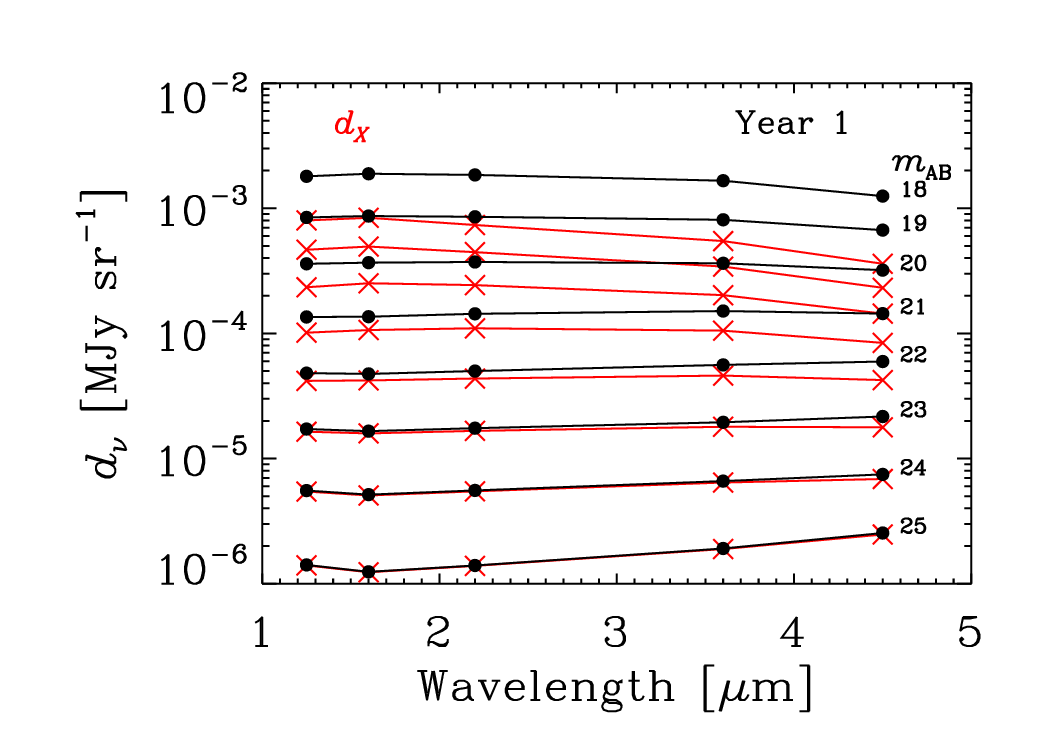}
   \caption{Characterization of Galactic stellar contributions. Top: The SKY model evaluated at $K$ band for stars with $m > 20.0$. Intensities are in
   units of MJy\,sr$^{-1}$ and are masked at $|b| <20\degr$ (left) and for Year 1 of \Euclid (right). 
   Middle: The monopole amplitude of the SKY model
   as a function of wavelength and the bright magnitude limit. 
   Bottom: The dipole amplitude of the SKY model
   as a function of wavelength and the bright magnitude limit. 
   For the $|b| <20\degr$ mask (left) only the dominant $X$-component of the dipole is plotted.
   For the Year 1 mask (right) the total dipole is shown in black, and the $X$-component is shown in red.
   At these magnitude cuts and for these masks, 
   the brightness and dipole of the model are dominated by the halo component, which is
   a spherical distribution of stars, peaked towards the Galactic center ($+X$).}
   \label{fig:fsm}
\end{figure*}


To assess the possible dipole arising from Galactic stars, if they are incompletely excluded from catalogs used to construct the IGL, we 
evaluated the SKY model
\citep{Wainscoat:1992,Cohen:1993,Cohen:1994,Cohen:1995} as implemented by \cite{Arendt:1998},
at a variety of wavelengths, and with cuts imposed to exclude stars brighter 
than chosen magnitude limits.
Figure~\ref{fig:fsm} shows the sky brightness predicted by the SKY model, with
masking applied generically for $|b| < 20\degr$ (left column) and specifically 
for \Euclid Year 1 (right column). We ran the HEALPix routine {\tt remove\_dipole}
on these masked models. The second row 
of Fig.~\ref{fig:fsm} shows the derived monopoles.
For the $|b| < 20\degr$ mask, the third row shows the $X$-components of the
dipoles ($Y$- and $Z$-components are orders of magnitude smaller).
For the Year 1 mask, the third row shows both the total dipole amplitudes, and the $X$-components only. For this masking, the $X$-component 
is only dominant at fainter magnitudes because the $X$ direction (towards 
the Galactic bulge) is not well sampled. 
The dipole amplitudes confirm that if stars are not excluded efficiently to
faint magnitudes, then they may contaminate the IGL with a significant dipole.
To probe the IGL dipole at the levels from Fig. \ref{fig:dipole}, we need to eliminate either $>99\%$ of the 
stars or choose sufficiently faint $m_0$, while keeping enough galaxies to 
ensure good ${\rm S/N}$. Figure \ref{fig:ngal_jwst} shows that even if strict magnitude cuts are needed
to exclude stars, there should be sufficient numbers of galaxies.

At the high latitudes of the Wide Survey, Gaia DR3 thoroughly samples the stellar disk
populations and reaches into the Galactic halo. On the basis of Gaia DR3 proper motions, it will be possible to reliably exclude Galactic stars to Gaia's
$G\sim21.4$\,mag or $\sim$9 $\mu$Jy \citep{Vallenari:2022}. 
Stars will only be a minority of all \Euclid detections fainter than this limit. The \Euclid pipeline will provide a flag in the final MER catalog indicating whether a detected source is a Galactic star, with a 1\% error rate. Thus by combining Gaia and standard
pipeline products it will be possible to reduce the level of stellar contamination by the required amount. In addition, star contamination can be entirely eliminated if one restricts the galaxy sample to the one that will be used for the weak lensing \Euclid\ measurements, that is objects with \IE\ size larger than $1.2 \, {\rm PSF_{FWHM}}$ \citep{Laureijs:2011}.

To estimate the extinction using different galaxy subsets as proposed here
(Sect. \ref{sec:ext_corr}), the subsets must be drawn from the same area of the sky such that
the dipole due to extinction, 
$\boldsymbol{D}_A$, is unchanged, but there is no requirement that the subsets be
complete in terms of source morphology. So while including some stars in the IGL calculation 
would generate systematic errors, there is no systematic error if the exclusion of Galactic stars
is conservative and some galaxies are excluded because they are mistaken for stars.

\subsubsection{Dipole contribution from clustering: \texorpdfstring{$m_0$}{TEXT}}

\begin{figure*}[ht] 
   \centering
   \includegraphics[width=7in]{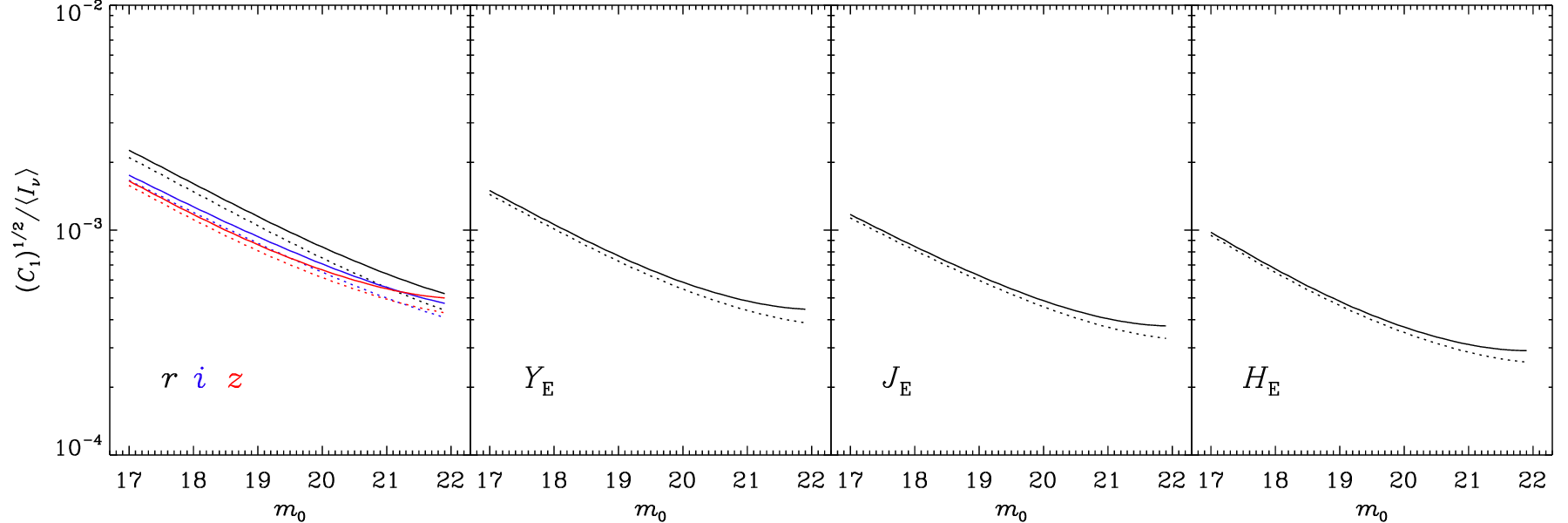} 
   \caption{Dimensionless dipole from clustering evaluated from the HRK reconstruction as described in \cite{Kashlinsky:2022}. The left panel includes the three photometric bands used by HRK that are covered by the wide \IE\ channel. Solid lines correspond to $m_1=24.5$ for $r$, $z$, and $i$, and $m_1=24$ for \YE, \JE,  and \HE\ bands, and dotted lines are for 0.5 magnitude fainter. This is to be compared to ${\sim}\,(4.75$--$5.75)\times10^{-3} (V/V_{\rm CMB})$ as displayed later in Fig. \ref{fig:velocity}.}
   \label{fig:dipole_clustering}
\end{figure*}

The lower limit on the magnitude of galaxies selected for this measurement is dictated by the requirement that the contribution to the probed dipole from their clustering is sufficiently lower than the one from the Compton--Getting effect produced by our motion, which is expected to be ${\sim}\,(4.75$--$5.75)\times10^{-3} (V/V_{\rm CMB})$ as displayed later in Fig. \ref{fig:velocity}. We have evaluated the dimensionless amplitude, $\sqrt{C_1}/\langle I_\nu\rangle$, of the clustering dipole using the HRK reconstruction as described in \cite{Kashlinsky:2022}, which is shown in Fig. \ref{fig:dipole_clustering}. The figure shows that for that term to be comfortably below the Compton--Getting terms one would want to select galaxies at $m_0\gtrsim 19$ in the VIS sample and $m_0\gtrsim 18$ for the NISP galaxies.

In real situation, post-launch we will compute the power from the {\it dipole-subtracted} IGL maps, then extrapolate from higher (say, $\ell\gtrsim 10$--$20$) harmonics to $\ell=1$ using (after verifying) the Harrison--Zeldovich spectrum ($C_\ell \propto \ell$). For now we already have 1/8 of the sky from Flagship2.1 (via CosmoHub) simulations. Uniformity of $m_0$ across the sky also can be tested via the uniformity of the shot-noise power component in source-subtracted CIB achieved from the source-subtracted CIB studies with \Euclid \citep{Kashlinsky:2018}.

\subsubsection{Choice of \texorpdfstring{$m_1$}{TEXT}}

\begin{figure}[ht] 
   \centering
   \includegraphics[width=3.5in]{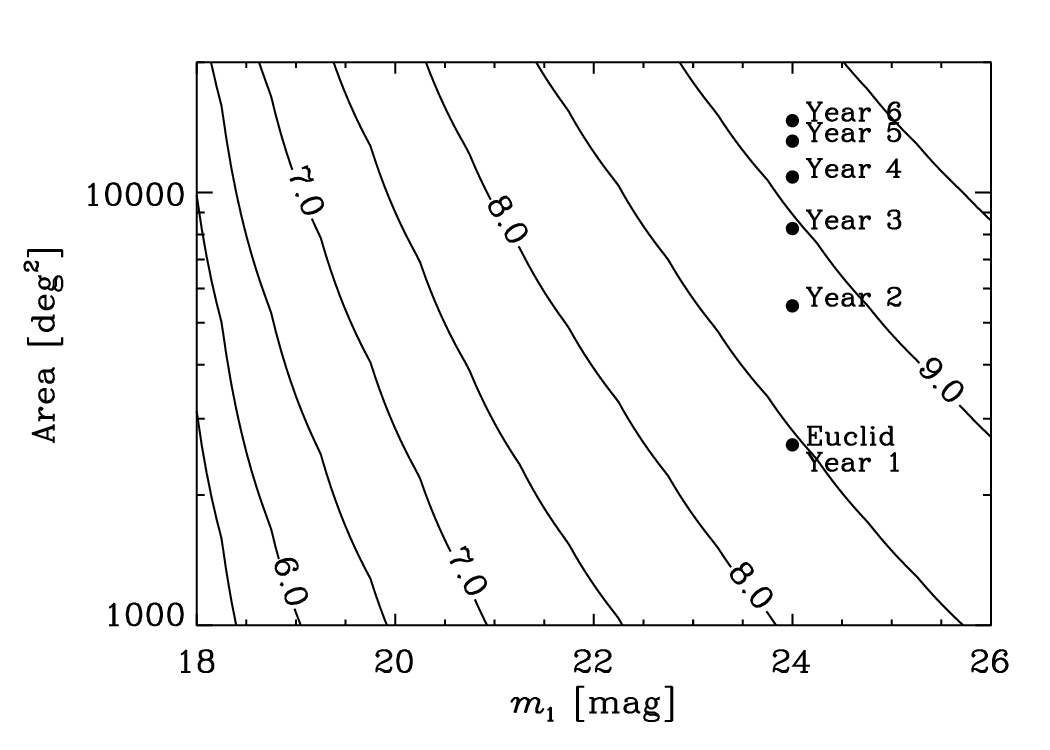} 
   \caption{Contours showing the decimal log of the expected number (\JE\ band) of galaxies [from Fig. 11 of \cite{Windhorst:2023}] as a function of $m_1$ and area, for $m_0 = 18$.
   The dots mark the nominal $m_1 = 24$ and survey area expected after each addition year of the Wide Survey.}
   \label{fig:ngal_m1}
\end{figure}

\cite{Scaramella-EP1} show that the Euclid Wide Survey is expected to reach 
its intended limiting magnitudes of $\IE = 24.5$ ($10 \sigma$ extended source),
and $\YE,\ \JE,\ \HE = 24$ ($5 \sigma$ point source) \citep{Laureijs:2011}. 
So we will assume these values for $m_1$.
In practice, a brighter limit for $m_1$ may be helpful for better source flux 
accuracy and potentially more reliable star galaxy separation. Conversely,
it should be possible to choose a fainter limit for $m_1$ (up to ${\sim}\,1$ mag), 
though at the cost of limiting the analysis to a smaller fraction of the sky. 
These more sensitive 
regions (due to low zodiacal and Galactic foregrounds) will be 
covered in the earlier years of the survey. However in general, given the relatively
shallow slope of galaxy counts at $m_1 \sim 24$, it is better to choose larger 
rather than deeper areas in order to maximize the number of sources used
for computing the IGL, and maximize the ${\rm S/N}$ of its dipole measurement.
The expected galaxy counts as a function of survey area and $m_1$ are shown in Fig. \ref{fig:ngal_m1}.

\subsubsection{The overall magnitude range required here}
\label{sec:m0m1-final}

Thus we concentrate on galaxies in the magnitude range of
\begin{equation}
    19 \leq m \leq 24.5 \hspace{2cm} {\rm VIS}\,,
\label{eq:mag_range_vis}
\end{equation}
\begin{equation}
   18 \leq m \leq 24 \hspace{2cm} {\rm NISP}\,.
\label{eq:mag_range_nisp}
\end{equation}
In the following sections we will select galaxies from the available simulation catalog according to Eqs. (\ref{eq:mag_range_vis}) and (\ref{eq:mag_range_nisp}).

\begin{figure}[ht!] 
   \centering
   \includegraphics[width=3.5in]{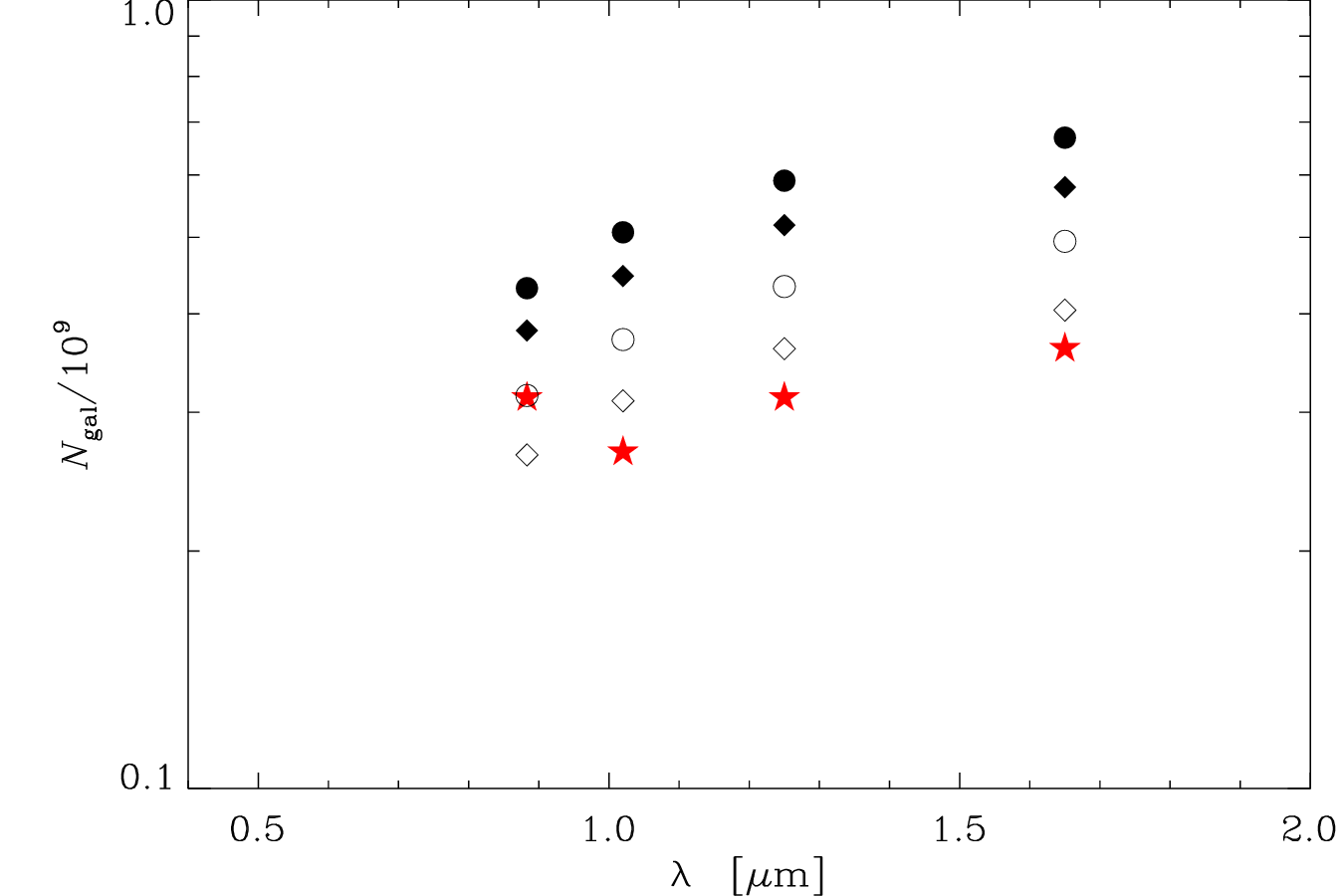} 
   \caption{Number of galaxies after Year 1 of $\num{2550}$ deg$^2$ coverage for $m_0\leq m \leq m_1$. Circles correspond to $m_0=18$ and diamonds to $m_0=22$. Filled/open symbols correspond to $m_1=25/24.5$. Red asterisks show the numbers for the range in Eqs. (\ref{eq:mag_range_vis}) and (\ref{eq:mag_range_nisp}) using the JWST counts \citep{Windhorst:2023}.}
   \label{fig:ngal_jwst}
\end{figure}

Figure \ref{fig:ngal_jwst} shows the number of galaxies expected around the required magnitude range using the JWST counts from \cite{Windhorst:2023}. Given that the statistical uncertainty is $\propto1/\sqrt{N_{\rm gal}}$, we expect to have only minor variations in the uncertainty as the magnitude range is refined if necessary.

\subsection{Understanding details of extinction in the measurement}
\label{sec:extinction}

\subsubsection{Extinction}


Extinction from dust in our own Galaxy would imprint apparent structure on an 
otherwise isotropic extragalactic background, whether directly measured as the CIB or reconstructed from 
observed galaxy brightnesses. Since the extinction is most simply a function 
of Galactic latitude, the strongest effect is expected in the quadrupole. 
One typically achieves $\epsilon_A\sim 0.1$ per band with photometric measurements and $\epsilon_A\sim 0.01$ with spectroscopic measurements. However
the Galactic ISM is highly and irregularly structured, so a dipole component 
to the extinction will be present as well. Due to Galactic structure the effects are expected to be smallest for the $Z$-component \citep[e.g.][]{Gibelyou:2012}. The extinction dipole goes as $\lambda^{-2}$ from \IE\ to \HE\ in the opposite trend than that of the IGL. 

\begin{figure*}[tp] 
   \centering
   \includegraphics[width=7in]{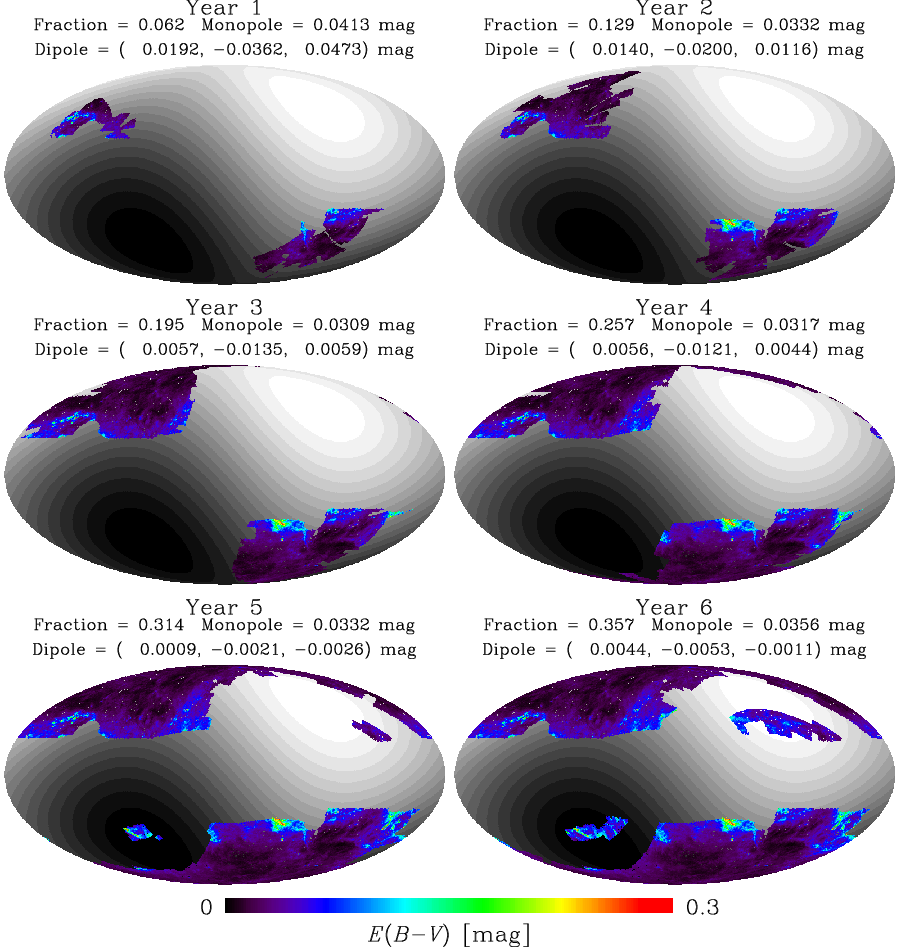} 
   \caption{Cumulative coverage of the SFD $E(B-V)$ reddening map in Galactic coordinates. $A_V \sim 3.1 E(B-V)$ and $A_{\rm near-IR} \sim E(B-V)$.
   For each year, annotations list the fraction of sky covered, the monopole, $\langle E(B-V)\rangle$, and the three dipole components.
   The uncovered regions of the maps show contours of the normalized CMB dipole
from -1 to 1 in steps of 0.1, to show that even the Year 1 coverage will 
sample a significant range of dipole intensity.  Our measurements of the CIB dipole will not assume the CMB prior direction and will measure the former directly, so the CMB dipole contours are displayed only for illustrative purposes. }
   \label{fig:sfd0}
\end{figure*}

\begin{figure*}[ht] 
   \centering
   \includegraphics[width=3.8in]{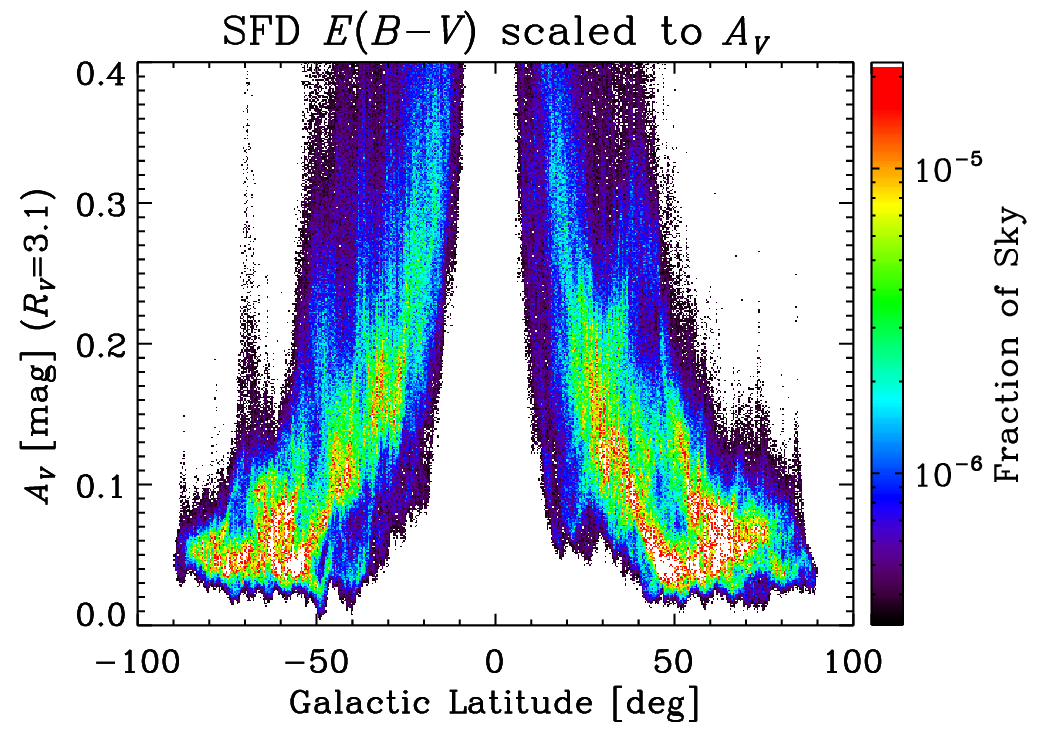} 
   \includegraphics[width=3in]{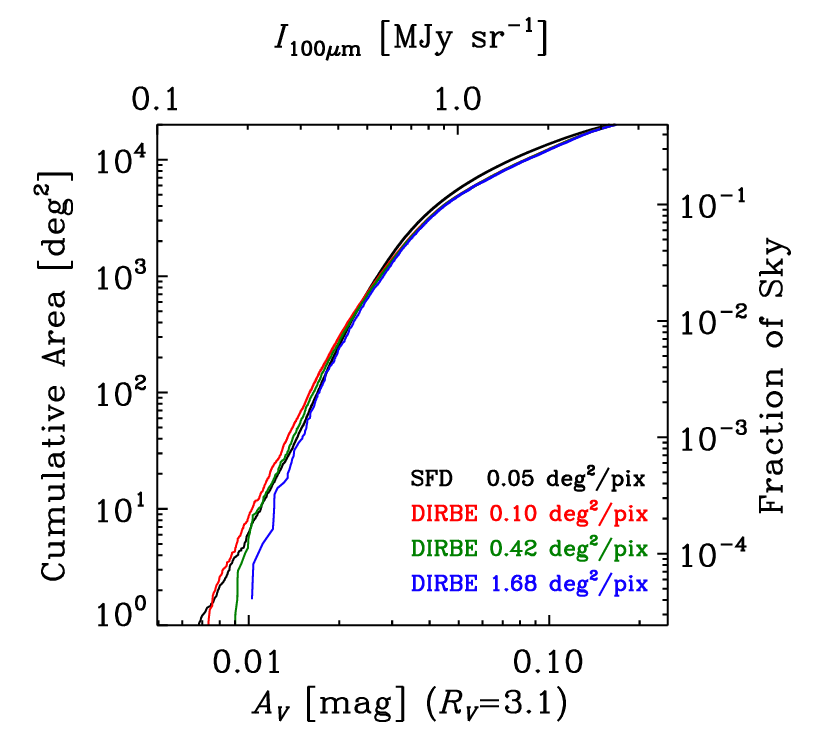} 
   \caption{Characterization of Galactic extinction. Left: $A_V$ as a function of Galactic latitude derived by 
   converting SFD $E(B-V)$ results using $R_V = 3.1$.
The color excess (reddening) is $E(B-V) = (B-V)_{\rm observed} - (B-V)_{\rm intrinsic}$ and
$R_V = A_V/E(B-V)$. The color bar adjacent to the right of the panel shows the fraction of the sky at $A_V$ given in colors from the main plot. Within the range of the \Euclid\ Wide Survey $A_V\ll 1$.
Right: The cumulative area where the SFD extinction is less than $A_V$ or where
the COBE/DIRBE 100 $\mu$m intensity (shown at 3 resolutions) is less than
$I_{100\ \mu{\rm m}}$.
}
   \label{fig:extinction1}
\end{figure*}

Figure \ref{fig:sfd0}
shows maps of Galactic reddening, $E(B-V)$ from \citep[][hereafter SFD]{Schlegel:1998}.
The reddening map is masked by the cumulative coverage of the Euclid Wide Survey
for each of the six years of the mission. 
Each of these masked images is fit for monopole and dipole components 
using the HEALPix routine {\tt remove\_dipole}, and the resulting amplitudes (in magnitudes)
of each component are listed.
Figure \ref{fig:extinction1} (left) shows the 
reddening converted to extinction [using $A_V = R_V\ E(B-V)$] and plotted as a function of 
latitude. At a fixed latitude there can be large variations in extinction. However 
relatively low extinction regions may be found at latitudes as low as $|b| \approx 40\degr$.
Figure \ref{fig:extinction1} (right) shows the total area of the sky that has extinction lower
than a given $A_V$ (i.e. below a horizontal line in the left panel), or equivalently the area
where the 100 $\mu$m emission is below $I_{\rm 100 \mu m}$, as 
$E(B-V) \approx 0.016\ I_{\rm 100 \mu m}$. The 100 $\mu$m results are shown for binning 
of the DIRBE measurements at three different angular scales.

\begin{figure*}[ht] 
   \centering
   \includegraphics[width=7in]{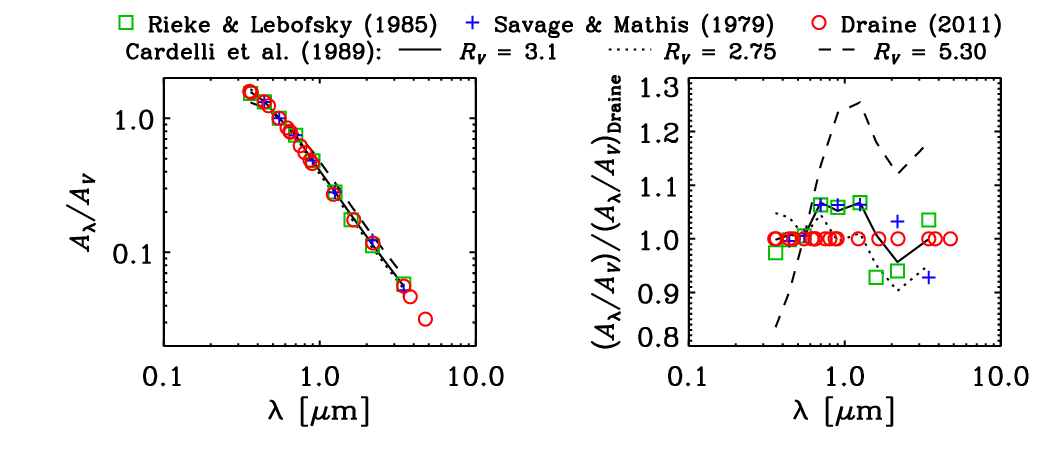} 
   \caption{Comparison of Galactic extinction laws. Left: Galactic extinction laws as reported by \cite{Draine:2011} (red), 
   \cite{Rieke:1985} (green), and \cite{Savage:1979} 
   (blue). The solid, dotted, and dashed lines are the \cite{Cardelli:1989} extinction law
   evaluated for $R_V = 3.1$, 2.75, and 5.30, respectively. Right: The same extinction laws are plotted normalized to the \cite{Draine:2011}
   extinction law to show that differences of several percent exist, and that variations of $\gtrsim 10\%$ can arise from $R_V$
   variations.}
   \label{fig:extinction2}
\end{figure*}

Maps of reddening or extinction are detailed but are subject to systematic errors. Most commonly
used ones originate in observations of far-IR emission \citep[e.g.][]{Schlegel:1998}. Models are required to convert the 
emission to a dust column density, then into a reddening $E(B-V)$, then into extinction $A_V$, and 
finally to extinction at the desired wavelength $A_\lambda$.  Factors that influence these steps 
are dust temperature and composition, the ratio of total to selective extinction $R_V$, and the 
reddening law $A_\lambda$/$A_V$ (see Fig. \ref{fig:extinction2}). All of these are known to vary as a function of line of sight, 
but it is common (sometimes necessary) to simply adopt standard mean values for $R_V$ 
and the reddening law. This means that there will be some imprint of extinction even on an 
IGL background that is constructed from extinction corrected source magnitudes. The imprint
will be that of the errors in the extinction correction, which   will not necessarily have the same 
pattern on the sky as the extinction itself. Ultimately this does not matter for the method that is presented below which will remove the extinction dipole contribution, or the dipole from the residual extinction corrections. 
The example of the SFD extinction maps used here shows that extinction introduces a non-negligible diffuse dipole component, although it may differ in some detail from other Galactic extinction maps \citep[e.g.][]{Schlafly:2011,Planck-Collaboration:2014,Delchambre:2022} and their wavelength dependence \citep[e.g.][]{Predehl:1995,Draine:2011}. In what follows we introduce methodology to remove the dipole contribution from extinction (in the limit of low extinction, $A_V \ll 1$) independent of any estimated extinction, or extinction correction, map.


\subsubsection{Managing the extinction contributions}\label{sec:ext_corr}

While the Galactic extinction can interfere with probing the intrinsic IGL dipole, its interference can be removed with the method proposed in this section. Let us say that the extinction magnitude correction, $A_\nu\ll 1$, is known to within relative accuracy $\epsilon_A$. Then, after the extinction correction the flux in Eq. (\ref{eq:igl}), $\tilde{I}_\nu\equiv 10^{-0.4\epsilon_A A}I_\nu$, becomes
\begin{equation}
\tilde{I}_\nu(l,b)=\left[1-0.4\,(\ln{10}) \epsilon_AA_\nu\right]I_\nu(l,b) \equiv I_\nu(l,b) + \Delta I_\nu(l,b).
\label{eq:igl_extinction}
\end{equation}
If no extinction corrections are made then one should read $\epsilon_A=1$. This leads to the extinction uncertainty contribution to the measured IGL of
\begin{equation}
\Delta I_\nu(l,b)=\pm0.92\epsilon_A A_\nu\, I_\nu(m_0,m_1){|_{(l,b)}}\,.
\label{eq:igl_correction}
\end{equation}
We now write $A\equiv \langle A\rangle+\delta A, I\equiv \langle I\rangle+\delta I$. 
Assuming that the extinction map in Fig. \ref{fig:extinction1} has a dipole $\boldsymbol{D}_A$ we can write taking $\epsilon_A=$const across the sky leads to the following uncertainty in the IGL dipole, Eq. (\ref{eq:igl_correction}), from extinction corrections
\begin{equation}
\begin{split}
\tilde{\boldsymbol{d}}_\nu =~&\boldsymbol{d}_\nu\left[1-0.92\epsilon_A\langle A_\nu\rangle\right]\\  & -0.92 \epsilon_A\boldsymbol{D}_A \langle I_\nu\rangle+ O[{\rm dipole}(\delta A\,\delta I)]\,.
\end{split}
\label{eq:decompose}
\end{equation}
Here $D_A$ is the dipole of the extinction map to be evaluated from the selected sky region in Fig. \ref{fig:extinction1} (left). (More generally, if the relative extinction correction, $\epsilon_A$, varies across the sky, the $\epsilon_A\boldsymbol{D}_A$ term should be understood as the dipole of the $\epsilon_A A$ product). The spherical harmonic expansion is defined via: $I_\nu = \sum_{\ell m} d_{\ell m} Y_{\ell m} \; ; \; A_\nu = \sum_{\ell m} D_{\ell m}Y_{\ell m}$. Over the full sky higher-$\ell$ harmonics do not couple to lower-$\ell$ ones. The extinction component of the dipole due to Eq. (\ref{eq:igl_correction}) decreases with wavelength 
($\propto\lambda^{-2-\alpha_\nu}$) 
and, very generally, has a very different dependence on $\lambda$ from that of the IGL, $d_\nu$, shown in Fig. \ref{fig:dipole}. This should enable component separation in Eq. (\ref{eq:igl_extinction}). 

With Eq. (\ref{eq:decompose}) the task of minimizing the extinction contribution to the probed IGL dipole is reduced to: 1) finding a region large enough to contain many galaxies [Table 1 in \cite{Kashlinsky:2022} shows that $N_{\rm gal}\gtrsim \num{1000}$--$\num{2000}$ deg$^2$  could be enough] that has small $\langle A_\nu\rangle$ and 2) where $D_A$ is sufficiently close to zero.
For the \Euclid bands the dimensionless IGL dipole is $(3-\alpha_\nu)V/c \simeq 0.5\%$ if the measured CMB dipole is entirely kinematic (if not, then it would be ${\simeq}\,0.5\%V/V_{\rm CMB})$. Hence in the absence of further corrections discussed below, one would want to select the sky regions with $D_A/\langle A\rangle \lesssim 0.2 \epsilon_A^{-1}$ in \IE, and significantly more relaxed in \HE, in order to probe, in each of the bands, the IGL dipole with the expected dimensionless amplitude of ${\simeq}\,0.5\%$ per Fig. \ref{fig:velocity}.
The above hinges on three assumptions: 1) $D(\delta I\delta A)\ll \langle I\rangle D_A, \langle A\rangle d_\nu$, 2) $\langle A\rangle, A \ll 1$, and 3) the coupling with higher $\ell$-order terms can be neglected (for now). Choosing a region where $D_A$ is minimal can be accomplished by selecting regions of the sky where $E(B-V)$ is roughly constant.
Figure \ref{fig:sfd0} illustrates that while the total sky observed by \Euclid after any year contains a dipole, 
selection of an area of sky with near constant $E(B-V)$ can reduce the dipole amplitude by roughly an order of magnitude.

However, a more efficient technique would work as follows: Fig. \ref{fig:extinction1} shows that the additive term, when multiplying $\boldsymbol{d}_\nu$  in the RHS of Eq. (\ref{eq:decompose}), is $0.92 \langle A_\nu\rangle \ll 1$ and can be neglected. It further affects only the precise conversion of the dipole into the velocity amplitude (at less than a few percent level), not its direction.  Assuming that the last term on the RHS is negligible, a firmer way to eliminate the extinction terms in Eq. (\ref{eq:decompose}) would be to 1) divide the large \Euclid galaxy sample into two groups with very different  spectral/morphological properties and 2) use one to eliminate the extinction dipole in the other. For example, if we define a subsample ``a'' with, say $\alpha_\nu^{\rm a}\gtrsim 1$, and magnitude range $m_0^{\rm a}<m < m_1^{\rm a}$,  one would get for its dipole
\begin{equation}
\frac{\tilde{\boldsymbol{d}}_\nu^{\rm a}}{\langle I_\nu^{\rm a}\rangle}=(3-\alpha_\nu^{\rm a})\frac{\boldsymbol{V}}{c}-0.92 \boldsymbol{D}_A(\lambda)\,,
\label{eq:subsample1}
\end{equation}
and similarly for the subsample ``b''. Then subtracting the two would lead to the residual wavelength-dependent dipole from extinction corrections determined as
\begin{equation}
\begin{split}
\boldsymbol{D}_A(\lambda)= & \frac{(0.92)^{-1}}{(\alpha_\nu^{\rm a}-\alpha_\nu^{\rm b})}\\
 & \times\left[ 
(3-\alpha_\nu^{\rm a})\frac{\tilde{\boldsymbol{d}}_\nu^{\rm b}}{\langle I_\nu^{\rm b}(m_0^{\rm b},m_1^{\rm b})\rangle} -(3-\alpha_\nu^{\rm b})
\frac{\tilde{\boldsymbol{d}}_\nu^{\rm a}}{ \langle I_\nu^{\rm a}(m_0^{\rm a},m_1^{\rm a} )\rangle}\right]\,.
\end{split}
\label{eq:extinction_final}
\end{equation}

Eq. \ref{eq:extinction_final} would give the systematic contribution to the kinematic CIB/IGL dipole from the residual extinction magnitude corrections at each wavelength. If they turn out to be non-negligible, the wavelength-independent Compton--Getting velocity term will be determined from:
\begin{equation}
(\alpha_\nu^{\rm a}-\alpha_\nu^{\rm b})\frac{\boldsymbol{V}}{c}=\boldsymbol{\Delta}_{\rm ab}\equiv -[\boldsymbol{{\cal U}}_{\rm a}-\boldsymbol{{\cal U}}_{\rm b}]\,,
\label{eq:v_final}
\end{equation}
where we further defined
\begin{equation}
\boldsymbol{{\cal U}}_{a,b}\equiv \frac{\tilde{\boldsymbol{d}}_\nu^{\rm a,b}}{\langle I_\nu^{\rm a,b}(m_0^{\rm a,b},m_1^{\rm a,b})\rangle}
\label{eq:u4v}
\end{equation}
to be used later. Note that the rms in Eq. (\ref{eq:v_final}) depends on the relative difference in $\alpha_\nu$ for the subsamples, rather than the absolute Compton--Getting amplification, $(3-\alpha_\nu)$.

The values of $\boldsymbol{D}_A$ determined empirically here will help verify the accuracy of the \Euclid extinction corrections. The errors resulting from this procedure are discussed in the next section using simulated \Euclid catalogs. Naively speaking the statistical error on $D_A$ determined here from sample ``b'' would be 
$\sigma_{D_A}\sim D_A ({N^b_{\rm gal})}^{-0.5}$.

We will have plenty of galaxies to do this per Table 1 of \cite{Kashlinsky:2022} and we want to choose subsample galaxies with, say, $\IE-\YE<0$, $\YE-\JE<0$, $\JE-\HE <0$ (in the observer frame) so that their IGL has  positive $\alpha_\nu$, ideally closer to $\alpha_\nu^{\rm RJ}=2$ and $m_0$ faint enough to ensure that the dipole from the clustering component is small.  \cite{Bisigello:2020} show in their Figs. 6 and 7 that one can choose a substantial subsample of such galaxies. Since $I_\nu/\nu^3$ is Lorentz-invariant for each subsample its $\alpha_\nu$ is independent. Hence we can select subsamples with a significantly positive $\alpha_\nu$ from \IE\ to \HE\ bands and the main sample from galaxies with $\alpha_\nu$ substantially negative to compensate for the reduction in the overall $N_{\rm gal}$. Then we would choose the optimal $(m_0^{\rm sub}, m_1^{\rm sub}, m_0, m_1)$ to enable sufficient reduction in the dipole from clustering. Then use Eq. (\ref{eq:v_final}) and propagate the errors; with this in mind choose areas where $\tilde{d}_\nu$ (and $D_A$) are minimal. The resultant velocity, Eq. (\ref{eq:v_final}), must be the same when derived at all bands and the absorption dipole, $\boldsymbol{D}_A$ must point in the same direction and have the corresponding wavelength dependence decreasing with $\lambda$.  This technique can be straightforwardly generalized to more subsamples with sufficiently different $\alpha_\nu$.

\subsection{Accounting for the Earth's orbital motion}
\label{sec:orbital}

The overall cosmological information was not yet available when the Compton--Getting effect was introduced almost 100 years ago as a way to probe the orbital motion of the Earth from cosmic ray observations \citep{Compton:1935}. The motivation is reversed here, but the now well-known Earth's orbital motion is important to account for in any precision measurement.

Dipole measurements having ${\rm S/N} \gtrsim 10$ will be sensitive to 
(require correction for) the Earth's (or spacecraft's) orbital motion around the Sun at $V_\oplus\sim 30$ km s$^{-1}$. 
Mapping the data in a coordinate system that rotates to keep the Sun fixed [i.e. $(l_{\rm ecl}-l_{\sun}, b_{\rm ecl})$ 
rather than $(l_{\rm ecl}, b_{\rm ecl})$], can be used to determine the ``Solar anisotropy'' \citep[e.g.][]{Abbasi:2012}. 
In these coordinates the dipole can be fit, and then its brightness can be subtracted from each of the observations 
as mapped into a standard fixed coordinate system.
In principle $(V_{\oplus}/c)\cos\Theta$ is already known, but this fitting provides an empirical measure of $\langle I \rangle$, which is needed for the 
subtraction. 
For \Euclid this dipole is poorly sampled in Year 1 because all observations are towards the ecliptic poles, perpendicular to the Earth's orbital motion dipole. However, 
for the same reason, the effect of this dipole will be correspondingly small in the Year 1 data. As the years progress and lower ecliptic latitudes 
are observed, sensitivity to the Solar anisotropy dipole increases.
In this fixed-Sun coordinate system, the cosmological kinematic dipole on the sky will be a noise term that averages down 
for \Euclid as more sky is sampled.

A more ideal solution is to modify dipole fitting routines (e.g. HEALPix {\tt remove\_dipole}) 
to include input specifying the time (or Solar elongation) of observation for each pixel and then 
subtract off the ``Solar anisotropy'' dipole as part of the fitting. In this case the fitting does 
not involve any additional free parameters, since the time (or Solar elongation) of the observations 
is a known quantity. 

\subsection{Photometry and magnitudes}
\label{sec:photometry}

Because galaxies are extended sources without cleanly defined edges,
there are multiple ways to define the magnitude of a galaxy.  Standard
techniques include using aperture photometry within a radius set by
the shape of the galaxy profile (e.g., Petrosian photometry), and
fitting a generic profile shape to galaxies to measure something akin
to a total magnitude [such as the CModel magnitudes developed by the
Sloan Digital Sky Survey \citep{Abazajian:2004}, and being used by the Hyper Suprime-Cam
survey and the {\it Rubin} Observatory \citep{Bosch:2018}].  
  Such measurements
will have Malmquist-like biases (i.e., where more faint galaxies
scatter into the sample at the faint magnitude limit than the
converse), which will depend on the sources of noise.  If those
sources of noise are not symmetric across the sky, this could imprint
a false dipole moment into the inferred magnitude range.  Given that
the Euclid Wide Survey will have quite uniform coverage, and the location
of \Euclid at L2 means that the imaging depth should have little
temporal variation, we anticipate that this will be a subdominant
effect in the IGL dipole measurement, but we will have to measure it.  The
sky background of the measurements will not be uniform across the sky,
adding to the noise of the galaxy photometry, and possibly leading to
additional systematic errors.  This varying sky background is 
dominated by zodiacal light, with a possible contribution 
at low Galactic latitudes from
reflected starlight from diffuse dust in the Milky Way (optical
cirrus), although the latter is again likely to be subdominant. 



The availability of multi-band measurements of our galaxies allows us
to carry out the measurements of the dipole from multiple independently
defined subsamples, which should allow us to test the robustness of
the results to many 
of the systematic errors we
have described in this paper.  In particular, we envision measuring
the dipole on subsets of galaxies divided in the following ways: 
\begin{itemize}
  \item Dividing up the galaxies into different regions of color-color
    space, selected to identify galaxies at different (photometric)
    redshifts and different physical properties, and measuring the
    dipole for each;
  \item Dividing the galaxies into bins of magnitude;
  \item Dividing the sky by the season or year in which each patch was
    observed, to test for systematic biases in photometric zero points
    with time or position of the spacecraft;
    \item Dividing the galaxy sample by the morphology of the galaxy,
      as measured, e.g., by S\'ersic index. 
\end{itemize}
  Seeing consistent dipoles among all these divisions of the data will
  give us confidence in the robustness of the results, and
  will allow us to constrain any residual systematic effects. 





\subsection{Systematic corrections}
\label{sec:systematic}

The dipole from the configuration by Eq. (\ref{eq:igl}) will have additional contribution arising due to the Galaxy participating with the bulk motion $\Delta m=-(2.5\logten {\rm e}) [1-\beta(m)]\frac{V}{c}\cos\Theta$ for CIB sources with SED $f_\nu\propto \nu^{-\beta}$ \citep{Ellis:1984,Itoh:2010}. The correction to the dipole from this effect can be evaluated by modifying the upper and lower limits on the integration in Eq. (\ref{eq:dndm}). 
Since $V\ll c$ the dipole due to Eq. (\ref{eq:dipole_alpha}) will be modified by the $m_0$, $m_1$ variation to

\begin{equation}
\boldsymbol{d}_\nu (m_0<m<m_1)= [(3-\alpha_{\nu,m_0<m<m_1})+ \Delta\alpha_\nu]\frac{\boldsymbol{V}}{c} \langle I_\nu(m_0<m<m_1)\rangle\,,
\label{eq:dipole_final}
\end{equation}
with
\begin{equation}
 \Delta\alpha_\nu=Q_\nu(m_1)[1-\langle\beta(m_1)\rangle]-Q_\nu(m_0)[1-\langle\beta(m_0)\rangle]\,,
\label{eq:dipole_correction}
\end{equation}
where we have defined
\begin{equation}
    Q_\nu(m)\equiv \frac{\frac{\diff I_\nu}{\diff m}|_{m}}{\langle I_\nu(m_0\!\!<\!\!m\!\!<\!\!m_1)\rangle}\,.
    \label{eq:q_nu}
\end{equation}

It was argued that the correction $\Delta\alpha_\nu$ in general is small \citep{Kashlinsky:2022}, but here it must be evaluated explicitly for the \Euclid configurations. This correction will affect only the (systematic) conversion of the measured IGL dipole into the effective velocity amplitude.

\begin{figure}[ht] 
   \centering
   \includegraphics[width=3.5in]{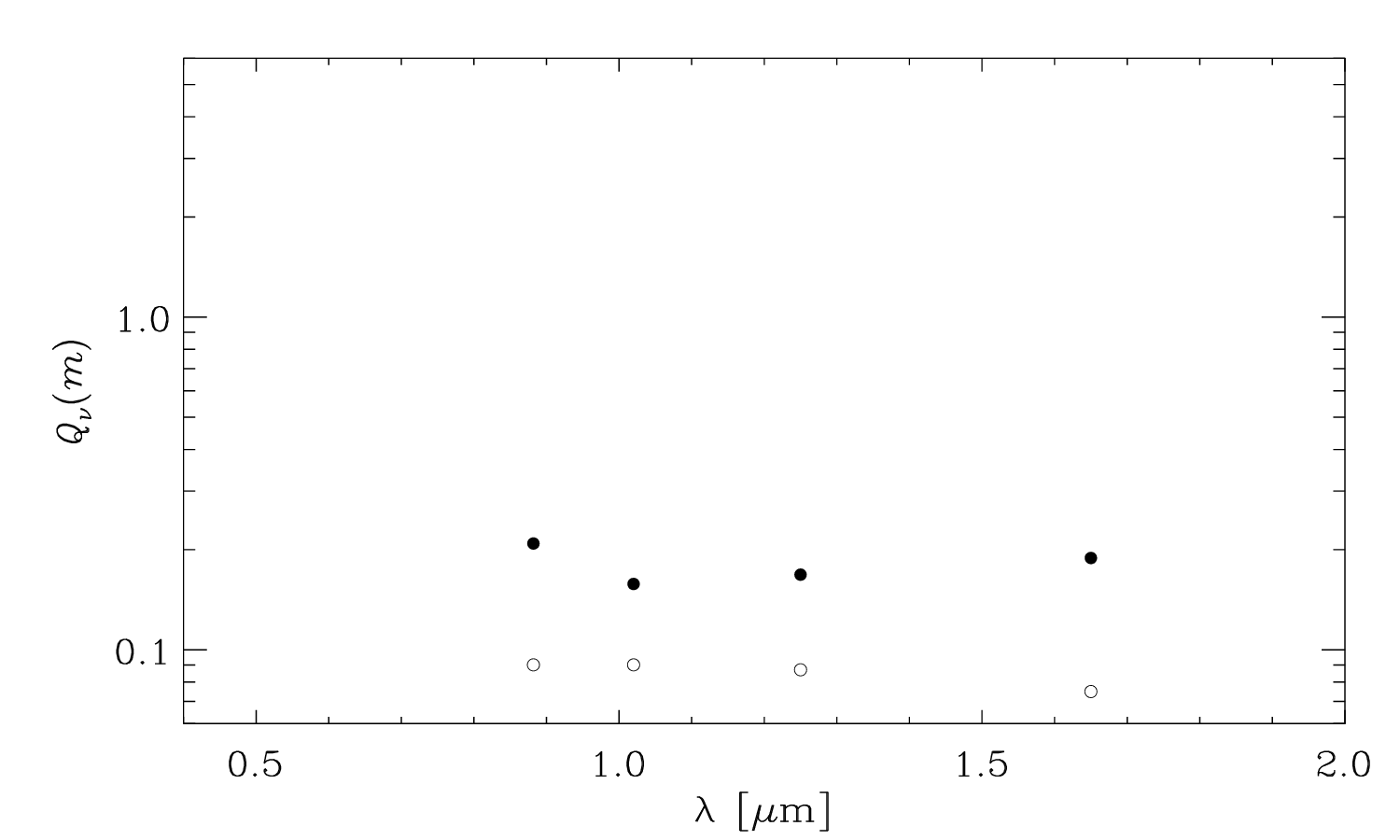} 
   \caption{Values of $Q_\nu(m_0)$ (Eq. \ref{eq:q_nu}) vs. $\lambda$ are shown with filled circles and $Q_\nu(m_1)$ with open circles at the \Euclid four bands using the observed JWST galaxy counts from \cite{Windhorst:2023}. The values of $(m_0,m_1)$ were selected per Eqs. (\ref{eq:mag_range_vis}) and (\ref{eq:mag_range_nisp}). The displayed vertical range reflects the Compton--Getting amplifications, $(3-\alpha_\nu)\sim 4$--$5.5$, reachable here.}
   \label{fig:q_nu}
\end{figure}

Figure \ref{fig:q_nu} shows the values of $Q_\nu(m_0)$ and $Q_\nu(m_1)$ for the range of magnitudes defined by Eqs. (\ref{eq:mag_range_vis}) and (\ref{eq:mag_range_nisp}) at the four \Euclid bands with the vertical axis deliberately plotted to compare with  the expected near-IR Compton--Getting amplification of $(3-\alpha_\nu)\simeq 4$--$5.5$ at these bands. We used the latest JWST observations of galaxy counts shown in Fig. \ref{fig:counts}. The expected overall contribution, $\Delta\alpha_\nu$, after accounting for the SED/color terms will be incorporated from \Euclid simulations in the next section. 

The Euclid Wide Survey will have measurements at an effective $\lambda$, within the four filters, from which we will need to reconstruct the genuine high-precision $I_\nu$, Eq. (\ref{eq:dndm}) and its derivative, $\alpha_\nu$ together with their uncertainties. This is discussed and quantified later in Sect. \ref{section5.3}. To aid with this we will have photometric $z$ for the entire sample and spectroscopic $z$ for many of them. In particular, as discussed in \cite{Scaramella-EP1,Laureijs:2011} we would expect upward of $5\times10^6$ galaxies with spectroscopic redshifts already in Year 1 providing a good sample for such estimates.

\subsection{Photometric zero points}
\label{sec:zeropoints}

A precision measurement of the intrinsic dipole in galaxy counts
requires precision photometric calibration over the full survey
footprint.  \Euclid's photometric calibration is good, but is not
perfect, and any dipole component in the error in photometric
calibration will translate directly into a false dipole signature.  In
particular, the dipole moment of a calibration error $\Delta m_{\rm dipole}$ will
behave exactly like that described in
Eq.~(\ref{eq:dipole_final}), and the (false) bulk motion $V_{\rm false}$
inferred from this dipole is given by
\begin{equation}
  \Delta m_{\rm dipole} = -(2.5 \logten {\rm e})[1 - \beta(m)] \frac{V_{\rm false}}{c}.
\end{equation}
For reasonable values of $\beta$, a $V_{\rm false}$ of 300 km/s (i.e., the
amplitude of the signal we are trying to measure) would correspond to a
0.5\% $\Delta m_{\rm dipole}$.  

The Euclid Wide Survey strategy is described in detail in
\cite{Scaramella-EP1}, and the calibration of the photometric system is
described in \cite{Schirmer-EP18}.  Most of the roughly $\num{15000}$
deg$^2$ of sky will be observed only once, meaning that strategies
that use overlaps to tie photometric calibration together 
\citep[e.g.][]{Padmanabhan:2008,Burke:2018} will be of only limited use.  Rather,
every 25--35 days, \Euclid will observe a self-calibration field within
its continuous viewing zone near the north ecliptic pole, to
redetermine the photometric calibration of the bands.
Section 6.3.3 of \cite{Schirmer-EP18} predicts that temporal changes
in the photometric calibration will be determined to an accuracy of
1--2 milli-mag, although the formal requirement on the calibration is
far looser, 1.5\% for the NISP instrument.  It is less clear what the
dipole moment across the 
sky of this calibration error will be.  If the calibration error is
uncorrelated across the sky in each field that is observed, the
dipole will be of order the error per independent calibrated patch,
divided by the square root of the number of patches. 
In this context, the number 
of patches is determined by how often the calibration is checked.
This means $\sim12$ patches per year of survey operations.
This would
suggest a residual dipole error due to calibration uncertainties that
is small relative to the expected signal.  However, if the calibration
errors are position-dependent in some way (e.g., somehow matching the
scanning pattern of \Euclid, or dependent on the angular separation of
any given field from the self-calibration field), the systematic error
on our measurement may be considerably larger, and we will need to
work with the \Euclid calibration team to explore the systematic
calibration errors.

  We anticipate that any calibration errors, if they are due to, e.g.,
  time-dependent changes in \Euclid's throughput, will be correlated
  between \Euclid's different bands.  This means that comparisons of
  the inferred dipole between bands is unlikely to be a panacea to
  such effects.  Similarly, calibration problems will be likely to affect
  the photometry of different galaxy populations in the same way, so
  splitting galaxies by, e.g., measured color, will not be
  informative.  




\section{Applying to the upcoming Euclid Wide Survey}
\label{section5}

\subsection{Evaluating overall statistical uncertainties}
\label{subsection5.1}

\begin{figure}[htbp!] 
      \includegraphics[width=3.5in]{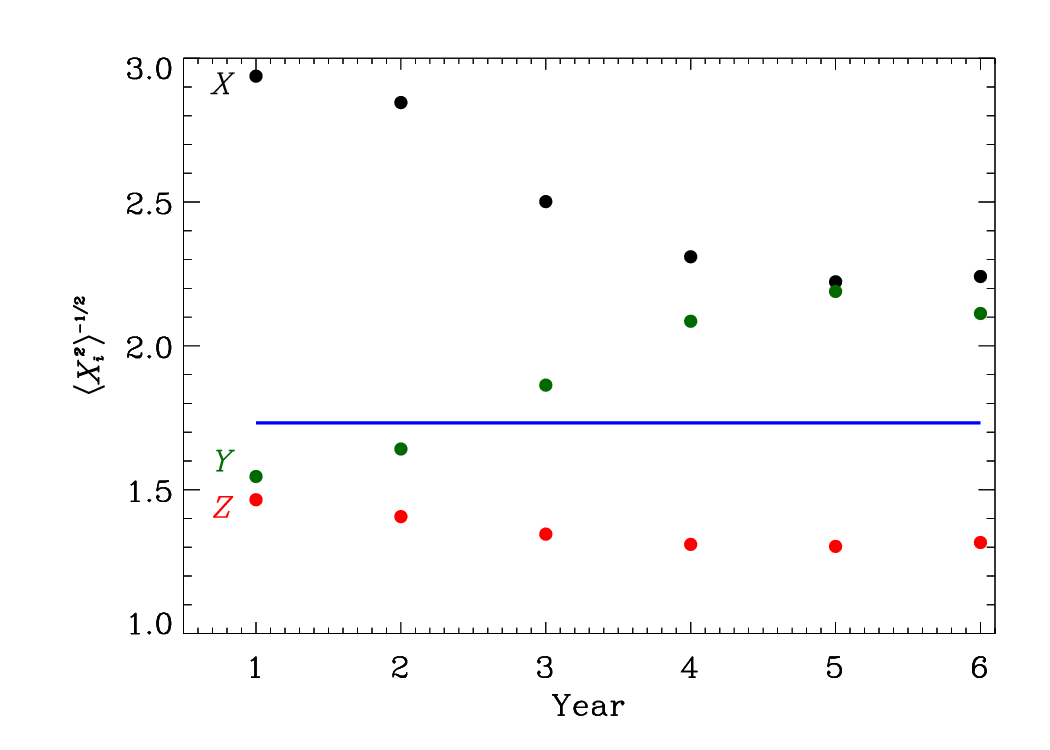} 
   \caption{
   Variation of the Poisson errors over the cumulative sky coverage of Fig.~\ref{fig:sfd0} as a function of mission years. Black, green and red correspond to the $X$, $Y$ and $Z$ direction cosines. The blue straight line shows $\langle X_i^2\rangle^{-1/2}=\sqrt{3}$, the value for a full sky coverage.
   }
   \label{fig:direction_cosines}
\end{figure}
The statistical (Poisson) errors on each dipole component are a function of the number density of galaxies and the area of the sky covered by the data. To compute these statistical uncertainties for the magnitude ranges of Eqs.~(\ref{eq:mag_range_vis}) and (\ref{eq:mag_range_nisp}) we used the number density of galaxies from the JWST counts, denoted by red asterisks in Fig.~\ref{fig:ngal_jwst}. The number densities of galaxies on an area of $\num{2550}$ deg$^2$, the size of the observed region in the first year of integration were $N_{\rm gal}=[313,270,313,362]\times 10^6$ for the \Euclid  \IE, \YE, \JE, and \HE\ filters, respectively. The uncertainty in each component of the dipole $d_{\nu,i}$ is 
\begin{equation}
    \sigma_i=\frac{1}{\sqrt{\langle X_i^2\rangle}} N_{\rm gal}^{-1/2}\,,
    \label{eq:sig_akeke}
\end{equation}where $X_i=(X,Y,Z)$ denotes the components of the dipole, $N_{\rm gal}$ is the total number of galaxies and $\langle X_i^2\rangle$ is the square of the $i$th component of the direction cosine averaged over the observed region \citep{Atrio-Barandela:2010,Kashlinsky:2022}. Note that already the Year 1 \Euclid\ data will cover significantly more sky area than combined from all the pre-COBE measurements as summed up in Table 1 of \cite{Lineweaver:1997}. Our measurements of the CIB dipole will not assume the CMB prior direction and will measure the former directly, so the CMB dipole directions are used only for illustrative purposes.

In Fig.~\ref{fig:direction_cosines} we plot $\langle X_i^2\rangle^{-1/2}=\sigma_iN_{\rm gal}^{1/2}$, which measures
the variation of the Poisson errors due to the increment of the sky coverage by mission years shown in Fig.~\ref{fig:sfd0}. Black, green and red dots
show this magnitude for the $X$, $Y$, and $Z$ direction cosines, respectively. The blue solid line shows
the same magnitude for a full sky coverage. In the first two years, the mission will observe preferentially close to the
ecliptic poles and the $Y$ and $Z$ components are reasonably well measured. As the mission progresses, the satellite will observe regions located away from the $Y$ axis and $\langle Y^2\rangle$
decreases, increasing its Poisson error. The dipole amplitude will be better sampled as shown
in Figs.~\ref{fig:confidence_levels_vis} and \ref{fig:confidence_levels_nisp} below. At the end of the mission,
the $Z$ component will have the smallest error bar since the areas near the Galactic poles will be the
regions best observed by \Euclid. A different scanning strategy will lead to different errors. 
The sky observed each year is shown in Fig.~\ref{fig:ecliptic_area_coverage} (see also \cite{Scaramella-EP1} Figs. 45 and 46).
Notice that, while the error on $Z$ is smaller than $\sqrt{3}$, the other two components are measured with
an error larger than $\sqrt{3}$ and the error on the dipole amplitude is, as expected, larger than for a full sky coverage; see \cite{Atrio-Barandela:2010} for extensive discussion.
\begin{figure*}[ht!] 
   \centering
   \includegraphics[width=\textwidth]{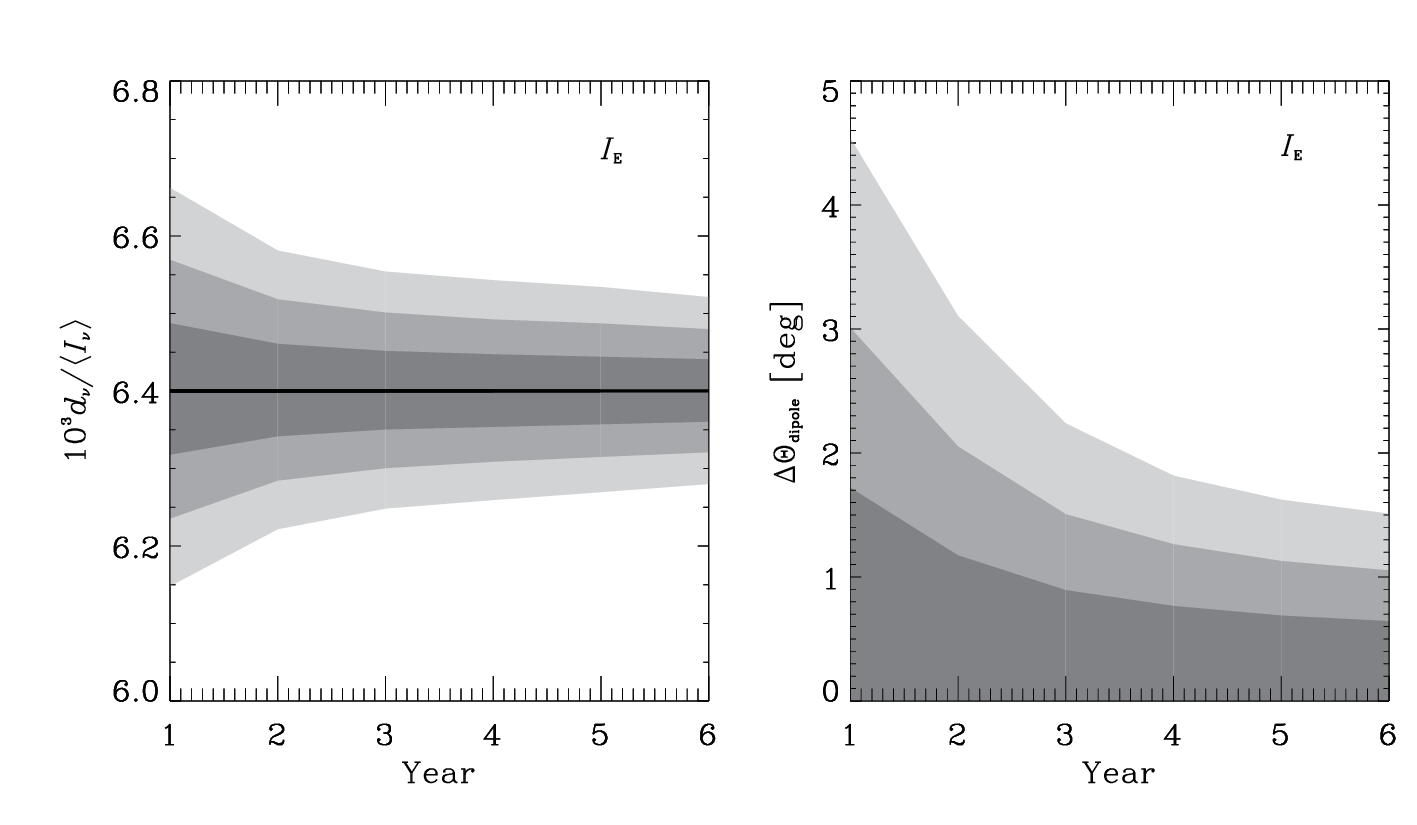} 
   \caption{
   Confidence level measured from the region of the sky after 1 to 6 years of observations with the \IE\ band.Regions of 68\% (dark grey), 95\%  (middle grey) and 99.75\% (light grey). The left panel shows the uncertainty of the dipole amplitude, with the horizontal thick line indicating the expected amplitude for a velocity of $V=370$ km s$^{-1}$. The right panel shows the uncertainty on the CMB dipole direction determination. The dipole is assumed to be in the direction of the Solar dipole, $(l,b)=(263\fdg85,48\fdg25)$ in Galactic coordinates. 
   }
   \label{fig:confidence_levels_vis}
\end{figure*}

\begin{figure*}[ht!] 
   \centering
   \includegraphics[width=\textwidth]{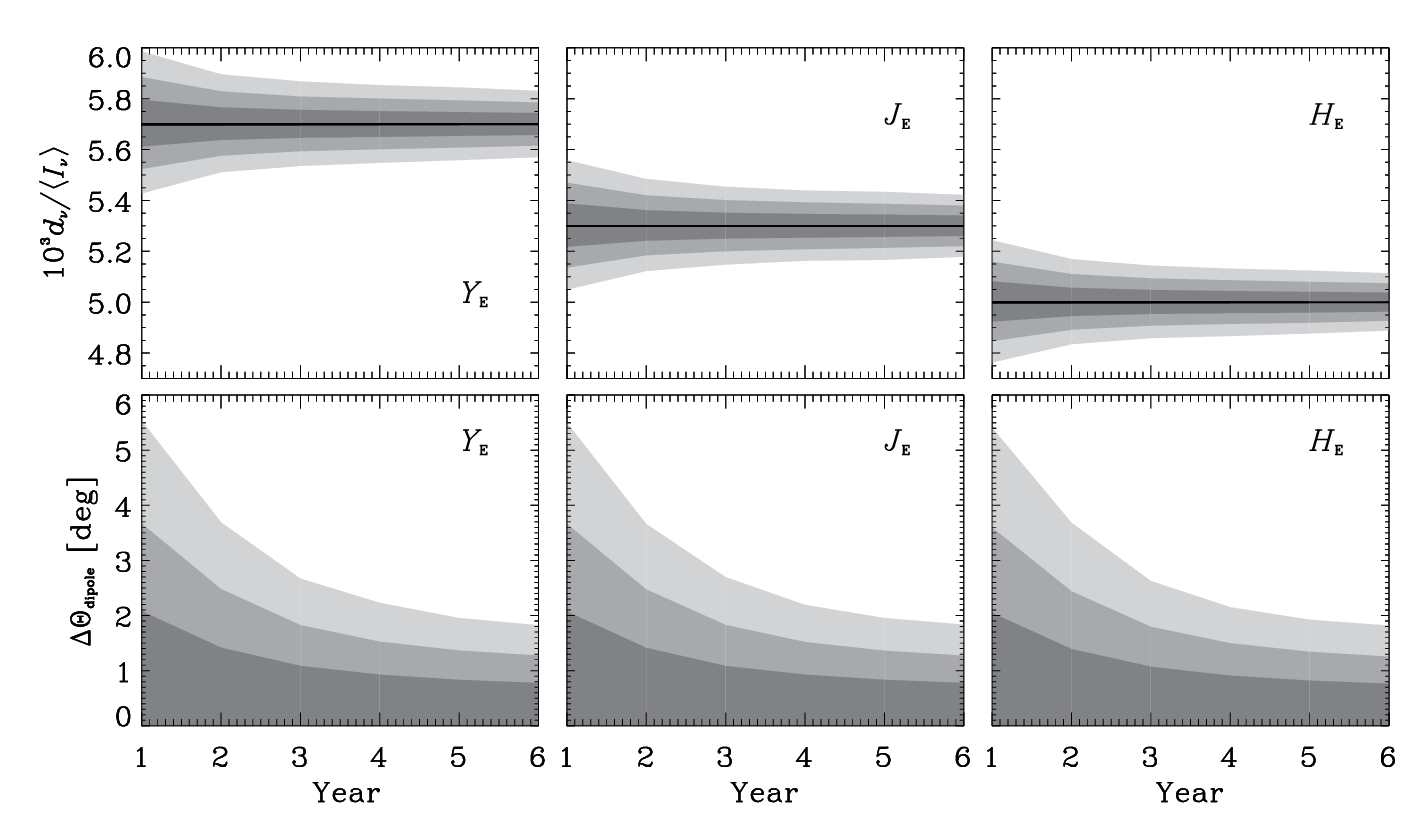} 
   \caption{
   Same as in Fig.~\ref{fig:confidence_levels_vis} for the NISP filters. In the top three panels the uncertainty of the dipole amplitude is represented, with the horizontal thick line showing the expected amplitude. The lower panels show the uncertainty on the CMB dipole direction. Left, middle and right panels correspond to the \YE, \JE\ and \HE\ NISP \Euclid filters, as indicated.
   }
   \label{fig:confidence_levels_nisp}
\end{figure*}

In Figs.~\ref{fig:confidence_levels_vis} and \ref{fig:confidence_levels_nisp} we present the expected confidence contours on the dipole amplitude and direction for the VIS and NISP Euclid Wide Survey data obtained from the Poisson errors $\sigma_i$. We assumed the dipole is in the direction of the CMB Solar dipole $(l,b)=(263\fdg85,48\fdg25)$. For each filter and year of observation we generated $10^5$ Gaussian distributed 
random errors around the CMB measured components $(d_{\nu,X},d_{\nu,Y},d_{\nu,Z})$ with rms deviations given by Eq.~(\ref{eq:sig_akeke}). We computed the random dipole amplitudes and their angular separations with respect to the CMB dipole direction. The confidence levels shown in Fig.~\ref{fig:confidence_levels_vis} were defined as the regions that enclose the 68\%, 95\%, and 99.75\%  of all simulated amplitudes and directions. The left panel displays the contours in the amplitude and the right panel in the direction. The darkest and lightest colors correspond to the 1 and 3 $\sigma$ contours, respectively. The horizontal line in the left panel represents the dipole amplitude with the Solar System moving at $V=V_{\rm CMB}=370$ km s$^{-1}$ with respect to the CMB frame.  
In Fig.~\ref{fig:confidence_levels_nisp}, the top three panels represent the dipole amplitude and  the bottom three panels the dipole direction and their  uncertainties with the same notation as in Fig.~\ref{fig:confidence_levels_vis}. The left, center and right panels correspond to the NISP filters \YE, \JE, and \HE, as indicated. The $(3-\alpha_\nu)(V/c)$ coefficients used to estimate the statistical significance are given later by Eq.~(\ref{eq:amp_final}) in Sect.~\ref{section5.3} below.

If the CMB dipole is entirely kinematic the statistical significance will be dominated by the $Z$ component as shown in Fig.~\ref{fig:direction_cosines} due to a larger number of observations close to the Galactic poles. As the figure shows, for the assumed cumulative sky coverage, the $Y$ component is more efficiently probed in the first two years, when the observations are closer to the ecliptic poles. The $X$ component will always be ill sampled and since its value is close to zero, its ${\rm S/N}$ will be always negligible.

The number of galaxies given separately by Eqs. (\ref{eq:mag_range_vis}) and
(\ref{eq:mag_range_nisp}) and the amplification factor with respect to the CMB dipole are different for each of the \Euclid bands although the overall ${\rm S/N}$ is very similar for the three NISP filters. The measured amplitude and direction on each of the filters must be consistent with the uncertainties given in Figs.~\ref{fig:confidence_levels_vis} and \ref{fig:confidence_levels_nisp}. Since the IGL dipole is purely kinematic by construction after eliminating the dipole clustering per Fig. \ref{fig:dipole_clustering}, larger differences in amplitude and direction of recovered dipole from different \Euclid bands will be an indication of possible systematics from extinction correction, star contamination, etc. For instance, extinction will be a stronger contaminant for \IE\ than for the \HE\  filter. In the following section we use a galaxy catalog obtained from the Euclid Flagship Mock Galaxy Catalogue to apply the methodology developed in the previous section for  eliminating extinction dipole contributions remaining in the Euclid Wide Survey.

The high precision measurement of the IGL/CIB dipole would allow subdividing the galaxy sample by narrow magnitude bins within the range of Eqs. (\ref{eq:mag_range_vis}) and (\ref{eq:mag_range_nisp}) in order to probe any dependence of the velocity on these parameters. Likewise, given the expected large sample of the Euclid Wide Survey galaxies with spectroscopic redshifts, one can divide the galaxies by $z$ in order to probe the behavior of $V$ in different $z$-shells.



\subsection{Uncertainties on the IGL/CIB kinematic dipole after extinction corrections}
\label{subsection5.2}


To incorporate the proposed above method for removing residual dipole from extinction corrections we used 
the Euclid Flagship Mock Galaxy Catalogue (version 2.1.10) (Castander et al., in prep.)
This catalog contains photometry in the \Euclid bands (and many other UV -- near-IR bands), and has major 
emphasis on modelling the clustering and shapes of galaxies in the range $0<z<3$, as needed for dark 
energy studies. 
This catalog does include Galactic extinction and could be used to calculate $\boldsymbol{D}_A$, but its
${\sim}\,4.8$ billion sources are distributed only over ${\sim}\,\num{5000}$ deg$^2$ in the general direction of the 
north Galactic pole. 
We used CosmoHub \citep{Carretaro:2017,Tallada:2020} to download a 1/128th fraction of the 
catalog (${\sim}\,38$ million galaxies) including photometry (with extinction applied) at (Subaru) $b$, 
(\Euclid) \IE, \YE, \JE, \HE, and (WISE) W1 and W2 bands. Ancillary parameters downloaded were the 
galactic coordinates $(l,b)$, the value of the color excess [$E(B-V)$], and the redshift ($z$) of each source. The downloaded catalog contains 38 million galaxies, a factor of ${\sim}\,1.6$ larger than expected from the observed galaxy counts and the HRK reconstruction. This is a known issue that was communicated to \Euclid by K. Helgason (private communication), but has no consequences for our goal here of testing the separation of galaxies by the resultant $\alpha_\nu$ (the logarithmic slope of the CIB with $\nu$). The catalog provides individual apparent galaxy fluxes, $F$, in each band in units of erg Hz$^{-1}$ s$^{-1}$ cm$^{-2}$, which were also converted into AB magnitudes as $m=-2.5\logten F -48.6$. We then select a conservative subset of galaxies satisfying {\it simultaneously} both Eqs. (\ref{eq:mag_range_vis}) and (\ref{eq:mag_range_nisp}).

From the downloaded catalog we removed galaxies at VIS/NISP bands $m<m_0=19/18$ and $m>m_1=24.5/24$. 
From the overall sample covering total area ${\cal A}$ sr at $m_0\leq m \leq m_1$ we computed the net CIB flux density in MJy\,sr$^{-1}$ as $I_\nu=\sum F_i/{\cal A}$ at each frequency $\nu$ and used the data to evaluate its logarithmic derivative $\alpha_\nu$ between 0.4 and 5 \micron\ for each subsequently selected subsample aiming to divide into at least two groups in Eq. (\ref{eq:v_final}). As discussed in the previous section this achieves two goals: 1) verifying the same $\boldsymbol{V}(\alpha)$ at each $\alpha_\nu$ as well as $\nu$ for every pair of sample+subsample, and 2) refining the overall $\boldsymbol{V}$ via $\chi^2$ from finer sample binning with each having more uniform $\alpha$.





We now turn to several specific examples of binning to effectively apply the method to isolate and separate the extinction term from the kinematic IGL/CIB dipole. We define the effective slope between two adjacent \Euclid wavelengths (1 and 2) for each Flagship catalog source contributing individual fluxes $F_{1,2}$ to IGL as
\begin{equation}
\beta_{1\rightarrow2}\equiv \frac{\ln (F_1/F_2)}{\ln(\lambda_1/\lambda_2)}\,.
\label{eq:source_color}
\end{equation}

In the first example we select all sources in the 2 individual subsamples where each color satisfies: 1) 
\begin{equation}
[\beta_{\IE\rightarrow \YE},\beta_{\YE\rightarrow \JE}, \beta_{\JE\rightarrow \HE}]\leq \beta
\label{eq:extinction_by_beta}
\end{equation}
for the extinction, and 2) 
\begin{equation}
[\beta_{\IE\rightarrow \YE},\beta_{\YE\rightarrow \JE}, \beta_{\JE\rightarrow \HE}]\geq \beta
\label{eq:dipole_by_beta}
\end{equation}
for the IGL dipole.
\begin{figure*}[t!] 
   \centering
   \includegraphics[width=7.1in]{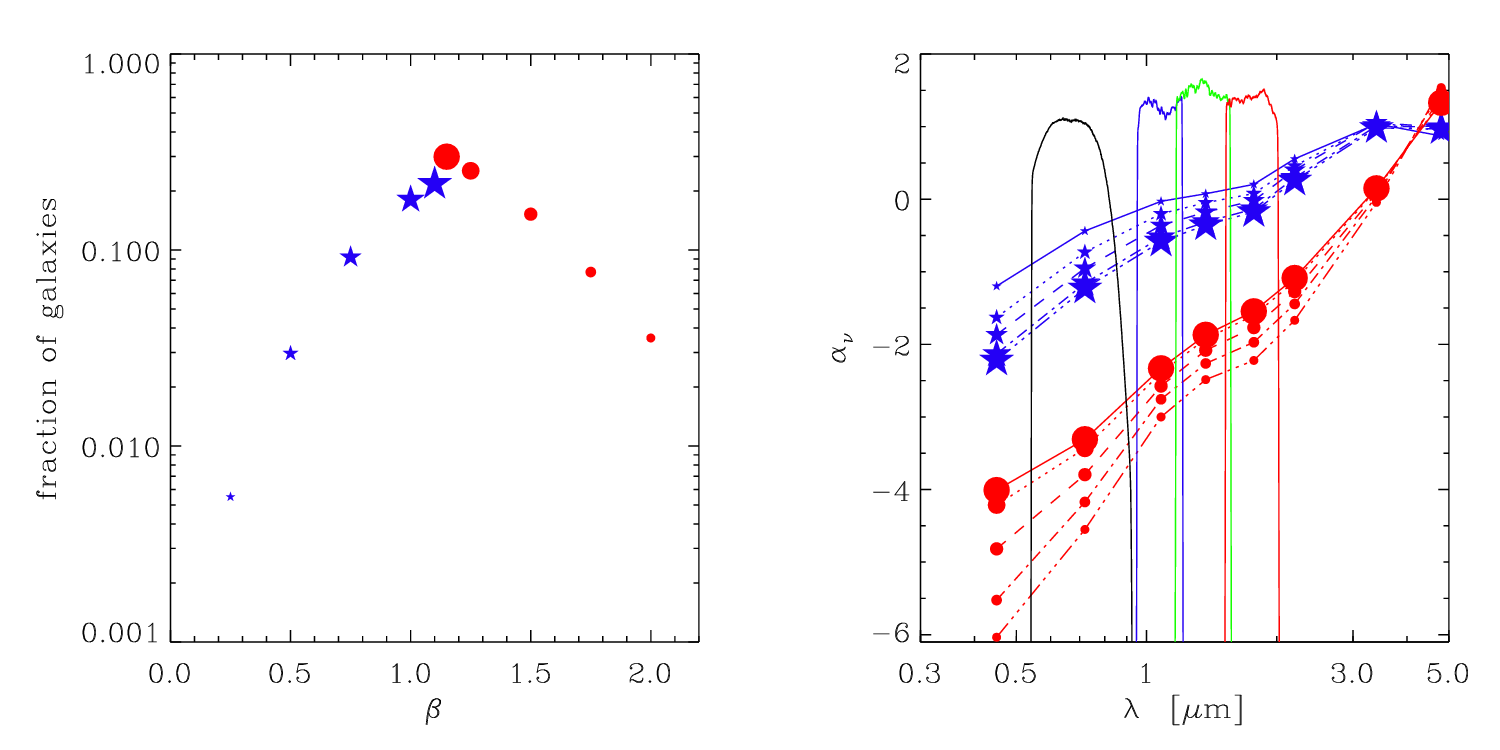} 
   \caption{Subsample size and spectral index. Blue asterisks mark selection according to Eq. (\ref{eq:extinction_by_beta}), and red circles according to Eq. (\ref{eq:dipole_by_beta}). The left panel shows the fraction of the Flagship catalog galaxies vs. $\beta$ according to Eqs. (\ref{eq:extinction_by_beta}) and (\ref{eq:dipole_by_beta}). Symbol size increases with the sample size. The right panel shows the resultant $\alpha_\nu$ vs. wavelength for the two subsamples following the notations in the left panel. The panel demonstrates the separability when Eq. (\ref{eq:v_final}) is applied at the marked \Euclid bands of \IE (black), \YE (blue), \JE (green), and \HE (red).}
   \label{fig:sec5}
\end{figure*}
Figure \ref{fig:sec5} (left) shows the fraction of galaxies vs. $\beta$ for each category in this case marked with blue asterisks for Eq. (\ref{eq:extinction_by_beta}) and red circles according to Eq. (\ref{eq:dipole_by_beta}). The plot shows that one could select samples of sufficient size in order to achieve high statistical accuracy when separating extinction contributions.
The right panel of the figure shows the corresponding $\alpha_\nu$ for the IGL from sources in each subsample. This demonstrates a clear difference between the Compton--Getting dipole amplification in the two subsample as required for good separation according to Eq. (\ref{eq:v_final}). The figure illustrates the desirable separability of IGL by $\alpha_\nu$ and show that one robustly recovers $\Delta\alpha_\nu\equiv |\alpha_\nu^a-\alpha_\nu^b|\gtrsim 2$ at the four \Euclid four bands. 

\begin{figure*}[t!] 
   \centering
   \includegraphics[width=7.1in]{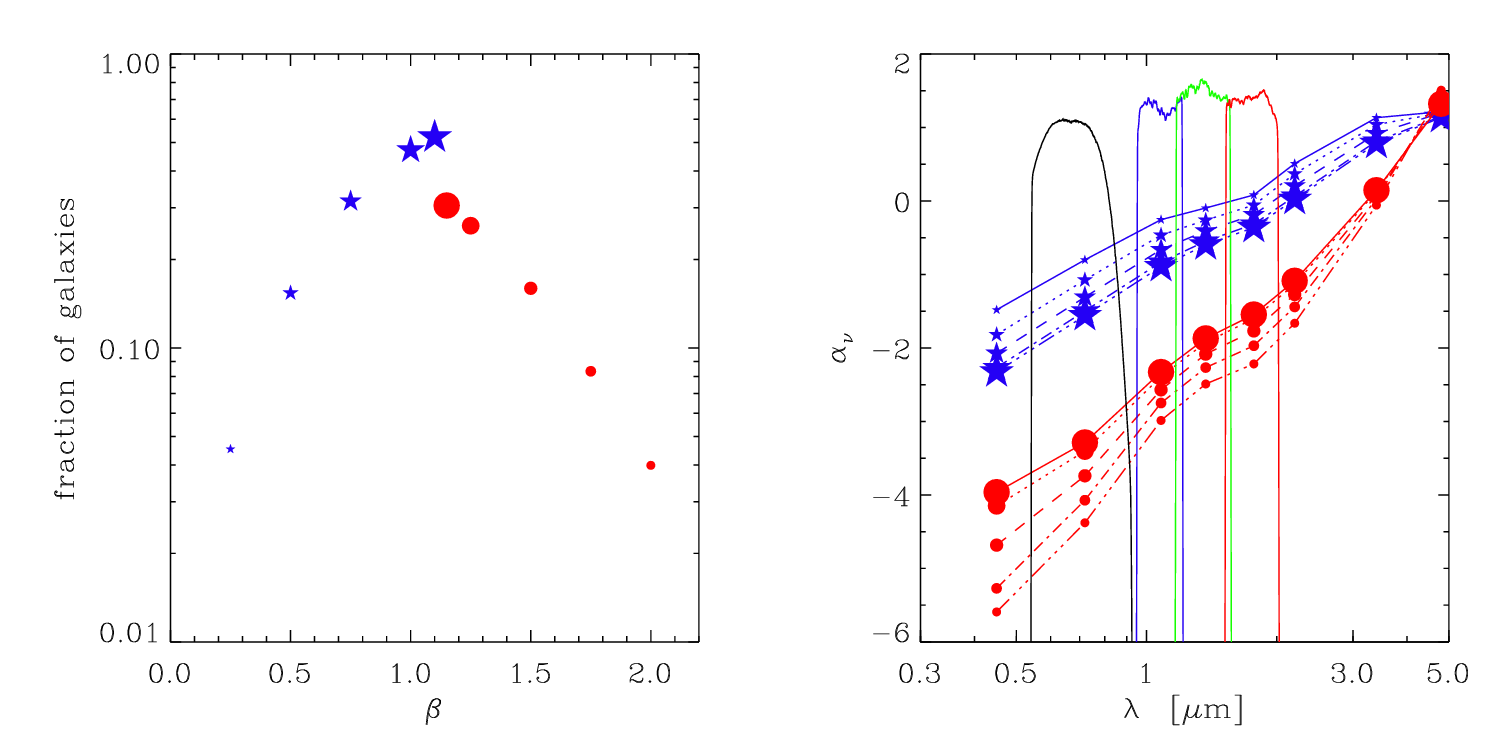} 
   \caption{Subsample size and spectral index. Blue asterisks mark selection according to Eq. (\ref{eq:extinction_by_beta_nisp}), and red circles according to Eq. (\ref{eq:dipole_by_beta_nisp}). The left panel shows the fraction of the Flagship catalog galaxies vs. $\beta$ according to Eqs. (\ref{eq:extinction_by_beta_nisp}) and (\ref{eq:dipole_by_beta_nisp}). Symbol size increases with the sample size. The right panel shows the resultant $\alpha_\nu$ vs. wavelength for the two subsamples following the notations in the left panel. The panel demonstrates the separability when Eq. (\ref{eq:v_final}) is applied at the marked \Euclid bands of \IE, \YE, \JE, and \HE.}
   \label{fig:sec5_nisp}
\end{figure*}

\begin{figure*}[t!] 
   \centering
   \includegraphics[width=7.1in]{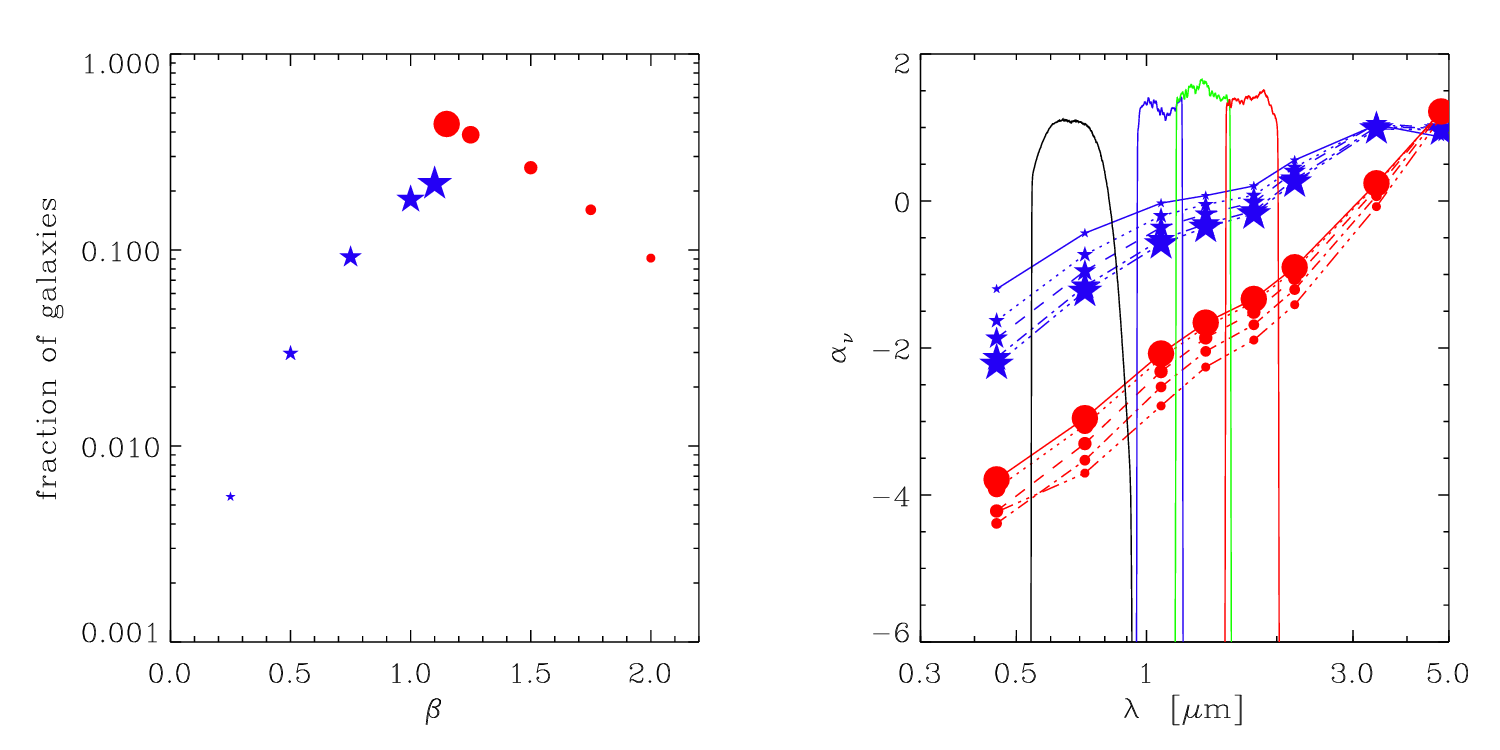} 
   \caption{Subsample size and spectral index. Blue asterisks mark selection according to Eq. (\ref{eq:extinction_by_beta_vis}), and red circles according to Eq. (\ref{eq:dipole_by_beta_vis}). The left panel shows the fraction of the Flagship catalog galaxies vs. $\beta$ according to Eqs. (\ref{eq:extinction_by_beta_vis}) and (\ref{eq:dipole_by_beta_vis}). Symbol size increases with the sample size. The right panel shows the resultant $\alpha_\nu$ vs. wavelength for the two subsamples following the notations in the left panel. The panel demonstrates the separability when Eq. (\ref{eq:v_final}) is applied at the marked \Euclid bands of \IE, \YE, \JE, and \HE.}
   \label{fig:sec5_vis}
\end{figure*}

\begin{figure*}[t!] 
   \centering
   \includegraphics[width=7.1in]{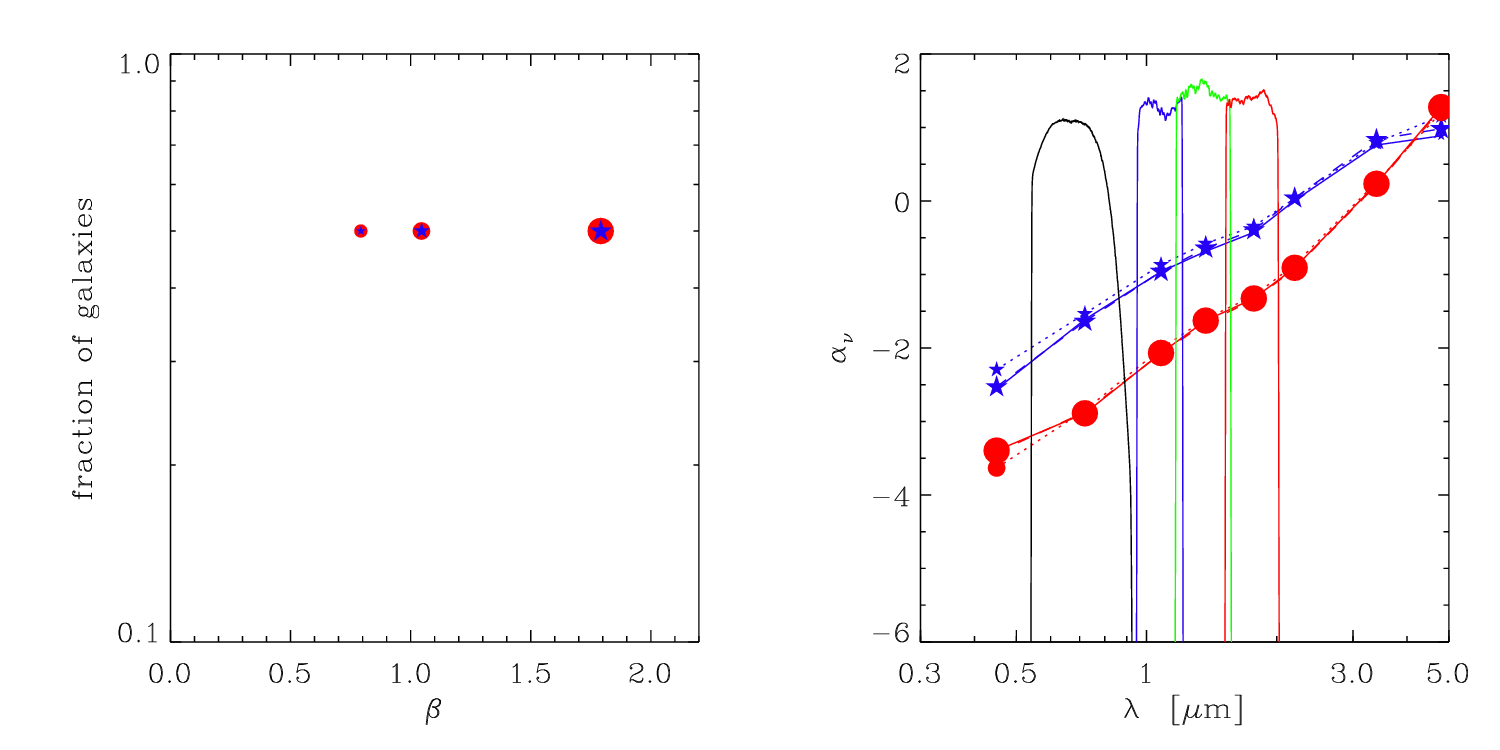} 
   \caption{Subsample size and spectral index. Blue asterisks mark selection according to Eq. (\ref{eq:extinction_by_beta_vyj}), and red circles according to Eq. (\ref{eq:dipole_by_beta_vyj}). The left panel shows the fraction of the Flagship catalog galaxies vs. $\beta$ according to Eqs. (\ref{eq:extinction_by_beta_vyj}) and (\ref{eq:dipole_by_beta_vyj}). Symbol size increases from (\IE,\YE), through (\YE,\JE), to (\JE,\HE). 
   The right panel shows the resultant $\alpha_\nu$ vs. wavelength for the two subsamples following the notations in the left panel. The panel demonstrates the separability when Eq. (\ref{eq:v_final}) is applied at the marked \Euclid bands of \IE, \YE, \JE, and \HE.}
   \label{fig:sec5_vyj}
\end{figure*}

In Figs. \ref{fig:sec5_nisp}--\ref{fig:sec5_vyj} we present similar examples using alternate criteria for subsample selection.
Subsample selections for Fig. \ref{fig:sec5_nisp} include only NISP colors (omitting the bluest, \IE, band):
\begin{eqnarray}
[\beta_{\YE\rightarrow \JE}, \beta_{\JE\rightarrow \HE}] & \leq & \beta\,,\\
\addtocounter{equation}{-1}\label{eq:extinction_by_beta_nisp}\addtocounter{equation}{1}
[\beta_{\YE\rightarrow \JE}, \beta_{\JE\rightarrow \HE}] & \geq & \beta\,.
\label{eq:dipole_by_beta_nisp}
\end{eqnarray}
Subsample selections for Fig. \ref{fig:sec5_vis} omit the reddest, \HE, band:
\begin{eqnarray}
[\beta_{\IE\rightarrow \YE},\beta_{\YE\rightarrow \JE}] & \leq & \beta\,,\\
\addtocounter{equation}{-1}\label{eq:extinction_by_beta_vis}\addtocounter{equation}{1}
[\beta_{\IE\rightarrow \YE},\beta_{\YE\rightarrow \JE}] & \geq & \beta\,.
\label{eq:dipole_by_beta_vis}
\end{eqnarray}
For Fig. \ref{fig:sec5_vyj}, we select all sources in the two individual subsamples where 
the $1\rightarrow 2$ color for each of the three pairs
(\IE,\YE), (\YE,\JE), and (\JE,\HE) {\it separately} satisfies
\begin{eqnarray}
\beta_{\rm 1\rightarrow 2} & \leq & {\rm median}(\beta_{\rm 1\rightarrow 2})\,,\\
\addtocounter{equation}{-1}\label{eq:extinction_by_beta_vyj}\addtocounter{equation}{1}
\beta_{\rm 1\rightarrow 2} & \geq & {\rm median}(\beta_{\rm 1\rightarrow 2})\,.
\label{eq:dipole_by_beta_vyj}
\end{eqnarray}

Now we can solve for $\boldsymbol{D}_A$ and $\boldsymbol{V}$ in Eqs. (\ref{eq:extinction_final}) and (\ref{eq:v_final}) and determine their uncertainties. This would be required if the magnitude of $D_A$ turns out to be non-negligible, say $D_A\gtrsim 0.1\%$. Equation (\ref{eq:v_final}) is sensitive to the difference in $\alpha$'s between the 
``(a,b)'' subsamples. Hence, to optimize the ${\rm S/N}$ for Eq. (\ref{eq:v_final}) we need to select 1) a more amplified subsample ``a'' (with more negative $\alpha_\nu^a$), then 2) select subsample ``b'' with much less negative $\alpha_\nu^b$, so ${\cal U}_{\rm b}^2\ll {\cal U}_{\rm a}^2$ (squares are since the uncertainties add in quadrature), while 3) keeping enough galaxies in the subsamples, so that 4) the final signal-to-noise ratio
\begin{equation}
     ({\rm S/N})_{\Delta_{\rm ab}} \simeq ({\rm S/N})
     \left[\frac{|\Delta\alpha_\nu|}{(3-\alpha_\nu^{\rm a+b})}\;\;\sqrt{\frac{f_{\rm a}f_{\rm b}}{f_{\rm a}+f_{\rm b}}}\;\right]
     \label{eq:sn_31}
\end{equation}
will still be high enough. In this expression, $f_{\rm a}$ and $f_{\rm b}$ are the fractions of ${\rm a}$ and ${\rm b}$ galaxies to the total number of galaxies, and $\alpha_\nu^{\rm a+b}$   being the spectral index of the IGL for the full sample, a+b.
Equation~(\ref{eq:sn_31}) shows that significant ${\rm S/N}$ can be achieved by dividing the galaxy population into  distinct samples with very different $\alpha_\nu$ and 
the figures in this section show that several of these samples are  possible.
We note that this procedure will be required only if we determine, from the actual data, that the extinction dipole, Eq. (\ref{eq:extinction_final}), is significant in {\it all the \Euclid bands}. If it turns out negligible, the method for correcting for the remaining extinction proposed here will not be required and we will proceed with the measurement per Eqs. (\ref{eq:igl}) and (\ref{eq:velocity}) directly.

\begin{figure}[ht!] 
\centering
\includegraphics[width=3.51in]{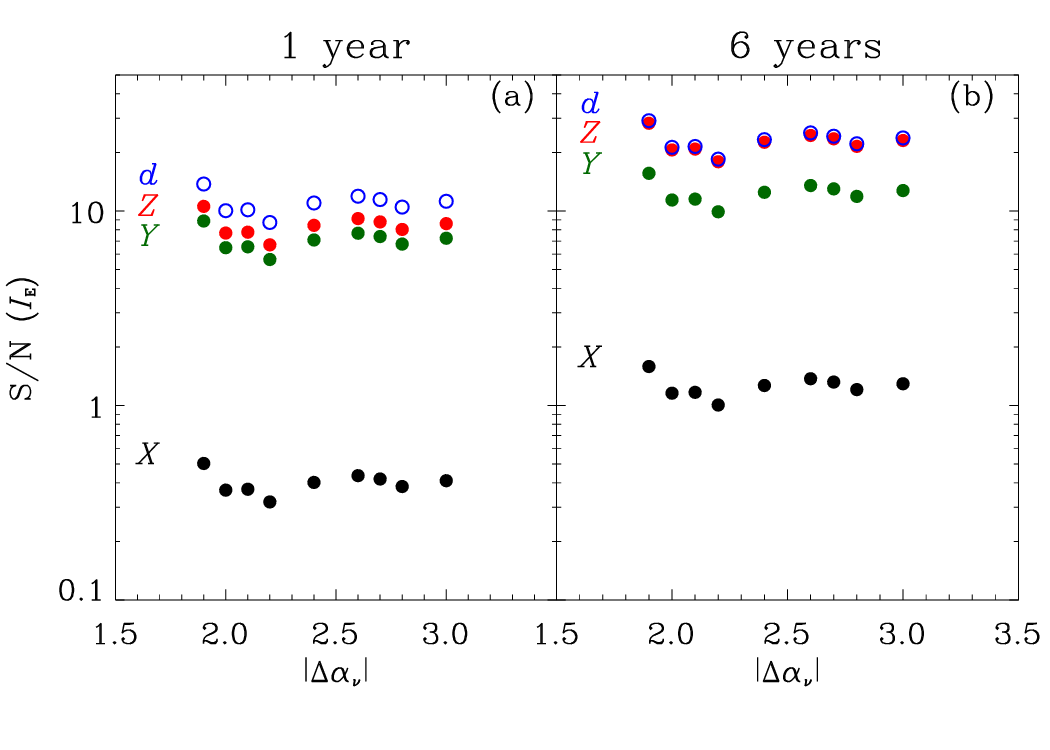} 
\caption{Statistical significance of the three dipole components and the dipole amplitude measured using
the \IE\ data and assuming the CMB dipole template, Eq. (\ref{eq:dipole_vec}). Black, green, red full circles correspond to the $X$, $Y$, and $Z$ dipole components; blue open circles to the overall dipole amplitude $d$. The left panel (a) shows the result after one year of integration and the right panel (b) at the end of the 6-year mission.
The fraction of the total number of galaxies in samples a and b are, from left to right, $f_{\rm a}=[0.5,0.2,0.1,0.2,0.2,0.2,0.15,0.1,0.2]$ and $f_{\rm b}=[0.5,0.3,0.3,0.25,0.2,0.2,0.2,0.2,0.1]$, respectively. For other configurations the ${\rm S/N}$ will increase according to Eq. (\ref{eq:sn_31}).}
\label{fig:sec5_eq31}
\end{figure}
If the extinction dipole, Eq.~(\ref{eq:extinction_final}), turns out to be important we will proceed as outlined in this section. This discussion suggests many possibilities to optimize the measurement after isolating the extinction. Equation~(\ref{eq:sn_31}) shows that this is achievable with the ${\rm S/N}$ loss of a factor of ${\sim}\,2$ if one concentrates on galaxy subsamples with $|\Delta\alpha_\nu|\sim 2$--3 while keeping the bulk of galaxies in both samples, so that $f_{\rm a}, f_{\rm b}\sim 0.3$--0.5. Additionally, the measurement in the four \Euclid bands with significantly varying extinction levels may lead to further clarity, since the \HE\ channel will have an order of magnitude lower extinction levels than \IE.

As the worst-case scenario in terms of the extinction contribution we consider the application of the presented formalism to the \IE\ band. Figure~\ref{fig:sec5_eq31} shows with full circles the ${\rm S/N}$ of the three IGL/CIB dipole components $(X,Y,Z)$ and the overall IGL/CIB dipole amplitude $d$ with open circles, for the first year and after six years of observations using the data of the \IE\ band alone and assuming $\boldsymbol{V}=\boldsymbol{V}_{\rm CMB}$. Since the dipole power follows a $\chi^2$ distribution, we define the uncertainty as half the width of the interval enclosing the 68\% confidence level. As indicated in Sect.~\ref{subsection5.1} the best measured component is always $Z$ since \Euclid will be observing preferentially around the Galactic poles, while the $Y$ component is better determined in the first two years of the mission, when the observations take place near the ecliptic poles. For the assumed dipole direction to be coincidental with the CMB dipole, the amplitude of the $X$ component is negligible so its statistical significance is always small; this would change if the observed CMB dipole has a non-kinematic component. 

Dividing the galaxy sample into two equally sized subsamples with $|\Delta\alpha_\nu|\gtrsim 2$ we can measure in \IE\ alone the amplitude of the IGL/CIB dipole with ${\rm S/N}\gtrsim 15$ after the first year and $\gtrsim 30$ at the end of the mission. After six years of observation, the $Z$ dipole component is determined about a factor of two better than  $Y$, dominating the statistical significance of the dipole amplitude, that is only a few percent better than  $Z$. Depending on the value of $|\Delta\alpha_\nu|$ four or six different samples could be constructed and optimized from the actual data, so the statistical significance could additionally increase. Similar results hold for the NISP filters, resulting in a further increment of a factor of $2$. However, the statistical significance would be ${\rm S/N}>100$ if the extinction dipole is small at least at the longest wavelengths and does not need to be subtracted off from the data. Comparison of the measured IGL/CIB dipole with and without the extinction dipole correction will further indicate the importance of this component at each frequency. Furthermore, juxtaposition between different frequencies will be a measure of systematic uncertainties. Large differences would indicate that extinction and/or other systematic uncertainties are present in the data and if such differences do not exist it will provide a strong vindication of the final result for the IGL/CIB dipole.

If the post-extinction-correction IGL/CIB maps at the four different \Euclid\ bands are found to be uncorrelated (e.g. the extinction dipoles, Eq. (\ref{eq:extinction_final}) are found to be widely different) one can potentially gain an improvement of up to a factor of 2 in the ${\rm S/N}$ over that shown in Fig. \ref{fig:sec5_eq31} by averaging over all four bands. 

To conclude this section, we have shown explicitly the many combinations that can be selected by color between the various \Euclid bands, in order to isolate the remaining extinction contribution and achieve the required IGL/CIB dipole measurement. Of course, when the data arrive, they will be dissected in more possible ways to fine-tune and optimize the measurement, and still more combinations of colors may be considered. The formalism developed here for isolating and removing the extinction contributions to the CIB dipole is mathematically precise and independent of any particular extinction model with the SFD maps used merely as an example of what the extinction dipole may look like. 

\subsection{Reducing systematic amplification uncertainties}\label{section5.3}
\label{subsection5.3}


Finally we concentrate on the systematic corrections needed to translate accurately the IGL/CIB dipole, for a well determined direction, into the corresponding velocity amplitude. The corrections below affect all 3 components of the velocity vector equally, leaving the direction intact. We will reconstruct the IGL/CIB from counts, evaluate as accurately as possible the value of $\alpha_\nu$ across the \Euclid spectrum, probe its dipole components and then deduce the effective velocity accounting for the systematics below. 

\begin{figure*}[ht!] 
   \centering
  \includegraphics[width=6.5in]{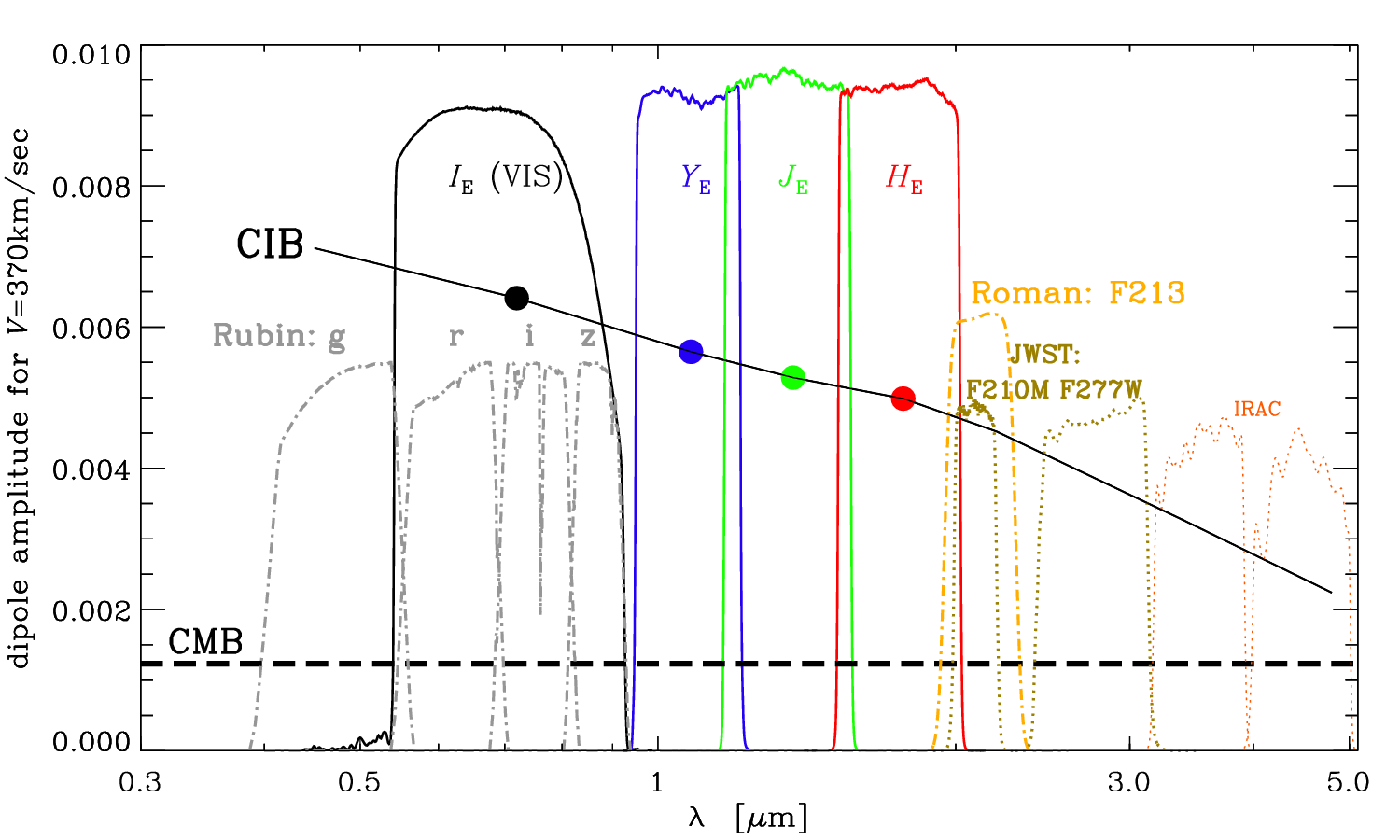} 
   \caption{Dimensionless CIB dipole. The solid line, marked ``CIB", shows the dimensionless IGL/CIB dipole from the Flagship2.1 catalog in the configuration used here combining the configuration given by Eqs. (\ref{eq:mag_range_vis}) and (\ref{eq:mag_range_nisp}). The figure assumes that $V=V_{\rm CMB}$, with the CMB dimensionless dipole marked with the thick horizontal dashed line. The \Euclid filters are shown with the filled circles showing the values at their nominal wavelengths. The {\it Roman} F213 filter covering photons just outside of the \HE\ NISP band, JWST F210M and F277W filters also near and longward of the \HE\ band, and {\it Rubin}'s $g$, $r$, $i$, and $z$ filters are also shown. With this additional coverage the value of $\alpha_\nu$ can be evaluated to good accuracy. The two shortest wavelength {\it Spitzer} IRAC filters at 3.6, 4.5 \micron\ are marked with pink dotted lines.}

   \label{fig:velocity}
\end{figure*}

Our task is to translate the measured CIB dipole into the equivalent velocity and compare to $V_{\rm CMB}=370$ km s$^{-1}$ in the precisely known direction $(l,b)_{\rm CMB}=(263\fdg85\pm0\fdg1, 48\fdg25\pm0\fdg04)$ \citep{Hinshaw:2009}. For precision measurement we must translate the measured dipole amplitude into the equivalent velocity with required accuracy when selecting galaxies with $m\geq m_0$ (to eliminate the contribution to dipole from galaxy clustering). Figure \ref{fig:velocity} shows the dimensionless Compton--Getting amplified IGL dipole amplitude for $V=370$ km s$^{-1}$ expected for Flagship 2.1 galaxies satisfying simultaneously Eqs. (\ref{eq:mag_range_vis}) and (\ref{eq:mag_range_nisp}). Specifically, if the CMB dipole is purely kinematic, one would recover at the four \Euclid bands, \IE, \YE, \JE, and \HE
\begin{equation}
    (3-\alpha_\nu) \frac{V_{\rm CMB}}{c} = [6.4, 5.7, 5.3, 5.0]\times 10^{-3}\,.
\label{eq:amp_final}
\end{equation}

The systematic correction, $\Delta\alpha_\nu$ discussed in Sect.~4, if uncorrected for would lead to velocity amplitude difference of $\Delta V/V\simeq \Delta\alpha_\nu/(3-\alpha_\nu)$, or 
\begin{equation}
\frac{\Delta V}{V}=\frac{Q_\nu(m_1)[1-\langle\beta(m_1)\rangle]-Q_\nu(m_0)[1-\langle\beta(m_0)\rangle]}{(3-\alpha_\nu)}\,.
\label{eq:dV}
\end{equation}
For the Flagship2.1 catalog the relative magnitude of this systematic correction, $\Delta V/V$, was evaluated to be $[0.04,0.05,0.07,0.17]$ in the [\IE, \YE, \JE, \HE] bands and is less than or comparable in magnitude to the correction due to Earth's orbital velocity. It does not affect the direction of the dipole and when converting the measured IGL/CIB dipole into  the equivalent amplitude for the velocity, the magnitude of $\Delta\alpha_\nu$ will be evaluated for each band directly from the \Euclid galaxy data and incorporated into the overall amplification per Eqs. (\ref{eq:dipole_final}) and (\ref{eq:dV}).
Moreover, we can gauge the effects of $\alpha_\nu$ by considering the galaxy SED estimated by the photometric redshift fitting.

The derived $\alpha_\nu$ may not be highly precise given the sparsity of points from \Euclid's four bands alone, particularly at \IE\ and \HE. To more accurately compute $\alpha_\nu$, Eq. (\ref{eq:alpha}), in each of the four \Euclid bands it would be useful to assemble a subset of the Euclid Wide Survey galaxies at wavelengths shorter than \IE\ and longer than \HE\ as e.g. shown in the figures in Sect. \ref{subsection5.2}. From such a subsample we would evaluate the net IGL over the range of \Euclid-selected magnitudes, Eqs. (\ref{eq:mag_range_vis}) and (\ref{eq:mag_range_nisp}), and then evaluate its logarithmic derivative at \IE\ and \HE. This can be achieved by using 
$ugriz$ data from various ground-based surveys that are providing complementary short wavelength data \citep[see][]{Scaramella-EP1}, 
and using past {\it Spitzer} and future {\it Roman} data at longer wavelengths.
The additional {\it Roman} data will be particularly useful here for accurately evaluating the Compton--Getting amplification at the \HE\ band. The data will be collected, even if over smaller area as currently envisaged, at F184 and F213 bands to magnitudes much deeper than the Euclid Wide Survey resulting in very rich sample of galaxies in the range of Eqs. (\ref{eq:mag_range_vis}) and (\ref{eq:mag_range_nisp}) as discussed in \cite{Akeson:2019}.




The compilation of multiband galaxy photometry in the COSMOS field by the COSMOS2020 team \citep{Weaver:2022} can be used to measure typical values for $\alpha_\nu$ in the \IE\ and \HE\ bands.  The COSMOS2020 compilation includes very deep multiband 
$YJHK_s$ imaging reaching $m>25$, $grizy$ photometry to 26--27 mag, and 3.6 and 4.5\,$\mu$m to $m>25.5$ covering the
2\,deg$^2$ COSMOS field.  The COSMOS $K_s$ imaging has been homogenized and is considerably deeper than the 
24\,mag requirement over the full field.  All the COSMOS2020 imaging is registered astrometrically to Gaia precision.  Given the combination of sensitivity and 
precision available for the publicly available COSMOS2020 catalog, we will easily be able to 
construct galaxy counts for of order \num{200000} galaxies detected in $J$ band, reaching 
24\,mag in the $J$, $K_s$, 3.6\,$\mu$m, and 
4.5\,$\mu$m bands \citep[Fig. \,11 of][]{Weaver:2022}. We can go similarly deep blueward of the \IE\ band, allowing us to make a high-precision estimate of $\alpha_\nu$ in the \Euclid \HE\ and \IE\ bands. The data from {\it Spitzer} at 3.6 and 4.5 \micron\ from e.g. \cite{Ashby:2013,Ashby:2018} would be further useful in refining the high(er)-precision evaluation of $\alpha_\nu$.


It would be sufficient to collect photometric data shortward and longward of the four \Euclid bands for only a small fraction of the \Euclid galaxies in the range covered by Eqs. (\ref{eq:mag_range_vis}) and (\ref{eq:mag_range_nisp}) to robustly probe the IGL and evaluate its $\alpha_\nu$ at the \Euclid filters \IE\ and \HE.  The additional filters particularly useful here are shown in Fig. \ref{fig:velocity}. Longward of \HE\ is the {\it Roman} F213 filter which probes emissions just outside of 2 \micron. The required magnitude coverage needed here is well within the {\it Roman} planned program currently scheduled to begin in late 2027, which nominally goes much deeper than \Euclid. Moreover the ongoing and future JWST surveys using its available NIRCam filters with central wavelengths between 2 and 3 \micron\ would provide suitable data for such calibration. The already completed observing JWST program using (among others) the F277W filter provides data on galaxies to $m\gtrsim 27$ over almost $\num{2000}$ arcmin$^2$ in the COSMOS field area \citep{Casey:2023} and an additional JWST observing program used the F210M filter over 10 arcmin$^2$ integrating to $m\gtrsim 28$--$29$ \citep{Williams:2023}; the data are already public. The {\it Rubin} $g$ band will add measurements shortward of \IE. At the same time, the additional $r$, $i$, and $z$ {\it Rubin} bands \citep{Ivezic:2019} shown here will add photometric measurements of the \Euclid galaxies at narrower intervals than the \Euclid \IE\ channel which will allow finer reconstruction of the IGL/CIB with wavelength and better accuracy in determining $\alpha_\nu$. {\it Spitzer} IRAC selected observations at 3.6 and 4.5 \micron\ present additional galaxy data that would be available for this task with galaxy samples going to sufficiently deep magnitudes, $m_{\rm AB}> 25$, from the various observing programs \citep{Ashby:2013,Ashby:2015,Labbe:2015} in the areas of the sky overlapping with the Euclid Wide Survey \citep[see Figs. 3 and 4 in][]{Moneti-EP17}. The advantage of the {\it Spitzer} IRAC galaxy data at 3.6 and 4.5 \micron\ is the negligible extinction compared to the shorter wavelengths, but the disadvantage is the larger separation in wavelength from the longest \Euclid NISP band as shown in Fig. \ref{fig:velocity} and the counts confusion at $m_{\rm AB}\gtrsim 21-22$ by the IRAC beam \citep{Fazio:2004}.

The task of probing the IGL/CIB dipole at high statistical significance with the Euclid Wide Survey will be accomplished in the first 1 (if the extinction corrections prove negligible in the dipole evaluation) to 2 years of the survey's start, which will happen in early 2024. Then the IGL dipole can be converted into the well determined velocity amplitude in the measured -- from the IGL/CIB dipole -- direction using the auxiliary data supplementing the \Euclid galaxies on both ends of the \Euclid bands. For this an additional sample of galaxies will be put together of much smaller $N_{\rm gal}$ to determine the $\alpha_\nu$ at each of the \Euclid bands including VIS and \HE. This can be done quickly using at most several square degrees from the {\it Rubin} and {\it Roman} measurements, which will become operational by that stage.

Additional uncertainty may arise from the cosmic variance effects of order a few percent due to clustering as shown in Fig. \ref{fig:mag24_histo_and_fit.pdf}. Although small, this may be reduced further by using auxiliary data at complementary wavelengths over small joint areas.

The effects of extinction corrections would not be significant when evaluating $\alpha_\nu$, which is required to translate the measured IGL dipole into the equivalent velocity $V$. Indeed the sky areas of relevance here have extinction $A_V < 0.1$ as shown in Fig. \ref{fig:extinction1} (left) and it would presumably be much smaller after extinction corrections are applied. While the extinction (correction) effects of order a few percent may be important for probing the IGL/CIB dipole of order ${\sim}\,0.5(V/V_{\rm CMB})\%$, here they would introduce a systematic correction in the Compton--Getting amplification, $\alpha_\nu$, of order $\simeq \epsilon_{A}$, which would affect the measured velocity amplitude (not direction, since $\alpha_\nu$ is the same for each velocity component) at the similar level of at most a few percent. For the COSMOS2020 area of $\simeq 1.82$ deg$^2$ one finds from the SFD maps that extinction is $A_V = 0.060\pm0.004$ with maximal/minimal values of 0.071/0.049. This would be about an order of magnitude lower at the NISP bands as shown in Fig. \ref{fig:extinction2} (left).  Thus, even with minimal extinction corrections, the systematic effects from the remaining extinction effects are expected to be less than a few km/sec for reasonable values of $V$.




Spectroscopic redshifts will be available for over $3\times10^7$ emission-line galaxies over the course of the mission. These will provide further help in reducing the systematics discussed here. 

After the velocity is well measured in both amplitude and direction, one would convert to the truly extragalactic frame by subtracting the well-known Sun's velocity around the Galaxy \citep{Kerr:1986}; this can be done as per Table 3 of \cite{Kogut:1993}.

\section{Summing up}
\label{section6}

In this paper we have presented the detailed tools and methodology required to probe at high precision the fully kinematic nature of the long-known CMB dipole with the Euclid Wide Survey. The method is based on measuring the Compton--Getting amplified IGL/CIB dipole as has been proposed recently \citep{Kashlinsky:2022}. This methodology will be applied to the forthcoming \Euclid data in the course of the NIRBADE and, as shown here, will measure the IGL dipole at high precision and identify any non-kinematic CMB dipole component. 

In this preparatory study we have identified the steps needed for the measurement to be done at high precision and the ways to eliminate the systematics that may potentially affect the results. The range of galaxy magnitudes to include in the final samples at each \Euclid band was determined from the requirement that the remaining clustering dipole be negligible. We then discussed the requirements from the star--galaxy separation in order to eliminate the Galactic star contribution to the measured signal. Extinction corrections, which are the largest at \IE\ and smallest at \HE, may present an additional obstacle and we designed a practical method to eliminate the extinction contributions and discuss the effects on the signal-to-noise ratio for the deduced IGL/CIB dipole. Additional systematics has been addressed together with ways for its elimination/reduction. 

We then evaluated the final results from the simulated Euclid Flagship2.1 catalog. First we do that for the overall data assuming that the a priori unknown extinction correction contribution turns out to be negligible, followed by applying the designed extinction separation method to the simulated catalog and showing the good efficiency of the proposed methodology. Finally we have addressed and quantified the additional amplification corrections required to convert the measured IGL/CIB dipole into the velocity amplitude. 

This study shows the excellent prospects for the high-precision probe by NIRBADE of the IGL/CIB dipole with the Euclid Wide Survey using the techniques developed here. Additionally,  such samples would enable us to bin galaxies by redshift enabling to probe the dependence of the measured velocity on cosmological distance.


Additional important developments for NIRBADE will come from {\it Roman}, currently scheduled for launch in late 2027. The extinction and systematics with {\it Roman} will be different and will provide a consistency check. {\it Roman}'s addition, if properly done, will increase the precision aspect of NIRBADE even further. However, a separate study is required to optimize {\it Roman}'s measurements for this experiment. The significant advantages will stem from 1) {\it Roman}'s longer wavelength filters, where extinction is substantially lower than in the \JE\ and \HE\ bands, and 2) {\it Roman}'s planned integrating to much fainter magnitudes ($m {\sim}\,26$) and hence more galaxies per square degree. On the other hand, {\it Roman} is currently planned to cover a substantially lower area of the sky of $\simeq \num{2000}$ ${\rm deg}^2$ in only the Southern hemisphere, although plans to extend the area are under consideration \citep{Akeson:2019}. The addition of the sky coverage, if done properly (see Sect. \ref{section5}), will be paramount for this measurement.

\begin{acknowledgements}
Work by A.K. and R.G.A. was supported by NASA under award number 80GSFC21M0002. Support from NASA/12-EUCLID11-0003 “LIBRAE: Looking at Infrared Background Radiation Anisotropies with Euclid” project is acknowledged. F. A.-B. acknowledges financial support from grant PID2021-122938NB-I00 funded by MCIN/AEI/10.13039/501100011033 and by “ERDF A way of making Europe” and SA083P17 from the Junta de Castilla y Le\'on. 
CosmoHub has been developed by the Port d'Informaci\'o Científica (PIC), maintained through a collaboration of the Institut de Física d'Altes Energies (IFAE) and the Centro de Investigaciones Energ\'eticas, Medioambientales y Tecnol\'ogicas (CIEMAT) and the Institute of Space Sciences (CSIC \& IEEC), and was partially funded by the ``Plan Estatal de Investigaci\'on Científica y T\'ecnica y de Innovaci\'on'' program of the Spanish government.
  
\AckEC  
\end{acknowledgements}

%
%

\bibliography{references,Euclid}

\begin{appendix}
\label{appendix}
\onecolumn
\section{The Compton--Getting effect for cosmic backgrounds}

The derivation of the Compton-Getting effect is elegantly, although implicitly, presented in the one-page paper by \cite{Peebles:1968} devoted to the then recently discovered CMB. In the notation used there, their photon number density, $n_\nu$, is directly proportional to the surface brightness intensity $I_\nu\propto \nu n_\nu$ used by us. Hence their Eq. (7) explicitly demonstrating the Lorentz invariance of $n_\nu/\nu^2$ is equivalent to the Lorentz invariance of our $I_\nu/\nu^3$ used in this paper.

\begin{figure*}[bh] 
   \centering
  \includegraphics[width=6.5in]{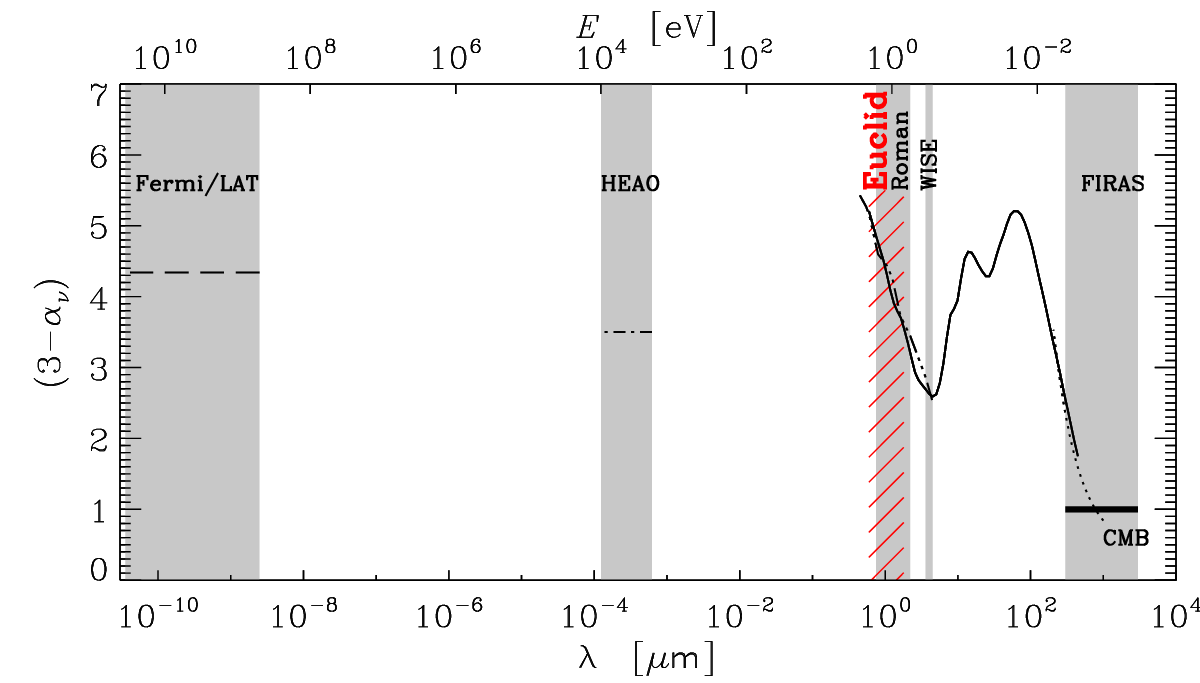} 
   \caption{The Compton-Getting dipole amplification, $(3-\alpha_\nu)$ shown for various wavelength data marked on the horizontal axis. For high-energy data the upper horizontal axis marks the corresponding energy, $E$. From left to right: dashes mark the amplification for the Fermi-LAT data, dashed-dotted line for the HEAO X-ray data, solid line for the mid- and near-IR data, dashed-triple-dotted line for the near-IR IGL at $m_0\geq 18$ and the dotted line shows the amplification the far-IR CIB determined from FIRAS. See text for details. The range of wavelengths covered by each line is marked with shadow rectangles. \Euclid is marked in red being the subject here.}
   \label{fig:appendix}
\end{figure*}

Fig. \ref{fig:appendix} shows the Compton-Getting amplification for cosmic backgrounds encompassing the wavelengths from mm to GeV energies. These were evaluated for the near-IR IGL/CIB using reconstruction from \cite{Helgason:2012} (dashed-triple-dotted line) which is consistent with the flux from the integrated counts of \cite{Driver:2016} (solid line) over the corresponding range, for the mid-IR CIB using the counts integration from \cite{Driver:2016}, in the far-IR the results of the CIB FIRAS analysis from \cite{Fixsen:1998} (dotted line), for the X-ray background (HEAO) from \cite{Boldt:1987}, and for the $\gamma$-ray Fermi/LAT background using observations from \cite{Ackermann:2015}. The figure shows the optimal windows where the amplification factor is $(3-\alpha_\nu)\sim 4-5.5$ over the CMB. However, in the mid- to far-IR the kinematic dipole is overwhelmed by Galactic foregrounds \citep[e.g.][]{Fixsen:2011} and in the Fermi-LAT range it turns out being  dominated by another component \citep{Kashlinsky:2024} possibly connected to the UHECRs observed by the Pierre-Auger Observatory \citep{Aab:2017}. This leaves the windows probed by \Euclid and {\it Roman}, where the analysis developed here applies, with the possible exception of X-rays. The figure shows the uniqueness of the \Euclid-Roman configurations in achieving the unprecedented high $S/N$ in the Compton-Getting probe for two reasons: 1) The dipole signal amplitude is amplified by a significant factor of $\sim(4-5.5)$, and 2) the statistical uncertainty in the measurement is greatly reduced by the overwhelmingly large galaxy samples to be used for the measurement.

The Compton-Getting \citep{Compton:1935} effect for diffuse backgrounds must be distinguished from the relativistic aberration effect proposed five decades later by \cite{Ellis:1984} for source counts of sources that have a uniform flux threshold and also well defined and homogeneous, and uniform, spectral properties across the considered sky. The magnitude of the relativistic aberration effect depends on the source counts' slope and their spectral index being {\it uniform} across the sky and known. It was applied to the appropriately suitable WISE and radio sources achieving ${\rm S/N}\sim 4$--$5$, which is statistically significant, but modest compared to what is planned here. Such ${\rm S/N}$ leads to angular uncertainties, Eq. (\ref{eq:direction}), of $\Delta\Theta\sim15^\circ$--$20^\circ$, clearly insufficient for the high precision purposes here. Also, the sources at \Euclid\ bands, with different morphologies, epochs and histories, have widely varying spectral properties as shown in Figs. \ref{fig:sec5}, \ref{fig:sec5_nisp}, \ref{fig:sec5_vis}, and \ref{fig:sec5_vyj}.

Similarly, the other currently suggested methods, such as probing the non-zero off-diagonal correlations between the CMB multipole moments at $\ell>2$ as proposed by \cite{Kosowsky:2011}, reach comparably low $S/N\sim 3-4$ with the subsequently poor directional accuracy \citep{Planck-Collaboration:2014a}, and a still larger directional uncertainty is achieved in the methodology later applied in \cite{Ferreira:2021}. These methods also do not appear to allow for high precision probe of any meaningfully interesting non-kinematic CMB dipole component.

\twocolumn

\end{appendix}

%

%
%
%
%
\end{document}